\documentclass[reprint,aps,prl,twocolumn,balancelastpage,a4paper,floatfix,superscriptaddress]{revtex4-2}

\usepackage[margin=2cm]{geometry}
\usepackage[T1]{fontenc}
\usepackage{graphicx}
\usepackage{kotex}
\usepackage{float}
\usepackage{amsmath,amssymb,bm}
\usepackage{xcolor}
\usepackage{booktabs}
\usepackage{enumitem}
\usepackage{dcolumn}% Align table columns on decimal point
\usepackage{hyperref}% add hypertext capabilities
\usepackage{hycolor}
\usepackage{setspace}
\hypersetup{colorlinks=true,citecolor=[rgb]{0,0.15,0.60}, 
urlcolor=[rgb]{0,0.15,0.60}, linkcolor=[rgb]{0,0.15,0.60}, final = true}
\usepackage[rightcaption]{sidecap}
\usepackage{physics}
\usepackage{mathtools}
\usepackage{relsize}
\usepackage{times}
\usepackage{siunitx}
\usepackage[scaled=0.86]{helvet}
\usepackage{setspace}
\setcitestyle{round,super}

  \newcommand{\ie}{\textit{i.e.}}
  \newcommand{\eg}{\textit{e.g.}}
  \def\fref#1{Fig.~\ref{#1}}
  \def\sb#1{\textbf{\textsf{#1}}}
  \def\nsb#1{\noindent\textbf{\textsf{#1~}}}

  \definecolor{YKB}{rgb}{0.00,0.18,0.65}

\def\Pi{\mathcal{P}}
\def\Rh{\mathcal{R}}

\def\rD{\textrm{Duration}}

\def\rDr{\textrm{Duration-ratio}}

\def\pM{\textrm{Pitch}}
\def\pC{\textrm{Chroma}}

\def\pMi{\textrm{Melodic-Interval}}

\def\AL{\mathcal{A}}
\def\HH{\mathcal{H}}
\def\TI{\mathcal{T}}
\def\GI{\mathcal{G}}
\def\OC{\mathcal{O}}

\def\MI{\mathcal{I}}
\def\IC{\mathcal{IC}}
\def\LL{\mathcal{L}}

\def\LN{\LL_{\textrm{NR}}}

\begin{document}

\title{
\includegraphics[width=0.33\textwidth]{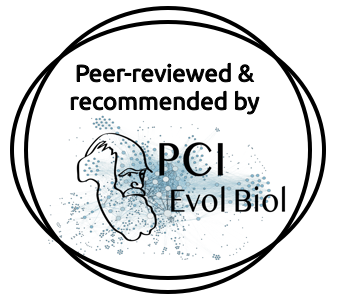}~ 
\\[1cm]
\sb{\larger[1] Information and motor constraints shape melodic diversity across cultures}
}

% \author[1,*]{John M. McBride}
% \author[1,2,*]{Tsvi Tlusty}

% \affil[1]{Center for Soft and Living Matter, Institute for Basic Science, Ulsan 44919, South Korea}
% \affil[2]{Departments of Physics and Chemistry, Ulsan National Institute of Science and Technology, Ulsan 44919, South Korea}
% \affil[*]{jmmcbride@protonmail.com, tsvitlusty@gmail.com}

  \author{John M. McBride}
    \affiliation{Center for Algorithmic and Robotized Synthesis, Institute for Basic Science, Ulsan 44919, South Korea}
    \email{jmmcbride@protonmail.com}
  \author{Nahie Kim}
    \affiliation{School of Business Administration, Ulsan National Institute of Science and Technology, Ulsan 44919, South Korea}
  \author{Yuri Nishikawa}
    \affiliation{Department of Molecular Life Science, Tokai University School of Medicine, Kanagawa, Japan}
  \author{Mekhmed Saadakeev}
    \affiliation{Department of Biomedical Engineering, Ulsan National Institute of Science and Technology, Ulsan 44919, South Korea}
  \author{Marcus Pearce} 
    \affiliation{Cognitive Science Research Group, School of Electronic Engineering \& Computer Science, Queen Mary University of London, London, United Kingdom}
    \affiliation{Department of Clinical Medicine, Aarhus University, Aarhus, Denmark}
  \author{Tsvi Tlusty}
    \affiliation{Center for Soft and Living Matter, Institute for Basic Science, Ulsan 44919, South Korea}
    \affiliation{Departments of Physics and Chemistry, Ulsan National Institute of Science and Technology, Ulsan 44919, South Korea}
    \email{tsvitlusty@gmail.com}

\begin{abstract}
%\textsf{
%{\setstretch{1.0}
%{\larger[0.5]
  \noindent The number of possible melodies is unfathomably large, yet despite this virtually unlimited potential for melodic variation, melodies from different societies can be surprisingly similar. The motor constraint hypothesis accounts for certain similarities, such as scalar motion and contour shape, but not for other major common features, such as repetition, song length, and scale size. Here we investigate the role of information constraints in shaping these hallmarks of melodies. We measure determinants of information rate in \num{62} corpora of Folk melodies spanning several continents, finding multiple trade-offs that all act to constrain the information rate across societies. By contrast, \num{39} corpora of Art music from Europe (including Turkey) show longer, more complex  melodies, and increased complexity over time, suggesting different cultural-evolutionary selection pressures in Art and Folk music, possibly due to the use of written versus oral transmission. Our parameter-free model predicts the empirical scale degree distribution using information constraints on scalar motion, melody length, and, most importantly, information rate. These results provide strong evidence that information constraints during cultural transmission of music limit the number of notes in a scale, and suggests that a tendency for intermediate melodic complexity reflects a fundamental constraint on the cultural evolution of melody.
  \\
  \\ This article was reviewed and recommended by Peer Community in Evolutionary Biology.~\cite{mesoudiRecommendation2025}
%}}}
\end{abstract}

\maketitle

%\section*{\sb{Significance Statement}}
%Melodies across cultures share common features such as a limited melodic range and length, preferential use of small intervals and arched contours, repetition, and the tendency to use scales with about 6 degrees. To add to this list, we have discovered a cross-cultural preference for melodies of intermediate complexity. We assimilate this preference with constraints on scalar motion and melody length into a model of melodies, which correctly predicts the number of scale degrees found in melodies.

\section*{\sb{Introduction}}

  \noindent Music is a fundamental component of cultures worldwide, fulfilling important social and individual functions.~\cite{brownUniversals2013,savageStatistical2015,mehrUniversality2019} Melody is a cross-culturally prominent
  characteristic of music and can be described as a sequence of sounds whose pitch and timing is drawn
  from a limited set (we call this set an alphabet) of pitches and durations, just as words in written English 
  consist of sequences of letters.~\cite{lerdahlGenerative1996} The space of possible melodies is
  uncountably vast, since it scales with melody length, $L$, and alphabet size $\AL$,
  as $\AL^{L}$. For example, counting only 10-note melodies in the major scale
  with the simplest isochronous rhythm ($\AL=7$) amounts to over 250 million unique melodies.
 
\begin{figure*}
\centering
\includegraphics[width=0.95\textwidth]{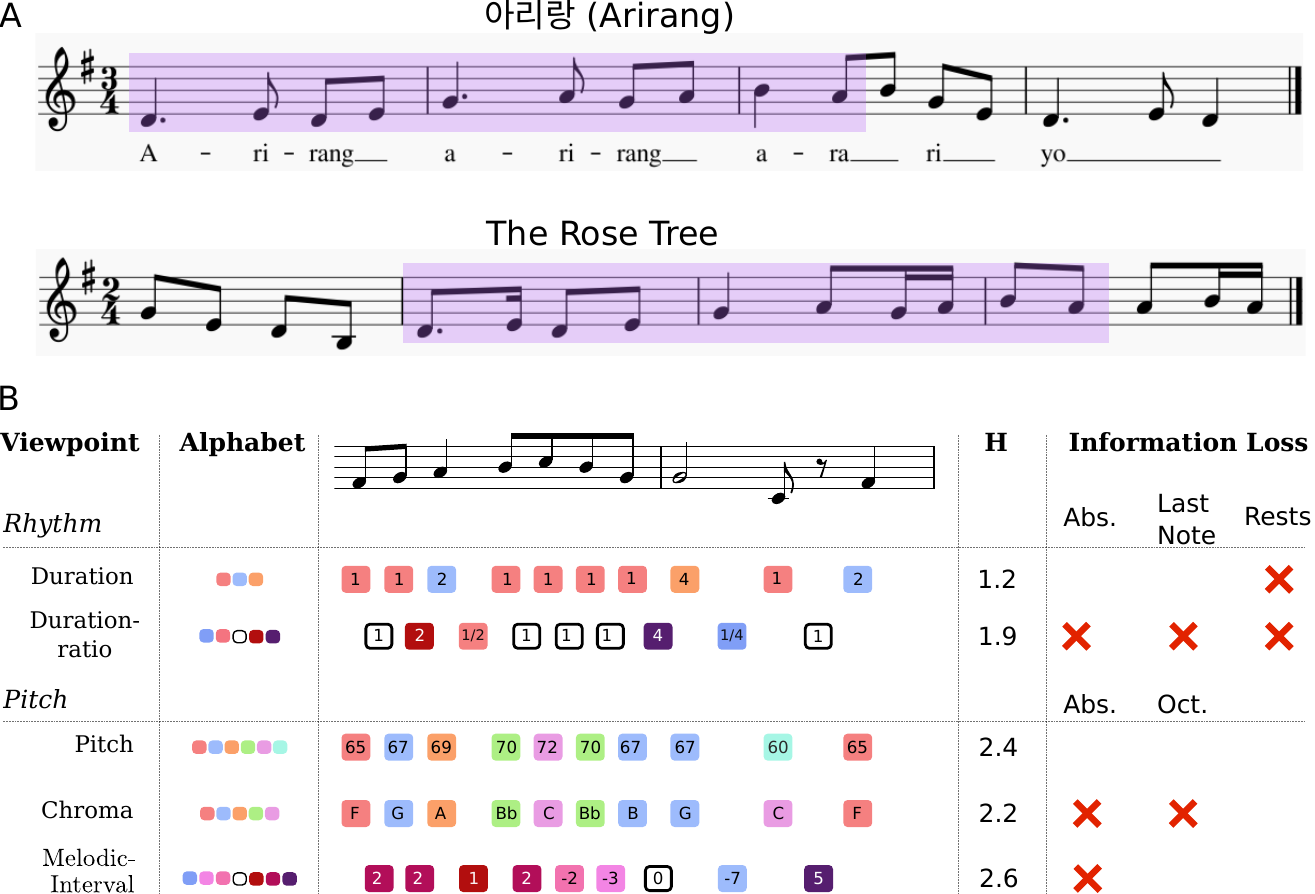}
\caption{\label{fig:fig1} 
  \textbf{Melodies, viewpoints and information.}
  A: The first four bars of the Korean folk song, `아리랑' (`Arirang') and the 
  Irish folk song, `The Rose Tree' (transposed to the key of G). 10-note sequences with identical pitch
  are highlighted. 
  B: Illustrative example of a melody, the different viewpoints (sequential representations of the melody that encode different information), and some of their information
  properties. An alphabet is a set of unique elements from which sequences can be composed; entropy, $\HH$,
  is a measure of information; different viewpoints differ in the information
  they contain, as indicated under `Information Loss' (Abs. indicates absolute rather than relative values for pitch and rhythm; Last note indicates whether the last note is represented; Rests indicates whether silences between notes are represented; Oct. indicates whether octave information is represented). 
  Duration describes the length of time a note is held; it ignores the value of rests.
  Duration-ratio is the ratio of consecutive Duration values; this is a time-invariant
  representation. Pitch describes the absolute pitch. Chroma is $\pM \mod 12$,
  shown here in \textit{solf\`{e}ge} notation; it is restricted to a single
  octave range so absolute pitch is lost, but it retains information about pitch
  position within an octave. Melodic-Interval is the difference in pitch between successive $\pM$ notes.
}
\end{figure*}

  Despite such potential for variation, melodies tend to be similar to each
  other.~\cite{eerolaStatistical2001,temperleyComputational2013,volkMelodic2012}
  This is evident in the classification of musical styles through shared characteristics, such as melodic patterns.~\cite{lomaxFactors1980,thompsonCrossCultural1993,temperleyCommunicative2004,dannenbergStyle2010,rodriguezzivicPerceptual2013,huangEvolution2017,weissInvestigating2019,singhDynamic2022,kulkarniInformation2024a,mossComputational2024}
  Even across cultures, melodies can be sufficiently similar to allow for consistent transmission of interpretable information.~\cite{trehubAdults1993,krumhanslTonality2000,eerolaPerceived2006,matsunagaCrosscultural2018,morrisonCultural2019,brinkmanCrossCultural2021,daikokuCrosscultural2023,klarlundWorlds2023,jacobyCommonality2024} 
  This is exemplified by comparing the traditional Irish polka, `The Rose Tree',
  and the national folk song of Korea, `아리랑' (`Arirang') (\fref{fig:fig1}A).
  These melodies share a 10-note melodic sequence, which occurs an estimated \num{200} million times more frequently than expected by chance (for details, see \textit{Melodic Similarity}), suggesting the existence of strong forces that drive melodies
  towards a specific niche within the vast landscape of possible melodies.

  Many common features of melodies may be explained by the vocal motor hypothesis, which proposed that they result from physiological constraints on production. Vocalization begins and ends at low
  sub-glottal pressure and low pressure produces low pitch,~\cite{sundbergScience1987}
  thus arch-shaped contours are common.~\cite{huronMelodic1996,tierneyMotor2011a,savageGlobal2017a,goldsteinExploring2023}
  Melodic range is limited by vocal range, meaning that large melodic pitch intervals tend to be followed by a change in interval direction (up vs. down) simply because they are likely to approach the limits of the range.~\cite{vonhippelRedefining2000}
  Phrase length is limited by lung capacity.~\cite{drewPhysiology1939}
  Scalar motion -- melodic movement using small pitch intervals -- costs
  less energy (through muscle contraction and relaxation)~\cite{sundbergScience1987}
  than melodies with large intervals, which are therefore more difficult to produce accurately.~\cite{wattFUNCTIONS1924,vosmp89,tierneyMotor2011a,
  savageStatistical2015,savageGlobal2017a,mehrUniversality2019,brinkmanCrossCultural2021} 
  However, there are some essential features of melody that are not explained by motor constraints:
  Melodies tend to use a small pitch alphabet, with typically 7 or fewer notes in a
  scale.~\cite{brownUniversals2013,savageStatistical2015,mehrUniversality2019,
  mcbrideCrosscultural2020a,mcbrideConvergent2023b,brownMusical2025}
  Motor constraints also fail to explain the establishment of \textit{differentiated}
  styles,~\cite{dannenbergStyle2010} the tendency towards repetition within songs,
  ~\cite{margulisRepeat2014,daiWhat2022} or limits on song length.
  Alternative explanations may include form-function relationships (\eg, lullabies
  should be soothing),~\cite{mehrForm2018,yurdumUniversal2023,labendzkiTemporal2025}
  and the emergence of styles from cultural-evolutionary processes
  of innovation through imitation.~\cite{savageSequence2022,hajicjr.Building2023,nakamuraComputational2023}
  Here we investigate the role of cognitive processes such as memory in constraining the features of melodies,~\cite{millerMagical1956a,cowanMagical2001,halpernMemory2010}
  through an interrogation of the information-theoretic properties of large and cross-culturally varied corpora of melodies.~\cite{youngbloodStyle1958,
  siromoneyStyle1964,knopoffEntropy1983,margulisMusical2008}

  We propose that the way information is encoded, stored in memory, and retrieved by the brain leads to constraints
  on the kind of melodies that are likely to be produced. We consider two information-theoretic quantities,
  whose determinants have been shown to affect memory in recall and recognition experiments in
  music and other domains:
  \textit{information rate}, which is primarily determined by sequence complexity and presentation rate;
  ~\cite{millerMagical1956a,kilgourMusic2000,akiva-kabiriMemory2009,oberauerWhat2016,barascudBrain2016,agresInformation2018,
  biancoMusic2019,biancoLongterm2020a,biancoPupil2020,plantonTheory2021,bakerFrequently2020}
  and \textit{total information}, which is the integral of the information rate over a sequence, and strongly
  dependent on sequence length.
  ~\cite{millerHuman1956,akiva-kabiriMemory2009,wardExamining2010,mathyWhat2012,barascudBrain2016,
  bakerFrequently2020,plantonTheory2021,lorchChunking2022}
  
  It has been hypothesised that verbal communication ought to be efficient,
  and thus occur at information rates close to the channel capacity, or in other words that speech should be almost as fast as possible while avoiding errors.~\cite{coverElements1999}
  This led to the uniform information density hypothesis which
  predicts that information rate should be stable along a spoken utterance,~\cite{levySpeakers2007,
  jaegerRedundancy2010,meisterRevisiting2021,meisterLocally2022} or musical sequence
  \cite{temperleyUniform2019,bjareTypicality2022}. Efforts have even been made to estimate the channel capacity finding trade-offs leading to
  cross-linguistic convergence in information rate.~\cite{pellegrinoAcrossLanguage2011,
  coupeDifferent2019,acevesHuman2024}
  
  Music and language however communicate different kinds of information, 
  and we do not assume that music needs to be especially efficient or operate at
  information rates close to the channel capacity. Instead, we consider evidence that human
  preferences for complexity in music~\cite{chmielBack2017,goldMusical2019,cheungUncertainty2019,
  chmielUnusualness2019,biancoPupil2020,streetRole2022,parmerEvolution2023} and art~\cite{tranEntropy2021} follow an inverted U-shaped curve, whereby an intermediate degree of complexity is preferred.
  We hypothesize that these preferences are partly determined by biological constraints on memory. Overly complex melodies may be difficult to learn and recall faithfully. Overly simple melodies may be understimulating (\ie, boring) according to hypotheses based on optimal arousal, cognitive fluency, and reward prediction.~\cite{berlyneAesthetics1971,reberProcessing2004,pearceChapter2025} For example, predictive coding theory proposes that prediction error is an indicator of learning, which can lead to prediction reward (\eg, release of dopamine), thus low prediction error would evoke little reward.\cite{stupacherBrief2022} If this hypothesis is true, then evidence of information constraints should appear in musical cultures where the melodic repertoire has evolved primarily through oral transmission, such as folk music.
  We thus collected \num{62} \textit{Folk} music corpora from a wide range of cultures, and compare these with music of greater and lower complexity. We use \num{7} corpora of music for children (\textit{Child})\, assuming that these are examples of low-complexity music.~\cite{labendzkiTemporal2025}
  To provide examples of what we assume is high-complexity music, we use 39 corpora of \textit{Art} music,
  which is typically transmitted through written notation and composed / performed by professional musicians. If memory imposes constraints on information in melodies, then we expect that melodies will differ in information rate and total information, with Child being the simplest, and Art being the most complex. If these constraints are especially strong, we expect to find similar levels of information across Folk societies. We do not make strong predictions about Art music, where constraints may differ due to the use of musical notation and the prevalence of professional composers -- for example, there is evidence of a ratcheting up of harmonic / tonal complexity in Western Europe throughout the common practice period and into the modern era.~\cite{gonzalez-espinozaQuantifying2023,buecheleCrystals2025} This kind of cumulative process may not be necessarily exclusive to Western music.~\cite{savageCultural2019}

  Information rate of human music perception cannot be assessed directly since it depends on the encoding mechanisms of the brain which are currently unknown.
  Instead we study several determinants or correlates of information rate:
  Entropy, $\HH$, is a useful, easy-to-calculate correlate of the true information rate. It tells you that the lowest possible information rate must be equal to or lower than this value.~\cite{coverElements1999}
  The number of distinct (temporal, or pitch) elements in a sequence, here referred to as the alphabet size, $\AL$, also sets an upper limit on how complex a sequence can be.~\cite{oberauerWhat2016} The more possible elements there are, the greater the uncertainty about what will come next in a sequence.
  It also matters whether the elements are uniformly distributed, or unevenly distributed. If one element is included \SI{90}{\%} of the time, the sequence is less complex than another sequence that has the same alphabet size but with equally distributed elements.
  Repeated motifs in pitch or rhythm allow learning of a more efficient
  coding where they are treated as chunks,~\cite{coverElements1999,lorchChunking2022} and as a result music that is stylistically familiar is
  easier to learn and has a lower effective information rate.~\cite{boltzStructural1991,rohrmeierArtificial2013,halpernMemory2010,demorestFMRI2010,hannonFamiliarity2012,demorestInfluence2016,daikokuPitchclass2016}
  We use a variable-order Markov model of melodic compression, Information Dynamics of Music (IDyOM),~\cite{pearceStatistical2018} alongside a methodology to control for differences across corpora,
  to estimate the degree of information reduction due to repetition within a melody. This model has proved useful in simulating expectation, memory, similarity, complexity, and pleasure in music perception.~\cite{pearceStatistical2018,goldPredictability2019,peajn17, saump19, harrisonPPMdecay2020}
  Thus while we cannot directly measure the information rate of human music perception,
  we can measure several determinants of information rate to understand its distribution across a large existing sample of melodies.

  We measure determinants of information rate and estimate the total information in melodies
  in 108 corpora (for details, see \textit{Melodic Corpora}, SI Section 1),
  ~\cite{sch95,shapi14,van19,sapis05,walpi14,abc22,rod22,vanjn10,gothamScores2018,kunst,bakerMeloSol2021,
  karaosmanogluTurkish2012,eerolaSuomen2004,eerolaPerceived2006,Www2023,nke63,allenSlave1867,
  lewinRock2000,folksongkorea69,chunhyangka77,sipos14,ives1141922,blackingVenda1967,
  mof04,mon07,nipponhosokyokaiRiBenMinYaoDaGuan1989,nishikawaCultural2022}
  primarily covering orally-transmitted folk music (62 corpora), notated art music (39 corpora), and music for children (7 corpora).
  Songs within societies can be more or less complex than the within-society average, but there are systematic differences when comparing societies by their average melodic properties.
  We find multiple trade-offs between the determinants of information rate for Folk music
  that all point to cross-culturally universal constraints on the average information rate in melodies: in melodies with larger pitch or duration alphabet sizes (increasing complexity) the pitch/duration elements tend to be less equally distributed (decreasing complexity); corpora with higher pitch entropy (increasing complexity) tend to have lower rhythm entropy (decreasing complexity); corpora with more complex
  songs (increasing complexity) tend to have more repetition between songs (decreasing complexity). Finally, we develop a parameter-free model of melodies informed by the empirical constraints on information rate, melody length and scalar motion, which quantitatively predicts the observed number of scale degrees.

\section*{\sb{Results}}

\nsb{Information in melodic viewpoints.~}
 Melodies can be described by two dimensions -- pitch and rhythm~\cite{monahanPitch1985} -- 
  and each dimension can be represented by different \textit{viewpoints} (\fref{fig:fig1}B), which are different representations of the pitch or timing of notes in a melody (\eg, scale degree or pitch interval).
  Each viewpoint describes the melody in a different way and differs in its information-theoretic properties: The word alphabet is used often in information theory to describe the set of possible distinct elements that can appear in a sequence -- for example, the alphabet of the $\pC$ viewpoint is equivalent to the scale of the melody.
  The number of unique
  elements is the alphabet size, $\AL$. Entropy is defined as $\HH=\sum\limits_{i}^{\AL} p_i \log p_i$, where
  $p_i$ is the probability of letter $i$. Entropy is a measure of the amount of information;
  in this context specifically, it is the mean information rate per note. As an example of what this means in musical practice, entropy increases through a progression of levels in singing instruction books (SI Fig. 9).~\cite{bakerMeloSol2021}
  
  There are many melodic viewpoints for both rhythm (duration, duration-ratio,
  inter-onset-interval [IOI], IOI-ratio) and pitch (pitch, chroma,
  scale degree, melodic interval, scale degree interval, contour), which differ
  in the information that is encoded and in efficiency. Converting between viewpoints can
  lead to information loss, but sometimes the information is redundant and the loss is superficial.
  For example, in the melodic example in \fref{fig:fig1}B, converting from 
  Duration to Duration-ratio results in information loss as we lose information about the duration of the last note; at the same time, the entropy increases from 1.2 to 1.9 bits, showing that there is an increase in information redundancy, thus it is a less efficient representation in both respects.
  We examined and compared each viewpoint in terms of the information loss and 
  efficiency (SI Section 2A). We find that information content in different
  viewpoints is often highly correlated (SI Section 2B-E), and that this
  can be quantitatively explained using models that encode basic constraints (scales, scalar
  motion, simple rhythms; SI Section 4). The interrelatedness between viewpoints leads to
  similar outcomes of information-theoretic analyses, so we chose a minimal set of
  viewpoints for the primary analyses that follow.

\begin{figure*}
\centering
\includegraphics[width=0.99\textwidth]{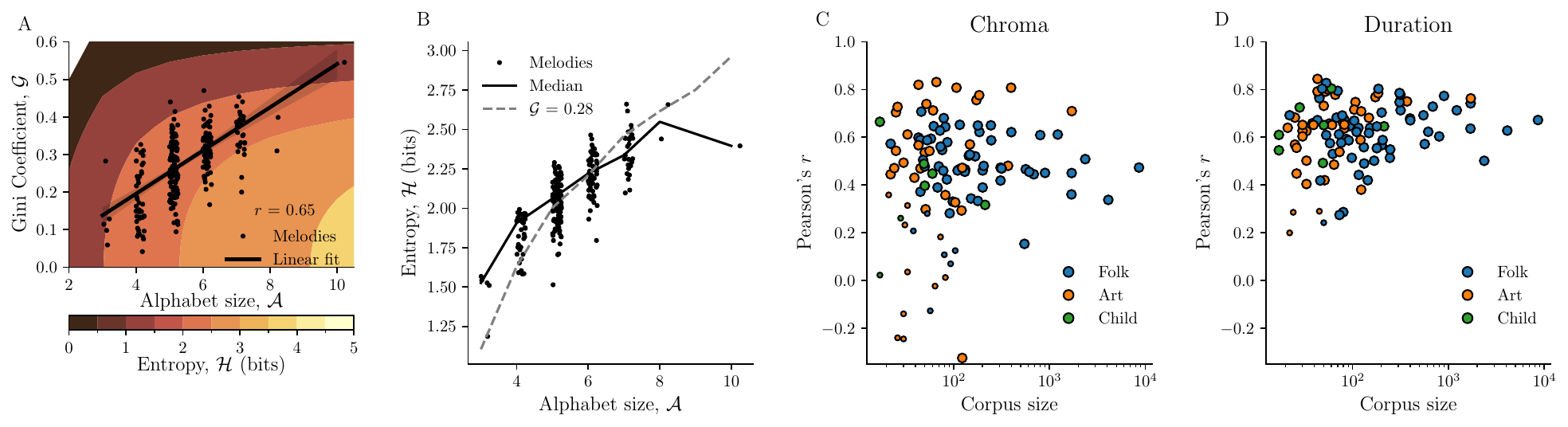}
\caption{\label{fig:fig2} 
  \textbf{Comparing alphabet sizes and distributional entropy within corpora.}
  A: Gini coefficient $\GI$ vs. alphabet size $\AL$ for the Sioux corpus ($\pC$ sequences). The contour indicates
  the entropy of a power-law distributed alphabet as a function of $\GI$ and $\AL$.
  Linear fit with shaded \SI{95}{\%} CI, and Pearson's $r$ are shown.
  B: Entropy $\HH$ vs. $\AL$ for the Sioux corpus. Solid line shows the median
  as a function of $\AL$; dotted line shows the expected value of $\HH$ for a power-law
  distributed alphabet with constant $\GI$.
  C-D: Pearson's correlation between $\AL$ and $\GI$ as a function of corpus size
  for each corpus, for $\pC$ (C) and $\rD$ (D) viewpoints. Colours indicate corpus type.
  Large circles indicate that $p<0.05$; the Benjamini-Hochberg
  procedure was used to control for multiple comparisons.
}
\end{figure*}

  We choose first and second order representations respectively of both
  rhythm and pitch. Duration denotes the amount of time a note is sounded, 
 ignoring periods of silence (for readers interested in why we chose Duration over IOI, see SI Section 2D).
  The second order rhythmic viewpoint is $\rDr$, the ratio between consecutive $\rD$ values,
  which is tempo-invariant and loses information about the duration of the last note.
  $\pC$ is octave-invariant pitch, calculated as pitch (absolute log-frequency)
  modulo 12, and often represented categorically using note names (e.g., A, B, \ldots, G); this representation
  loses information about the absolute pitch and does not distinguish between 
  octaves. However, we found that due to the predominance of small melodic intervals, this information is about \SI{95}{\%} recoverable, if one simply assumes that octave changes are more likely than large intervals (SI Section 2B). The second order pitch viewpoint is the $\pMi$, which
  is the difference between consecutive Pitch values in a melody; this is key invariant and
  loses information about the absolute pitch.
  We primarily study first order representations as they tend to be more efficient (SI Fig. 3).
  We use second order representations only when studying repetition between melodies, in which
  case it is important that the viewpoint is insensitive to temporal or key changes.

\nsb{Melodies with larger alphabets have less equal distributions.~}
  For any viewpoint, the entropy $\HH$ depends on the alphabet size $\AL$,
  and how evenly the elements are distributed. If the elements are uniformly
  distributed, entropy is at its maximum, $\HH = \log \AL$. Conversely, as a distribution
  tends towards maximum inequality (\ie, when only one letter is used) entropy tends to zero.
  The inequality of a distribution can be measured using the Gini coefficient, $\GI$,
  \begin{equation}
  \GI = \sum \limits_{i=1}^{\AL} \frac{\theta(p_i)}{(i/\AL)} - \frac{1}{2}~,
  \end{equation}
  where $p_i$ is the probability of the $i^{\textrm{th}}$ element in the alphabet
  arranged in order of increasing probability, and $\theta(p_i)$ is the cumulative
  probability function. $\GI$ ranges from zero for a uniform distribution to one
  for maximal inequality. Therefore an increase in the Gini coefficient typically leads to lower entropy, while an increase in alphabet size typically leads to higher entropy. Since these are opposing effects, when alphabet size is positively correlated with the Gini coefficient the entropy distribution should have lower variance (and vice versa for a negative correlation) than when they are uncorrelated. Thus a positive correlation should lead to lower variance in entropy across melodies.

  To illustrate how alphabet size, Gini coefficient and entropy interact in melodies we examine $\pC$ sequences from a Sioux Native
  American corpus (\fref{fig:fig2}A). The shaded contour in \fref{fig:fig2}A shows entropy as a function of alphabet size and Gini coefficient for 
  power law distributions, which closely corresponds to the behavior of empirical
  distributions (SI Fig. 10). The Gini coefficient and alphabet size are strongly correlated
  which means that as melodies use larger scales they also have more unequal pitch distributions.
  Consequentially, the variation of entropy across songs (\fref{fig:fig2}B, solid line) is lower than if $\GI$ was independent
  of $\AL$ (\fref{fig:fig2}B, dashed line). We find strong positive correlations between the Gini coefficient and alphabet size in most corpora,
  for both pitch (\fref{fig:fig2}C) and rhythm (\fref{fig:fig2}D). 
  This means that societies that use fewer notes in scales (or rhythmic categories) do not necessarily
  have less complex pitch (rhythm) sequences, if they compensate by using more equal pitch (rhythm) distributions.
  Although we do not see significant positive correlations in all cases, the majority of
  exceptions can be attributed in part to sample sizes. Ultimately, the effect of this trade-off
  is to reduce the overall variance in entropy between songs.

  We note that these findings are not surprising, and may have been deduced from previous results. Several studies have found long-tailed distributions for pitches and note durations,~\cite{manarisZipf2005,zanetteZipf2006,mehrUniversality2019} and one expects to find a correlation between alphabet size and Gini coefficient in such a case. What is new here is simply the framing, which explains how this leads to a narrower distribution of entropy compared to what would occur otherwise.

\begin{figure*}
\centering
\includegraphics[width=0.95\textwidth]{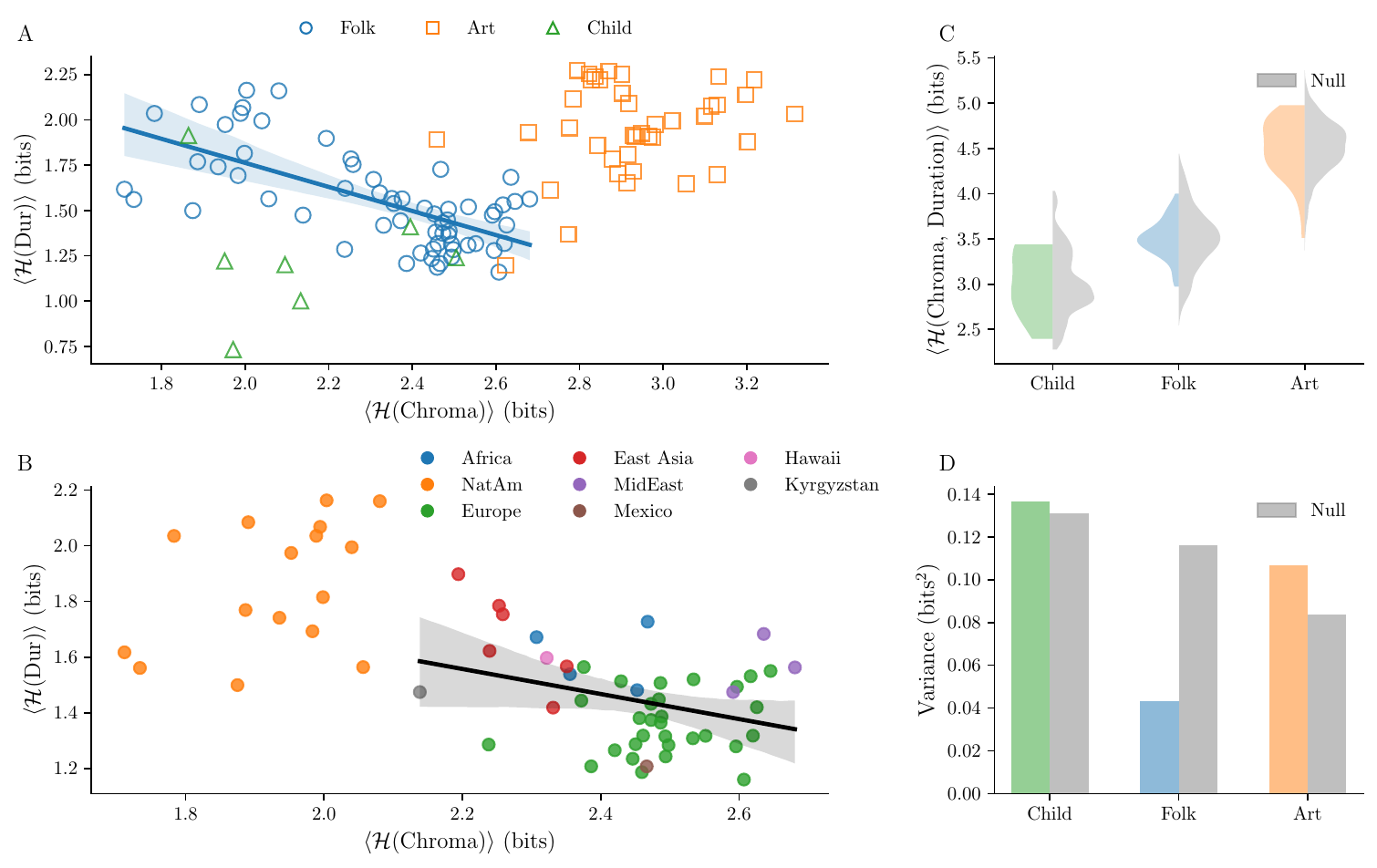}
\caption{\label{fig:fig3} 
  \textbf{Comparing pitch, rhythm and joint entropy across corpora.}
  A: Mean pitch entropy per corpus $\langle \HH(\pC) \rangle$ vs. mean rhythm entropy
  per corpus $\langle \HH(\rD) \rangle$;
  linear correlation is shown for Folk corpora (Pearson's $r=-0.66$, $p<10^{-8}$, $n=62$).
  B: $\langle \HH(\pC) \rangle$ vs. $\langle \HH(\rD) \rangle$ for Folk corpora,
  with geographic regions indicated by color; linear correlation is shown for
  corpora excluding the Densmore Native American corpora (Pearson's $r=-0.34$, $p<0.05$, $n=45$).
  C: Distributions of mean joint entropy per corpus $\langle \HH(\pC, \rD) \rangle$
  for different corpora types; distributions are also shown for a model 
  where pitch and rhythm entropy are uncorrelated (Null).
  D: Across-corpora variance of the empirical and null $\langle \HH(\pC, \rD) \rangle$ distributions.
}
\end{figure*}

\nsb{Societies differ via pitch-rhythm trade-off.~}
  Between Folk corpora there is a strong, negative correlation between
  pitch and rhythm entropy (\fref{fig:fig3}A, blue line), which is not observed for other types of corpora. Clustering of geographical regions (\fref{fig:fig3}B) suggests that melodic styles in different societies are influenced by neighbors, but also means that the observed correlation could be incluenced by sample balance and autocorrelation.
  For example, the correlation is heavily influenced by Native American music, which tends to be more
  rhythmically complex than the other Folk societies in our collection, but we still
  see a significant correlation if we remove these corpora (\fref{fig:fig3}B, black line).
  Likewise, we find significant correlations if we use a more general sub-sampling
  approach to decrease the influence of Native American and European corpora (SI Fig. 11).
  Within corpora we don't see such a strong link between pitch and rhythm complexity. Pitch and rhythm entropy tend to be positively correlated, although the effect sizes are small and mostly non-significant (SI Fig. 12). This means that within a society, songs can differ in overall complexity, and the trade-off we see only applies to the averages across societies.
  This suggests that different musical cultures \textit{specialize} in either rhythmic
  or pitch complexity.

% This is bounded at the lower end by
% corpora of music for children, and at the upper end by art-music traditions,
% which are perhaps evolved under different informational constraints due to
% the development of music as a form of prestige and employment. Art-music traditions do
% not conform to any simple pitch-rhythm trade-off.

\nsb{Similar low levels of pitch-rhythm covariance are found across corpora.~}
  Pitch ($\Pi$) and rhythm ($\Rh$) can co-vary in a way that reduces the entropy of the
  joint viewpoint. Such covariance between pitch and rhythm (or metrical stability, which differs from note durations) is when certain pitches (\eg, high vs low pitch, or specific scale degrees) co-occur with note durations or metrical position more or less than chance. This has been termed "tonal-metric hierarchy", and has been studied in Western art music,~\cite{brozeHigher2013,princeTonalMetric2014} and in cognitive science.~\cite{wenPerception2019} We can measure the covariance between pitch and note durations using the mutual information,
  $\MI(\Pi, \Rh) = \HH(\Pi, \Rh) - [\HH(\Pi) + \HH(\Rh)]$, 
  which quantifies how much can be inferred about pitch if just the rhythm is known
  and vice versa. To control for confounds we calculate
  $\MI^* = \MI - \MI_{\textrm{ran}}$, where $\MI_{\textrm{ran}}$ is the
  value of mutual information expected by chance (SI Section 5A).
  We find that $\MI^*(\Pi, \Rh)$ is in the approximate range \SIrange{0.05}{0.15}{bits},
  indicating that there is slightly higher covariance than expected by
  chance (SI Fig. 13A), with no clear dependence on corpus type or overall
  complexity.

  There are clear musical interpretations of this covariance (SI Section 5B):
  Covariance between $\pMi$ and $\rD$ is mainly due to the
  co-occurrence of long notes with large interval sizes (SI Fig. 13B),
  and this is consistent across most cultures (SI Fig. 14).
  This makes sense when you consider that transitions between
  sung notes are not instantaneous, and hence larger intervals
  need more time to transition between the two notes making up the interval.
  This could also result from the tendency for final notes in phrases to be longer along with the tendency for pitch intervals to be larger across phrase boundaries than within phrases.
  For $\pC$ and $\rD$ covariance, we find that the majority of melodies show evidence of a clear tonal center (the 'tonic', which is repeated more than other notes) and another note that is a fifth above the tonic. It has been hypothesized that these notes provide a stabilizing role in the scale.~\cite{krumhanslQuantification1979} As such, we find that these two notes tend to co-occur with longer durations compared to other tones in the scale (SI Fig. 13C-D, SI Fig. 14).

\nsb{Complexity is constrained in folk music.~}
  The joint pitch-rhythm viewpoint affords a better estimate
  of melodic complexity than either pitch or rhythm alone. We see clear differences
  in joint viewpoint entropy between Art and Folk corpora (\fref{fig:fig3}C).
  Child corpora tend to be the simplest, but do overlap with Folk corpora.
  This overlap may be in part due to the inclusion of songs that are child-directed but sung by adults (\eg, lullabies),~\cite{mihelacComputational2023} and also because children develop musical skills rapidly with age.~\cite{hannonMetrical2005,trehubInfant2006}
  We use the previous calculations of $\HH(\pC)$, $\HH(\rD)$ and $\MI(\pC, \rD)$,
  to generate a null model of what the joint entropy would be if we forced pitch
  and rhythm entropy to be uncorrelated (\fref{fig:fig3}C, Null; for details, see \textit{Joint entropy null model}).
  This shows that the correlation between pitch and rhythm for Folk music results in
  reduction in the variance of the joint entropy distribution by a factor of \num{2.6} (\fref{fig:fig3}D).

\begin{figure*}
\centering
\includegraphics[width=0.95\textwidth]{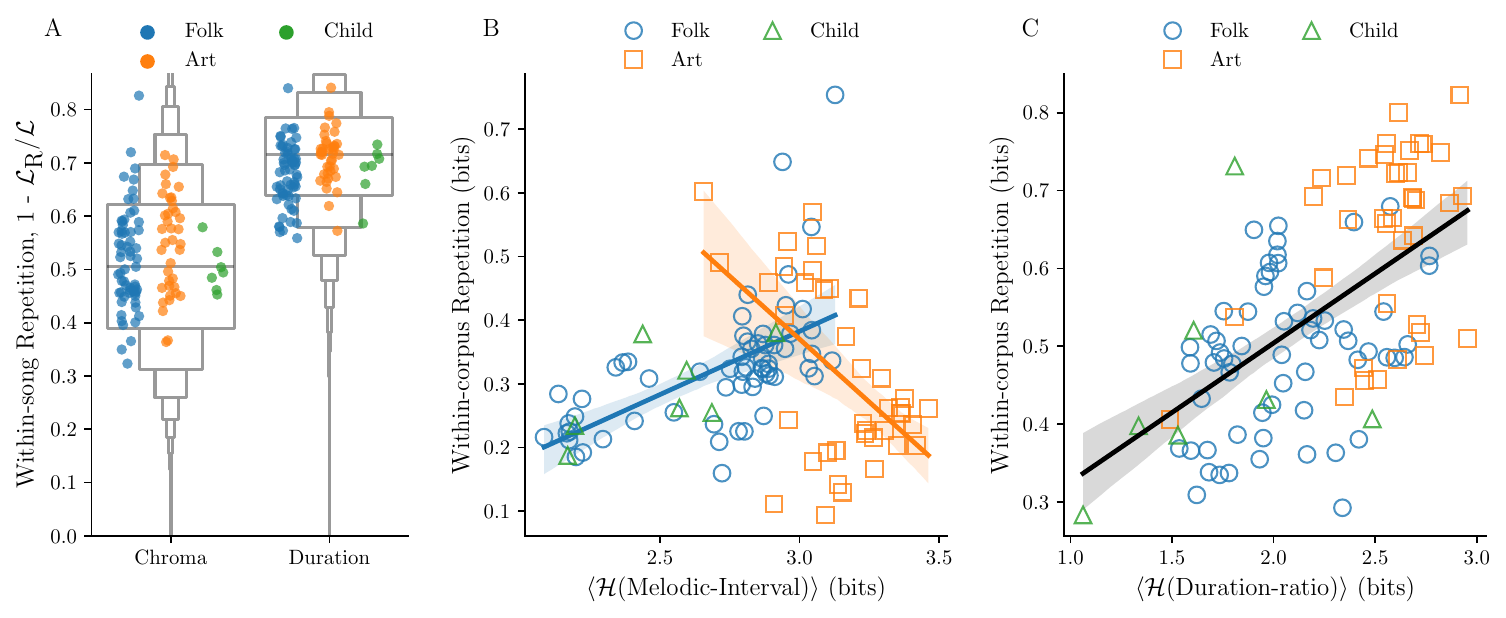}
\caption{\label{fig:fig4} 
  \textbf{Repetition within and between melodies.}
  A: Total fraction of a melodic sequence ($\pC$ or $\rD$) that is repeated; letter-value plot
  is for all melodies;  corpus averages are shown as circles colored by corpus type.
  B-C: Degree of repetition between melodies in a corpus vs. mean entropy per
  corpus for $\pMi$ (B) and $\rDr$ (C) second order viewpoints. Correlations are shown
  for Folk corpora (B, $r=0.59$, $p<10^{-6}$, $n=62$), Art corpora (B, $r=-0.54$, $p<0.005$, $n=39$),
  and for all corpora (C, $r=0.60$, $p<10^{-12}$, $n=108$).
}
\end{figure*}

\nsb{Rhythm is more repetitive than pitch within songs.~}
  Corpora can differ in the amount of repetition in ways that reflect choices
  made by the authors or collectors or the corpus rather than the musical tradition.
  Some only report melodic skeletons without variation or embellishments. Some repeat entire sections with small variation, while others save space by using repeat lines or by annotating melodic variation using polyphonic annotations. These differences stem from choices of the transcribers
  and collectors, and should not be interpreted as systematic differences in how musical traditions use repetition within songs.
  This means we can make within-corpus comparisons of relative degrees of repetition between pitch and rhythm, and approximately estimate degrees of within-song repetition,
  but we cannot draw conclusions from between-corpora differences in \textit{within-song} repetition.
  We estimate the amount of repetition in a melodic sequence by recursively
  removing repeated sub-sequences of length 2 or more, and count the total
  length of the remaining sequence, $\LN$ (for details, see
  \textit{Repetition within melodies}). The
  fraction of repetition in a sequence is then $1 - \LN / \LL$.
  We find (\fref{fig:fig4}A) that rhythm sequences have substantially more repetition (\SI{71}{\%},
  averaged over all melodies) than pitch sequences (\SI{51}{\%}).

\nsb{Folk corpora with more complex songs have more repetition between songs.~}
  So far we have presented the information properties of single melodies.
  It is also possible for cultures to differ in how information is distributed
  across melodies, through repetition of motifs and rhythms. By learning 
  the statistics of a corpus, one can efficiently encode frequently-occurring
  sequences thereby reducing the information rate through data compression.
  To study within-corpus repetition
  we use IDyOM, a machine learning tool that learns higher-order sequence statistics.
  The compression that IDyOM achieves depends on: (i) melody length; (ii) size of the corpus;
  (iii) alphabet size and letter distribution; (iv) and whether $\rD$ / $\pC$
  representations are normalized / transposed to a specific tempo or key.
  To estimate the degree of within-corpus repetition while controlling for each
  of these we:
  (i) truncate melodies at 50 notes; (ii) train IDyOM on only 10 melodies;
  (iii) compare the reduction in information content of original sequences to shuffled sequences;
  (iv) and we use second order viewpoints ($\pMi$ and $\rDr$) which do not depend
  on key and tempo. The results are not sensitive to the choice of melody truncation length or the number of melodies used to train IDyOM (SI Fig. 15-16). The final measure, "within-corpus repetition", is equivalent to the amount of information reduction (in bits) due to repetition, above what one can obtain from training IDyOM with a random set of sequences with matched alphabet size and letter distribution (ii). For more details, see \textit{Repetition between melodies}.

  \begin{figure*}
  \centering
  \includegraphics[width=0.95\textwidth]{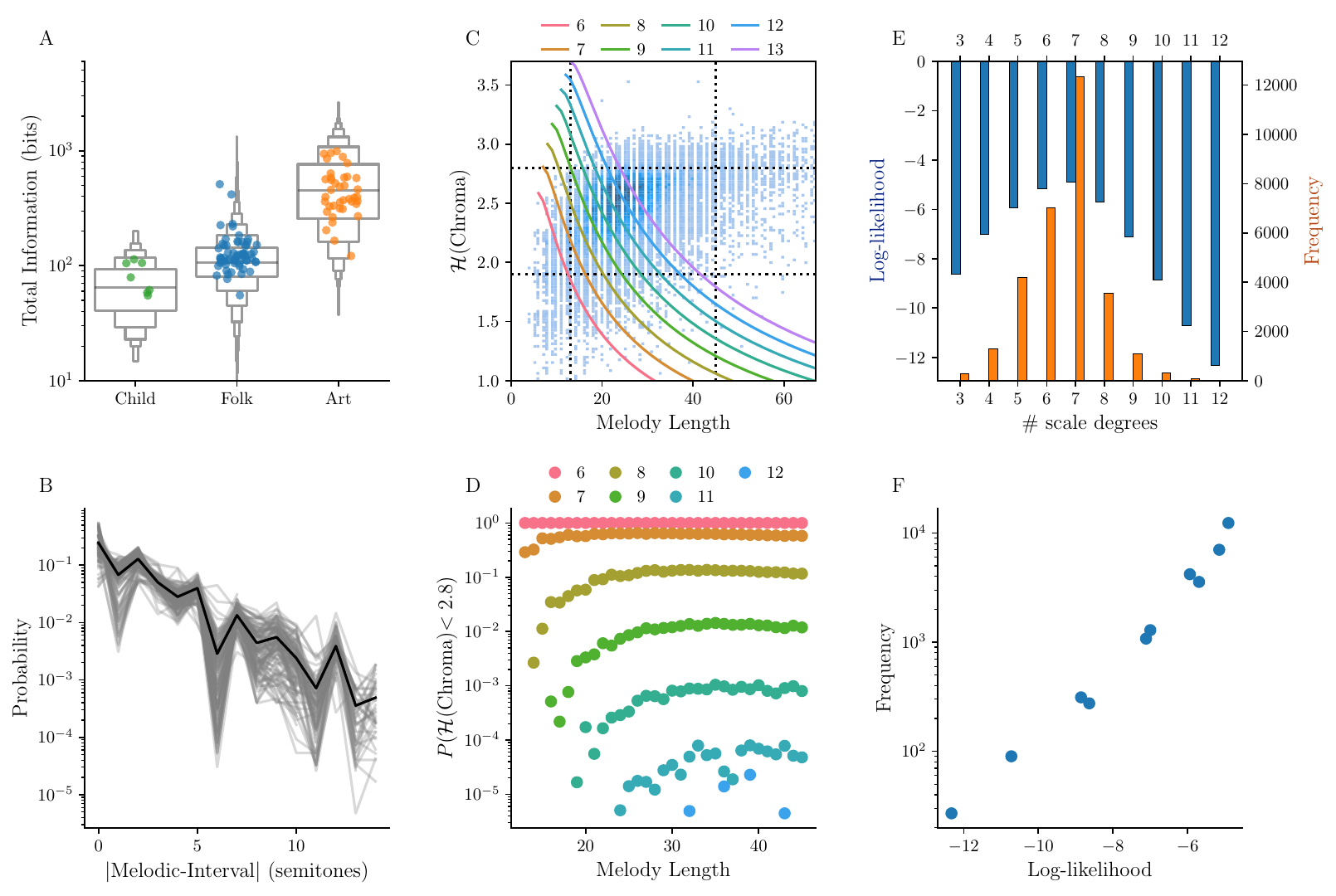}
  \caption{\label{fig:fig5} 
    A: Total information; letter-value plot is for all melodies; 
    corpus averages are shown as circles.
    B: Probability distribution of absolute values of $\pMi$ across all corpora (black line)
    and for all individual corpora (grey lines).
    C: Solid lines show the minimum $\HH(\pC)$ as a function of melody length ($\LL$), for different values
    of alphabet size ($\AL$). Histogram of $\HH(\pC)$ and $\LL$ for Folk
    corpora is shown in blue (darker indicates higher density).
    Dashed lines enclose a central region that contains
    \SI{99}{\%} of all Folk melodies.
    D: Probability that a melody generated through scalar motion has $\HH(\pC)<2.8$ bits,
    as a function of $\LL$, for different values of $\AL$ (different colours).
    E: Log-likelihood per melody (blue) that a set of pitches generated through scalar motion
    resulting in an alphabet size $\AL$ reproduces the $\HH(\pC)$ distribution
    observed in Folk melodies.
    Frequency of each number of scale degrees (orange) observed in Folk melodies.
    F: Log-likelihood correlates with Frequency ($r=0.99$, $p<10^{-6}$, $n=10$)
  }
  \end{figure*}

  For a quick, intuitive understanding of "within-corpus repetition", one can think of it as a proxy for repetition across melodies in a corpus. We find that Folk corpora with more complex songs tend to have greater
  within-corpus repetition for both pitch (\fref{fig:fig4}B) and rhythm
  (\fref{fig:fig4}C; the correlation shown is for all corpora combined; for only Folk corpora we obtain Pearson's $r = 0.33$, $p = 0.01$). 
  Greater repetition in corpora with more complex songs (and vice versa) effectively
  leads to a further reduction in the variance of the information rate, although due to the aforementioned dependencies we can only estimate relative not absolute reductions.
  We see the opposite trend for pitch in Art corpora,
  where composers writing more complex songs also repeat themselves less, suggesting a lack of constraint on complexity.
  These markers of complexity also correlate with composer birth year, 
  reflecting the historical trends in European art music (SI Fig. 17).
  Rhythmically, Art music follows the same trend as Folk music, in this case this may be due to a bias within European Art music towards increasing pitch complexity rather than rhythmic complexity, and may not reflect other Art musics.
  This points towards different types of constraints on Folk music,
  where information is constrained consistently across musical cultures, and Art music which lacks such informational constraints and exhibits
  a drive towards higher complexity over time.

\nsb{Orally-transmitted songs are limited in length.~}
  We expect to find large differences in total information
  between Child or Folk corpora, and Art corpora,
  given that the former are typically transmitted orally, while the latter
  are typically transmitted through written notation. Given the previously
  highlighted differences in handling of repetition within songs,
  we use the length of sequences after controlling for repetition, $\LN$, instead
  of the total length $\LL$; this precludes calculation of an entropy
  rate by taking into account high-order sequence dependencies, so we simply
  equate the unigram entropy to mean information rate per note. The total information
  is then $\TI = \HH(\pC, \rD) \times \LN$.

  Unsurprisingly, we find massive differences between Folk songs (interquartile range [IQR],
  $80 \leq \TI \leq 144$ bits) and Art songs (IQR, $245 \leq \TI \leq 673$ bits),
  indicating that Art songs indeed contain much more information than Folk songs
  as expected; although there are a few outliers, these differences are mostly consistent when looking at
  corpus means, shows that this is consistent across diverse societies (\fref{fig:fig5}A).
  Child songs also have lower total information (IQR, $40 \leq \TI \leq 93$ bits).

\nsb{Scalar motion dominates in melodies.~}
  In line with other reports we find that pitch movement in melodies primarily
  consists of small melodic intervals (scalar motion).
  ~\cite{wattFUNCTIONS1924,vosmp89,tierneyMotor2011a,savageStatistical2015,savageGlobal2017a,mehrUniversality2019,brinkmanCrossCultural2021}
  We find this consistently in every society (\fref{fig:fig5}B, SI Fig. 18).

\nsb{Multiple constraints limit the number of possible scale degrees.~}
  The number of scale degrees $\AL$ in a melody is first limited by
  the melody length $\LL$, as $\AL \leq \LL$. We reiterate that the entropy is has strict bounds, $0 \leq \HH \leq \log \AL$. Next
  consider that the range of possible $\HH$ values depends also on the
  melody length. If $\AL=\LL$, then every note is heard once and $\HH=\log \AL$.
  For $\AL < \LL$ the lower bound, $\HH_{lower}$, is achieved when one note
  is repeated and all other notes are only heard once,
  \begin{equation}
  \HH_{lower} = (1 - \frac{\AL}{\LL}) \log (\frac{\LL}{\LL - \AL + 1})
              + \frac{\AL - 1}{\LL} \log (\LL) ~.
  \end{equation}
  \fref{fig:fig5}C (solid lines) shows how $\HH_{lower}$ depends on $\AL$
  and $\LL$, with the central region in between dotted lines indicating where 
  \SI{99}{\%} of Folk melodies are found. This shows that it is
  technically possible to use 13 scale degrees, yet still produce a melody
  that stays within the empirical constraints on length and entropy.
  Even if we constrain melodies to follow scalar motion, it is possible 
  to achieve $\HH_{lower}$, but we are more interested in
  probability than possibility. Therefore, we use a model (for details, see \textit{Generative model of pitch sequences})
  to estimate the probability that a scale generated by scalar motion
  with an alphabet size $\AL$ and a length $\LL$ will achieve an entropy
  rate within the empirical Folk \SI{95}{\%} inter-quantile range of $1.7 \leq \HH(\pC) \leq 2.8$ bits.
  We find that for $\AL>8$, the probability is consistently lower than
  about \SI{1}{\%} (\fref{fig:fig5}D).

  What is the optimal number of scale degrees given constraints on scalar motion,
  melody length and information rate? To answer this we use a parameter-free model
  (for details, see \textit{Generative model of pitch sequences}) which generates melodies by sampling
  the melody length and melodic intervals directly from empirical
  Folk distributions, and evaluates optimality based on how well the melodies reproduce the empirical information rate distribution. For each number of scale degrees, $\AL$,
  we evaluate the log-likelihood that scales with $\AL$ scale degrees would
  produce $\HH(\pC)$ distributions consistent with the empirical $\HH(\pC)$
  distribution. By plotting the log-likelihood against the empirical
  $\AL$ distribution (\fref{fig:fig5}E), we see that there is a strong
  correlation (Pearson's $r =  0.99$, $p<10^{-6}$, $n=10$) between the likelihood and the actual probability distribution (\fref{fig:fig5}F).
  Thus, a plausible explanation for the observation that scales tend to
  have $\AL \leq 7$ scale degrees is that there are cross-cultural constraints
  on information rate in melodies.

\section*{\sb{Discussion}}

\begin{figure}
  \centering
  \includegraphics[width=0.45\textwidth]{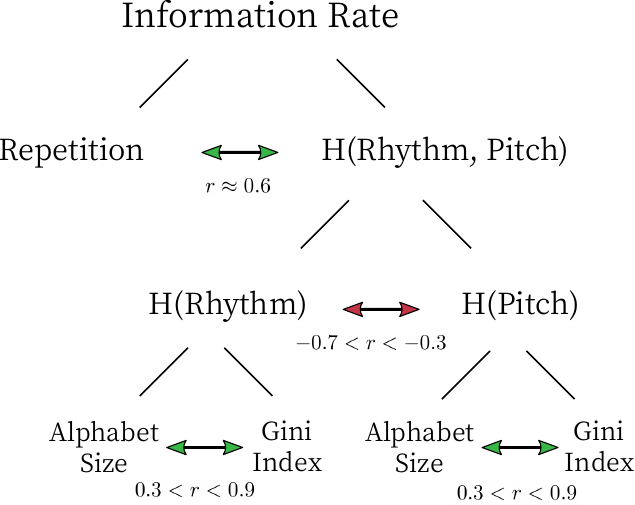}
  \caption{\label{fig:fig6} 
    \textbf{Hierarchy of determinants of information rate and their correlations in Folk corpora.}
    Higher entropy $\HH$ and larger alphabet size $\AL$ lead to higher information rate,
    while higher repetition and higher Gini index (letter distribution inequality)
    lead to lower information rate. Each of these correlations acts as a tradeoff to reduce the information rate: increasing one determinant of complexity reduces another (and vice versa).
  }
  \end{figure}

\nsb{Hierarchy of correlations implies constraints on information rate.~}
 We hypothesized that if melodies are constrained by memory, we ought to find some signature of this in the information properties of melodies across cultures, and between Child, Folk and Art music. As expected, we find differences in both information rate (specifically, the joint entropy; \fref{fig:fig3}C) and total information (\fref{fig:fig5}A) between (from low, to high) Child, Folk and Art music. This matches our intuition, at least for the music which the authors are familiar with. However, we lack intuition for what to expect from a cross-cultural comparison. It could be that some societies perform music to different degrees and that leads to different levels of information complexity. We find that
  the information rate is constrained cross-culturally through a hierarchy of correlations (\fref{fig:fig6}). To recap the determinants of information rate, information rate increases with
  alphabet size and entropy, but decreases with inequality of letter use
  (as measured by the Gini coefficent) and repetition. For Folk corpora we find significant
  positive correlations between alphabet size and the Gini coefficient (\fref{fig:fig2}),
  and between entropy and within-corpus repetition (\fref{fig:fig4}), while we find
  negative correlations between pitch and rhythm entropy (\fref{fig:fig3}). The effect
  of each of these correlations is to constrain the overall information rate.

  A constraint on information rate is supported by studies that
  find preferences for an intermediate degree of complexity in music,
  ~\cite{chmielBack2017,goldMusical2019,cheungUncertainty2019,chmielUnusualness2019,biancoPupil2020,streetRole2022,parmerEvolution2023}
  and similar results have been found for Western popular music.~\cite{parmerEvolution2023}
  The trade-off in pitch-rhythm complexity has also been observed at an individual level in
  perceptual experiments.~\cite{monahanPitch1985} Evidence of constraints on
  information rate in speech has also been reported,~\cite{levySpeakers2007,
  jaegerRedundancy2010,meisterRevisiting2021,meisterLocally2022} although communicative pressures may
  be different between music and speech. We have focused here on information
  rate per note since the corpora do not contain details of tempo, but future
  studies should focus on information rate measured in bits per unit time.
  We predict that songs with higher information rate per note will
  have lower note density per unit time.

  This constraint on information rate is clearly flexible. For example, we find that within corpora songs do vary in complexity (\eg, \fref{fig:fig2}A). We emphasise that it is the average complexity, or complexity distribution that seems to be consistent across the Folk music studied here. We have also shown how in Art music, these apparent bounds can be surpassed, as Western art music became increasingly complex over the years (SI Fig. 17). Thus, we expect that in some cultures where music is rarely performed,~\cite{shiltonWhy2025,singhLoss2025} music may be less complex than we have reported here. Indeed, our sample of corpora may be biased towards societies that have a strong tradition of melodic music. To explore this further, methods ought to be developed for the computation of pitch and rhythm entropy from audio recordings rather than transcriptions.

\nsb{Multiple constraints act to limit the size of scales.~}
  We find evidence of constraints on total information (\fref{fig:fig5}A; highly correlated with melody length),
  scalar motion (\fref{fig:fig5}B, SI Fig. 18), and information rate (\fref{fig:fig3}C-D).
  By inputting these empirical findings into a minimal, generative model of melodies,
  we find that they predict the observed distribution of scale size (\fref{fig:fig5}E-F).
  This is in line with mounting statistical evidence confirming the prevalence of scales
  with 7 or fewer notes,~\cite{savageStatistical2015,mcbrideConvergent2023b,brownMusical2025}
  additionally supported from recent iterated learning experiments.
  ~\cite{verhoefMelodic2021a,popescuCore2024,anglada-tortLargescale2023a}
  We now provide compelling evidence to support the hypothesis
  that this cross-cultural trend can be explained by information constraints. This should later be comprehensively compared to alternate hypotheses.~\cite{mcbrideMelody2024a} It remains to be seen whether these constraints can be explained by biological constraints on memory.~\cite{millerMagical1956a,cowanMagical2001,bradyCompression2009}
  While iterated-learning experiments are an excellent way of studying constraints
  on melody evolution, future work should study melodies longer than \num{15} notes (\fref{fig:fig5}C)
  and control for the effects of production variance.

\nsb{Scaling relations in pitch and rhythm representations.~}
  In our supplementary analyses we find several scaling relations between
  melodic viewpoints (SI Section 2, SI Fig. 3), and demonstrate that they can be reproduced using
  basic ingredients: scale structure and scalar motion; simple rhythms and entrainment to a beat (SI Section 4, SI Fig. 3).
  The first implication is that it can be sufficient to study a minimal set of
  viewpoints, as we have done here. First-order viewpoints were
  typically more efficient than second-order viewpoints, although the difference
  between Chroma and melodic interval entropy is not so extreme (SI Fig. 5).
  This may explain why notation systems appear to predominantly use first-order viewpoints,\cite{ellingsonNotation1992,bentNotation2001}
  with a few that use second-order pitch viewpoints (\eg, Byzantine neumes),~\cite{dalitzOptical2008} although another explanation for this is that mistakes are propagated using second-order viewpoints.
  These findings also raise the question of how melodies are encoded in the
  brain, and which viewpoints are most relevant -- our results suggest that
  there are multiple candidates for pitch, given similar levels of information efficiency.

\nsb{Rhythm is less complex than pitch in melodies.~}
  We find several indications that rhythm is under stronger constraints
  than pitch. Pitch entropy is higher than rhythm entropy in \SI{78}{\%}
  of corpora, although this could be due to over-representation of
  European corpora. The correlation between entropy of different
  rhythm viewpoints is much higher than for pitch viewpoints (SI Fig. 3, SI Fig. 5),
  suggesting that rhythmic constraints are more stringent than pitch
  constraints. Higher correlations are also observed
  between $\AL$ and the Gini coefficient for rhythm than for pitch (\fref{fig:fig2}).
  Compared to pitch, rhythm is also found to be much more repetitive within songs (\fref{fig:fig4}A),
  and exhibits a stronger correlation between within-corpus
  repetition and entropy (\fref{fig:fig4}C).
  However we note that we have focused on \textit{melodic} corpora, and there
  may be other sources of music that exhibit different effects.
  Likewise, there may be some systematic bias due to the use of symbolic
  notation, so alternate methods ought to be developed to investigate this using audio recordings.

\nsb{Channel capacity for music.~}
  Measuring the information rate may tell us something about the channel capacity
  of human melodic communication. For example, if we assume a mean tempo of 90 beats per minute,
  the average information rate in Folk music is approximately 6 bits per second,
  which is comparable to an estimate of the phonemic information rate in French.~\cite{giroudChannel2023}
  However this is at best a crude approximation, and the reality is much more complicated.
  On one hand, knowledge of a song's genre,
  mode (\eg, major / minor) or function (\eg, dance, love) will reduce the information rate,
  while on the other hand the notated music studied here neglects important
  details of pitch (vibrato, ornaments) and rhythm (accent, microtiming),
  and completely ignores other salient musico-linguistic dimensions (dynamics, timbre,
  lyrics). Estimating the complexity of music ought to take into account these
  different dimensions and degrees of detail,
  ~\cite{mauchEvolution2015,parmerEvolution2023,hamiltonTrajectories2024}
  and only then will the estimations of musical channel capacity bear relevance to musical practice.

  \nsb{Limitations.~}
  We have not differentiated between vocal and instrumental melodies. Most of our corpora contain only vocal melodies, some only contain instrumental melodies, and many melodies are not annotated with this information. In theory one should observe different constraints, since vocal motor constraints do not apply to instruments (although other physical constraints may).
  We do have one pair of corpora of Dutch songs~\cite{van19} that are separated into vocal and instrumental songs, which shows differences in scale range (vocal, \num{12.5} semitones; instrumental, \num{15.6} semitones), and some slight differences in Melodic-Interval distributions (octaves are more common and unison intervals are less common in instrumental compared to vocal music). This should be investigated more systematically in future work. 

  By studying symbolic notation, we have ignored intonation. One may rightly be concerned about potential Western influence. Thankfully our analyses of pitch mainly use the Chroma viewpoint, for which the results do not at all depend on the tuning system (SI Section S1A). However, results using the Melodic-Interval viewpoint may change based on the tuning system. Out of the corpora studied, the results from the Turkish Makam corpus are most likely to be affected by this.\cite{karaosmanogluTurkish2012} Estimating the size of such effects is worth further study.

  We separated corpora of music for adults based on whether it is primarily transmitted orally (Folk) or through notation (Art). This simplistic classification misses some nuances between musical traditions. For example, some orally transmitted music relies on vocables (\eg, Indian sargam). The use of this mnemonic device in singing practice may lead to different constraints on information properties. Additionally, many folk traditions have recently started to use notation which may have led to shifts in constraints. Studying these nuances will require specialized datasets from the relevant musical traditions.

  Lastly there is a question about the validity of studying symbolic notation derived from transcription. Within ethnomusicology it is known that one can transcribe different levels of detail for different purposes.\cite{hoodEthnomusicologist1971} It is difficult (perhaps impossible) to know how this has varied across the numerous sources we have included here. Furthermore, there are clear examples (\eg, the Korean corpora) where non-discrete vocal ornaments are notated in transcriptions, but such notation is culture-specific and is not amenable to the type of cross-cultural comparison undertaken here. Circumventing this issue would require a method of estimating complexity from raw audio recordings, and this method would have to avoid cultural bias. Lacking such a method, analyses of symbolic notation are a useful tool to guide further work.

  \nsb{Conclusion.~}
  
  We have focused in this study on the information properties of melodies,
  and with the ultimate aim of uncovering cognitive constraints that act across
  societies to shape melodies. We found cross-cultural empirical evidence of constraints on information rate, which appears to be sufficient (alongside constraints on melodic intervals and melody length) to reproduce the observed number of scale degrees. We hypothesize that these constraints stem from memory. An upper limit on the complexity may be expected as songs become too complex to follow or remember. Likewise, predictive coding theory suggests that simple songs may be 'boring' because they lead to low prediction reward. Ultimately we need to consider a range of theories and devise cognitive tests to fully test the hypothesis.

\section*{\sb{Methods}}

\nsb{Melodic Similarity.~}
  To estimate how similar melodies are across societies, we compare one Korean traditional
  song, Arirang, to a database of Irish folk songs, "thesession.org".~\cite{keithTheSessiondata} We take the first \num{10} notes
  from Arirang and convert them to a sequence of $\pMi$ values. We compare this sequence
  to all $n_{\textrm{mel}} = $ \num{37833} $\pMi$ sequences of Irish songs. The probability that two
  identical sequences are drawn is given by $\AL^\LL$. For Arirang $\AL = 5$,
  while in the Irish songs the mean value for \num{10}-note sequences is $\AL = 5.7$.
  We choose $\AL=5$ to get a conservative estimate of the probability of
  finding two identical \num{10}-note sequences,
  $p_{10} = 1/(5^{10}\times5^{10}) \approx 10^{-14}$. For a melody of length $n$,
  the probability of finding a specific \num{10}-note sequence is $p = p_{10}(n - 9)$.
  Thus the expected number of times to find a melody containing the Arirang sequence is
  $\sum\limits_{i}^{n_{\textrm{mel}}} p_i \approx 4\times 10^{-8}$,
  where $p_i$ is the probability for the $i^{\textrm{th}}$ melody.
  We find \num{8} melodies that include the Arirang sequence.
  Thus, we observe this sequence at a rate that is approximately \num{200} million
  times higher than chance. Since this calculation is limited to two cultures,
  it should be treated as an illustrative example rather than a general prediction.

\nsb{Melodic Corpora.~}
  We chose corpora with the aim of covering musical styles of different levels
  of complexity, and to cover geographically diverse societies (SI Section S1A).
  At the lower end of the complexity scale we have music for children (\textbf{Child}, 7 corpora).
  \textbf{Folk} corpora consists of music performed by non-professional musicians,
  and passed down orally (62 corpora). \textbf{Art} corpora are associated with professional
  musicians, and music that is transmitted with the aid of written notation (39 corpora).
  We also use a set of \textbf{Teaching} corpora that are used to teach singing at different levels (5 corpora).
  Only monophonic musical lines are considered; for a few
  polyphonic vocal works we extracted a single vocal line for analysis.
  In total we collected \num{113} melodic corpora from different musical traditions
  and societies (SI Table 1), amounting to about \num{36000} melodies.
  The Art corpora are all European except for one Turkish collection.
  The Folk corpora are skewed towards European (30), and indigenous North American (16) societies,
  but also includes other regions (16) such as Asia and Africa (SI Fig. 1).
  While the majority of the corpora were obtained from previously-published sources (SI Section S1B),
  we additionally coded \num{12} new corpora to bridge gaps (SI Section S1C).

\nsb{Joint entropy null model.~}
  We calculate the expected values of joint entropy, $\HH(\pC, \rD)$, if pitch and rhythm
  entropy are uncorrelated. We randomly sample pitch and rhythm entropy
  ($\HH(\pC)$, $\HH(\rD)$) and mutual information ($\MI(\pC,\rD)$) from the set of
  average values per corpus in a set of corpora, and calculate the joint entropy,
  \begin{equation}
  \begin{split}
  \HH(\pC, \rD) = \HH(\pC) + \HH(\rD) \\ - \MI(\pC, \rD)~.
  \end{split}
  \end{equation}
  We sample $10^4$ times with replacement to get a distribution (\fref{fig:fig3}C),
  and calculate the variance (\fref{fig:fig3}D).

\nsb{Repetition within melodies.~}
  Instead of examining cross-cultural differences in repetition we control for
  it by algorithmically removing repetition (SI Fig. 19, SI Alg.3, SI Fig. 20). 
  We take a melodic sequence $S$, and find all substrings of length $\LL > \LL_{min}$
  that repeat at least $N = 2$ times, where $\LL_{min}$ is a parameter that we set. The maximum $L$ is given by the floor of half the length of the full sequence. Out of all identified substrings, we find the substring $S_m$ that maximises $N \times \LL$
  and we remove all instances of it, separating the original sequence into a set
  of substrings $S'$. We then recursively repeat this process on all substrings $S'$ (and their substrings), until there
  are no more substrings for which $\LL > \LL_{min}$ and $N > 1$. The total combined
  length of all unique substrings is what we call the the length of non-repeated sequence, which is our proxy for melody length that controls for repetition.
  For an appropriate value of $\LL_{min}$, we calculate the typical
  length of repeated substrings in random sequences. To obtain this, we calculate the average length of non-overlapping
  substrings of randomly shuffled melodic sequences (SI Fig. 21). Since we find
  that this average length is $2 \geq \LL \geq 3$, we choose $\LL_{min}=2$.

\nsb{Repetition between melodies.~}
  To estimate the amount of repetition between melodies in a corpus,
  we use IDyOM (Information Dynamics of Music), a variable-order
  Markov model that predicts the $i^{th}$ note in a sequence;
  in particular, we use the long-term IDyOM model.\cite{pearceStatistical2018}
  IDyOM is first trained on a set of melodies from a corpus that does
  not include the target melody: n-grams up to order $n$ are counted
  and predictions from each order are combined in a variable-order model
  using the prediction-by-partial-matching (PPM) algorithm.\cite{harrisonPPMdecay2020}
  The trained model is then used to calculate the average information content
  of the target sequence, where information content is the log probability
  of each note in the sequence, $\IC = \log P(x_i|x_{i-1},\ldots,x_{i-n})$.

  Direct comparison of $\IC$ across different corpora is inadvisable,
  since the absolute value of $\IC$ depends on many factors, including
  alphabet size $\AL$, sequence length $L$, the number of training examples,
  and the unigram statistics. We control for the number of training
  examples by only training the model on \num{10} melodies (results do not depend on the size of the training set; SI Fig. 15); for each target
  melody the training melodies are randomly selected without replacement.
  We control for $L$ by truncating sequences at $L=50$ (results do not depend on the truncation length; SI Fig. 16).
  It is more difficult to control for $\AL$ and the unigram statistics, since some
  corpora have been transposed to a single key (decreasing $\AL$) while others have not.
  Thus, instead of reporting $\IC$ directly, we also calculate the information content, $\IC_r$,
  using a model trained on the same set of melodies but with the letters randomly shuffled,
  and report $\IC_r - \IC$. This measure approximates the reduction of information of a melody
  given knowledge of other melodies from the same corpus, in a way that controls for potential differences in unigram statistics.

\nsb{Generative model of pitch sequences.~}
  We generate pitch sequences by drawing $\LL$ melodic intervals from the overall distribution
  of melodic intervals across all Folk corpora (SI Fig. 18A), within a fixed pitch range, $\OC$.
  We generate $10^8$ sequences, convert them to the Chroma viewpoint by collapsing pitches onto a single octave. We calculate $\AL$ and $\HH$ for each Chroma sequence.
  We then separate the sequences into groups according to the number of scale
  degrees $\AL$. To achieve sufficient sampling of $\AL$, we choose
  values of $\OC \in \{0.5, 1, 1.5, 2\}$. To investigate how $\LL$ and $\AL$ affect the probability of generating
  scales with $\HH(\pC) < 2.8$ bits (the \SI{95}{\%} percentile of the empirical Folk $\HH(\pC)$ distribution),
  we repeat this process with different values of $13 \leq L \leq 45$ -- corresponding to the
  \SI{90}{\%} inter-quartile range of melody lengths after controlling for repetition (SI Fig. 20) --
  and examine $\HH$ as a function of $\AL$ (\fref{fig:fig5}D).

  To estimate the likelihood that scales using $\AL$ degrees would generate the empirical
  Folk $\HH(\pC)$ distribution, we compare this to the generated $\HH(\pC)$ distribution for each $\AL$.
  We estimate the probability density, $P(\HH)$, of $\HH(\pC)$ for all Folk melodies,
  using kernel density estimation (Gaussian kernel, we choose the bandwidth using
  Silverman's rule). To prevent zeros in $P(\HH)$, we add to $P(\HH)$ an uninformative prior to get
  $P'(\HH) = \alpha P(H) + (1-\alpha) / \beta$, where $1/\beta = [\int_0^5 d\HH]^{-1}$ is a uniform
  distribution over the range $0 \leq \HH \leq 5$ bits; we set $\alpha = 0.999$. We estimate the probability density, $Q(\HH)$,
  of $\HH(\pC)$ for all model-generated melodies of alphabet size $\AL$ using the same procedure
  for $P(\HH)$. The log-likelihood per melody that melodies of alphabet size $\AL$ generated the empirical distribution, $P(\HH)$,
  is $\log \mathcal{L}(\AL | P(\HH)) = \int Q(H) \log P'(\HH)$; in practice we evaluate this numerically
  using bins of width \SI{0.005}{bits}.

\nsb{Author Contributions.~} JM conceived the project; JM, NK, MS and YN collected the data; JM and MP analyzed the data; JM, NK and YN wrote the manuscript; TT and MP supervised the project; JM, YN, MP and TT revised the manuscript.\\

\nsb{Data and Code Availability.~} Data and code can be found at \href{https://zenodo.org/records/13338167}{https://zenodo.org/records/13338167}.

\nsb{Acknowledgements.~} This work was supported by the Institute for Basic Science, Project Code IBS-R020-D1. We acknowledge discussions wtih Prof. Andrew Killick on cross-cultural musical notation. We thank Prof. Tuomas Eerola for sharing a corpus of songs from South Africa. We thank Claire Arthur and two anonymous reviewers for reviewing this work.

% Bibliography
% \bibliographystyle{unsrtnat}
\bibliography{master}

\end{document}

\title{Supporting information for \\ ``Information \& motor constraints shape melodic diversity across cultures''}

  \author{John M. McBride}
    \affiliation{Center for Algorithmic and Robotized Synthesis, Institute for Basic Science, Ulsan 44919, South Korea}
    \email{jmmcbride@protonmail.com}
  \author{Nahie Kim}
    \affiliation{School of Business Administration, Ulsan National Institute of Science and Technology, Ulsan 44919, South Korea}
  \author{Yuri Nishikawa}
    \affiliation{Department of Molecular Life Science, Tokai University School of Medicine, Kanagawa, Japan}
  \author{Mekhmed Saadakeev}
    \affiliation{Department of Biomedical Engineering, Ulsan National Institute of Science and Technology, Ulsan 44919, South Korea}
  \author{Marcus Pearce} 
    \affiliation{Cognitive Science Research Group, School of Electronic Engineering \& Computer Science, Queen Mary University of London, London, United Kingdom}
    \affiliation{Department of Clinical Medicine, Aarhus University, Aarhus, Denmark}
  \author{Tsvi Tlusty}
    \affiliation{Center for Algorithmic and Robotized Synthesis, Institute for Basic Science, Ulsan 44919, South Korea}
    \affiliation{Departments of Physics and Chemistry, Ulsan National Institute of Science and Technology, Ulsan 44919, South Korea}
    \email{tsvitlusty@gmail.com}

\maketitle
\tableofcontents

\section{Melodic Corpora}
\subsection{Overview}

  We assembled a large set of melodic corpora that would not have been possible
  without the efforts of a large range of people, from those that collected
  and transcribed the original melodies, to those that digitized the collections
  in machine-readable formats such as MIDI, `kern', and XML (Extensible Markup Language),
  and ABC notation.~\cite{huronHumdrum1997,walpi14}
  We used the Humdrum Toolkit~\cite{huronMusic2002} and the Music21 python package~\cite{cuthbertMusic212010}
  to convert between data formats and parse the melodies.
  A summary of all corpora can be found in \tref{tab:tab1}.

\subsubsection{Inclusion Criteria}
  We chose corpora focused on melodies; most of these are
  monophonic, but some polyphonic choral corpora are included, in which case we only
  extract a single monophonic melody (the top voice). We ignored any harmonic or
  percussive accompaniment. Examples of corpora that we left out are orchestral 
  works \cite{neufd18}, or transcriptions of polyphonic instruments
  where melodies cannot be easily and unambiguously
  determined algorithmically \cite{charryMande2000}. We also restricted our search to corpora of
  symbolic notation, rather than working with higher resolution data
  such as audio, or fundamental frequency (F0) annotations;
  this type of data requires conversion to symbolic notation, a feat that is
  not easily achievable using algorithmic approaches, and otherwise requires manual
  transcription by experts in the associated musical tradition.

\subsubsection{Corpus Types}
  We can group the melodies by the type of performers, how the melodies are learned,
  and how complex they are: \textbf{Folk} corpora covers music performed by non-profession
  musicians, and is transmitted orally; \textbf{Art} corpora are associated
  with professional musicians, and depend much more heavily on notated music;
  \textbf{Child} corpora covers music for children;
  \textbf{Teaching} corpora refers to sets of melodies that differ explicitly
  in their difficulty, as they are used to teach singing.
  We hypothesized that if there are cross-cultural constraints on the information content
  of melodies, these would be most apparent in Folk, and Child corpora.
  We expected that Child corpora would provide a lower bound on the complexity
  of melodies. We expected that Art corpora would be less constrained than Folk corpora
  due to the performers being professional, full-time musicians, as opposed
  to amateur musicians that are historically associated with Folk music.
  Teaching corpora are included to serve simply as an example of how singing difficulty
  may correlate with information complexity.
  Accordingly, we aimed to get wide geographical coverage for Folk and Child corpora,
  to see if the hypothesized constraints are consistent across cultures.
  Most of the previously-published corpora consists of music from Europe and
  indigenous societies of North America, with a few corpora covering
  Turkey and China; only one collection of German children's songs were available.
  To supplement these, we sourced and digitized additional corpora (Section 1C).

\subsubsection{Considering different tuning systems}
  A superficial critique of studying symbolic notation is that it ignores
  the details of how instruments or voices are tuned. Indeed, tuning changes
  across geography and time, but this has little bearing on most of the
  information properties that we report. The same song, as read in symbolic
  notation, can be performed in many different tuning systems, but this will
  not change the number of scale degrees ($\AL$), or the entropy ($\HH(\pS)$).

  The one property that is affected is $\pMi$, as using an equidistant
  tuning system minimizes the $\AL(\pMi)$. For example, if there are
  $\AL(\pC)=5$ notes in a scale, there are $\AL(\pMi) = \AL(\pC)(\AL(\pC)-1)/2=10$
  possible intervals. If the scale is equidistant, then there are only
  $\AL(\pMi)=4$ possible interval, which ultimately reduces $\HH(\pMi)$.
  If the tuning system is not regular, it is possible that all $\pMi$
  values are unique, and $\HH(\pMi)$ would be much higher.
  
% Consistency checks:
% We automatically detected and removed any melodies with errors
% by checking that the number of note durations equals the number
% of pitches. We removed grace notes, since the duration / $\rI$ is unspecified;
% we ignore rests when using the $\rD$ viewpoint, since the pitch is unspecified. 

\subsection{Pre-existing corpora}
  We separated these collections into sub-corpora, in an attempt to
  avoid grouping melodies that come from distinct cultures.

  \begin{itemize}
    \item The Essen collection is split into 18 (15 from Europe, 3 from China)
      sub-groups according to geography or culture \cite{sch95,brinkmanExploring2020}. This includes 17 Folk and 1 Child corpora.
    \item The Densmore Native American collection includes 17 Folk and 1 Child corpora. They are separated according
      to how they were published, which can include more than one society grouped together \cite{shapi14}.
    \item The KernScores collection includes 9 Folk corpora and 1 Art corpus from Europe \cite{sapis05}.
    \item The ABC collection includes 6 Folk corpora (4 from Europe, 2 from the Middle East) \cite{abc22}.
    \item The Meertens Dutch collection includes two Folk corpora, one for songs and one for instrumental melodies \cite{van19}.
    \item The SymbTr Turkish collection was separated into one Folk and one Art corpus based on song annotations \cite{karaosmanogluTurkish2012}.
    \item The digital archive of Finnish songs is included as one Folk corpus \cite{eerolaSuomen2004}.
    \item A set of songs from South Africa is included as one Folk corpus \cite{eerolaPerceived2006}.
    \item One Mexican Folk corpus and one Hawaiian Folk corpus were obtained from "bethnotesplus.com" \cite{Www2023}. 
    \item The Josquin Research Project collection includes 8 Art corpora \cite{rod22}.
    \item The Lieder collection and a collection of French and German lieder contain overlapping
      sets of composers \cite{vanjn10,gothamScores2018}. When a composer was present in both collections, we took the set from the collection
      with the greater number of compositions. In total these include 27 Art corpora, separated by composer.
    \item A collection of vocal lines from Mozart opera was obtained from the KunstDerFuge website \cite{kunst}.
    \item The MeloSol collection is separated into 5 Teaching corpora by book \cite{bakerMeloSol2021}.
  \end{itemize}

  \begin{figure*}  \centering
  \includegraphics[width=0.90\linewidth]{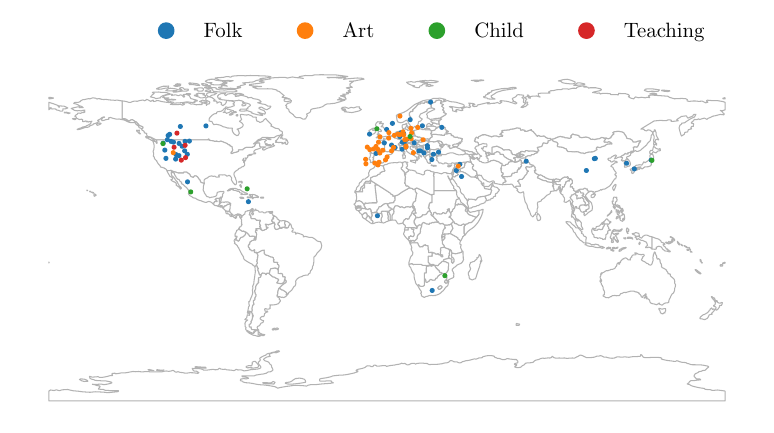}
  \caption{\label{fig:si_orld}
  \textbf{Geographic distribution of corpora} colored by corpus type. Geographical positions are placed by country with some jitter.
  }
  \end{figure*}

\subsection{Newly-coded corpora}

  We chose books based on the aforementioned selection criteria (Section 1A),
  and only digitized books if they had 20 examples.

  Each book was digitized by one or two primary coders, checked for inaccuracies
  algorithmically, and finally verified by a second or third coder.
  We mainly coded the melodies in kern format, with the help of \textit{Verovio
  Humdrum Viewer}~\cite{verov}; for Ghana, which included mainly call-and-response songs,
  we used \textit{MuseScore}~\cite{musescoredevelopercommunityMuseScore2023}
  since this made it easier to separate the solo and chorus
  parts. We used \textit{Audiveris} optical music recognition software
  to extract digital transcriptions~\cite{bitteurAudiveris2023}.
  We found that Audiveris produces transcriptions with
  many mistakes, so we these were only ever used as starting points for manual coding;
  for some sources (e.g. Charles Ives songs) the starting points were so bad that manually coding
  from scratch was faster.
  We provide all newly-digitized corpora in both kern and xml format for
  others to use. Although our primary concern in this paper is melodic content,
  we additionally annotated song type (\eg, work song, love song, lullaby) and
  phrase markings where available. We also found examples of non-standard use
  of symbols which lacked explanations in the text; while we can guess the meaning
  in some cases (\eg, glides, melisma), we ignored them all.

  \nsb{Folk Songs of Ghana}
  A collection of Ghanaian, orally-transmitted folk songs from the Akan people 
  \cite{nke63}. The author notes that the scale used is ``diatonic
  in character'', and although it does not correspond to equal temperament, it is
  easy to notate using staff notation. Pitch drift (gradual change of tonal center)
  was noted as occurring during performance, but eliminated from transcriptions.
  Most songs were sung in keys ranging from F to Bb, and all transcriptions
  were transposed to G. Songs are typically call-and-response between solos by
  one or more alternating leaders and a chorus. Chorus parts are typically annotated
  as two parts moving in parallel thirds. We only analyse the solo parts, which
  make up the bulk of the transcriptions.

  \nsb{Slave Songs of the United States.}
  A collection of songs from former slaves, recorded by a group of abolitionists
  just after the abolition of slavery in the USA \cite{allenSlave1867}.
  We considered whether to include this collection, given the ongoing debate about
  ``decolonisation'' of research \cite{sauveAntiColonial2023}. We acknowledge the colonial slave trade
  and the damage that was done, and its lasting legacy that is still felt in
  academia as in other domains. Despite the relatively (for the time)
  good intentions of the authors,
  they still occasionally write in a way that would not be acceptable today (\eg,
  dichotomies between civilised and savage behavior). The authors ultimately profited
  from the culture of ex-slaves, and although most of the collection was obtained 
  directly from singers, the singers / songwriters are almost never credited.
  However we still feel that it is appropriate to
  include this collection as it is an important part of the scholarly record.
  The authors note that the notated versions of the songs are representative
  examples out of a tradition that includes many variations. Occasionally
  some of the variations are notated in staff notation as simultaneous notes,
  in which case we take the high note.
  The authors also note characteristics of the songs that were
  not captured in notation: pitch slides and turns, vocal timbral changes,
  staggered chiming of chorus voices, and timing ``irregularities''.
  The songs are mostly spiritual songs, but also include work and boat songs amongst others.

  \nsb{Rock it Come Over: The Folk Music of Jamaica.}
  A collection of folk songs of Afro-Carribean people of Jamaica \cite{lewinRock2000}, that includes
  songs of diverse origins dating back to the years of slavery.
  The song types include work, dance, story, spiritual, love, and children's songs.
  We separate the collection into one Folk and one Child corpus. 

  \nsb{Folk Songs of Korea.}
  A collection of Korean folk songs released by the National Classical Music Institute
  of Korea \cite{folksongkorea69}. The collection was built up over many years
  of interactions with local singers and performers, and from numerous field recordings.
  The songs are from numerous provinces in South Korea
  (including Jeju island) and North Korea.
  The songs are often stories, but include songs used in rituals such as
  weddings, harvest festivals, ancestral rituals and community gatherings.
  The melodic style includes extensive use of vibrato and pitch bends, which
  is notated in transcriptions but not included in our analysis.

  \nsb{Chunhyangga.}
  Transcriptions of songs from Chunhyangga, a pansori folktale from Korea \cite{chunhyangka77}.
  These are a set of narrative songs about a love story,
  based on recordings of Kim So-hi from 1958. The narrative has been set to text
  since approximately three centuries ago (e.g. Chunhyangjeon), while the vocal tradition has been passed down through
  pansori masters, whose styles are imitated and elaborated on through improvisation.
  The style includes a lot of vocal ornaments (\eg, timbral changes, pitch glides,
  vibrato); some of these are notated in
  the transcriptions, but there is no description within book of the meanings of each symbol.

  \nsb{Survey of Japanese Folksongs: Okinawa-Amami Islands}
  A collection of Okinawa folk songs collected by the Japanese public broadcasting company Nippon Hoso Kyokai (NHK)\cite{nipponhosokyokaiRiBenMinYaoDaGuan1989,nishikawaCultural2022}. Okinawa is located in southwestern Japan and consists of numerous culturally diverse islands. We selected only songs from Okinawa island, the largest island, and separated the collection into one Folk and one Child corpus. The folk corpus includes ritual, work, and amusement songs. Lullabies are excluded from the corpora. Vibrato is occasionally noted in transcriptions but not included in our analysis.

  \nsb{Kyrgyz Folksongs.}
  A collection of 84 Kyrgyz folksongs (out of a total of 426) \cite{sipos14},
  chosen to be representative of a range of song types and forms.
  The songs include wedding songs, laments, lullabies,
  lyrical songs, and religious (Caramazan) songs.

  \nsb{114 Songs by Charles Ives.}
  A self-published collection of songs composed by Charles Ives \cite{ives1141922}.
  Although we have access to a lot of pre-existing corpora of Western art music,
  we chose to also add some of this collection as an extreme example of music that
  is tonally and rhythmically complex, with little repetition between songs.
  We specifically chose these songs by Ives since we wanted a collection of monophonic
  music from a modernist composer. 
  
  \nsb{Venda Children's Songs.}
  A collection of children's songs of the Venda people from northern South Africa \cite{blackingVenda1967}.
  The songs are categorized by social function, \eg, counting songs, action songs,
  boy / girl songs, mockery songs. The transcriptions include metrical information
  about beat accents that is not included in our digitizations.

  \nsb{British Nursery Rhymes.}
  A collection of British nursery rhymes dating from 16th to 18th centuries,
  gathered from oral sources across England \cite{mof04}.
  The collection includes lullabies, action songs, counting songs, animal rhymes,
  story rhymes and sing-alongs.

  \nsb{El Patio de mi Casa.}
  A collection of Mexican children's songs, curated from various sources \cite{mon07}.
  Many of these songs were passed down through family, while others were documented
  by folklorists and researchers such as Vicente T. Mendoza, and Francisco Moncada Garc\'{i}a.
  Out of this collection we digitized all songs except chants, which have constant pitch.
  The collection includes singing games, narrative songs and lullabies.

\section{Melodic Viewpoints}

  \begin{figure*}  \centering
  \includegraphics[width=0.90\linewidth]{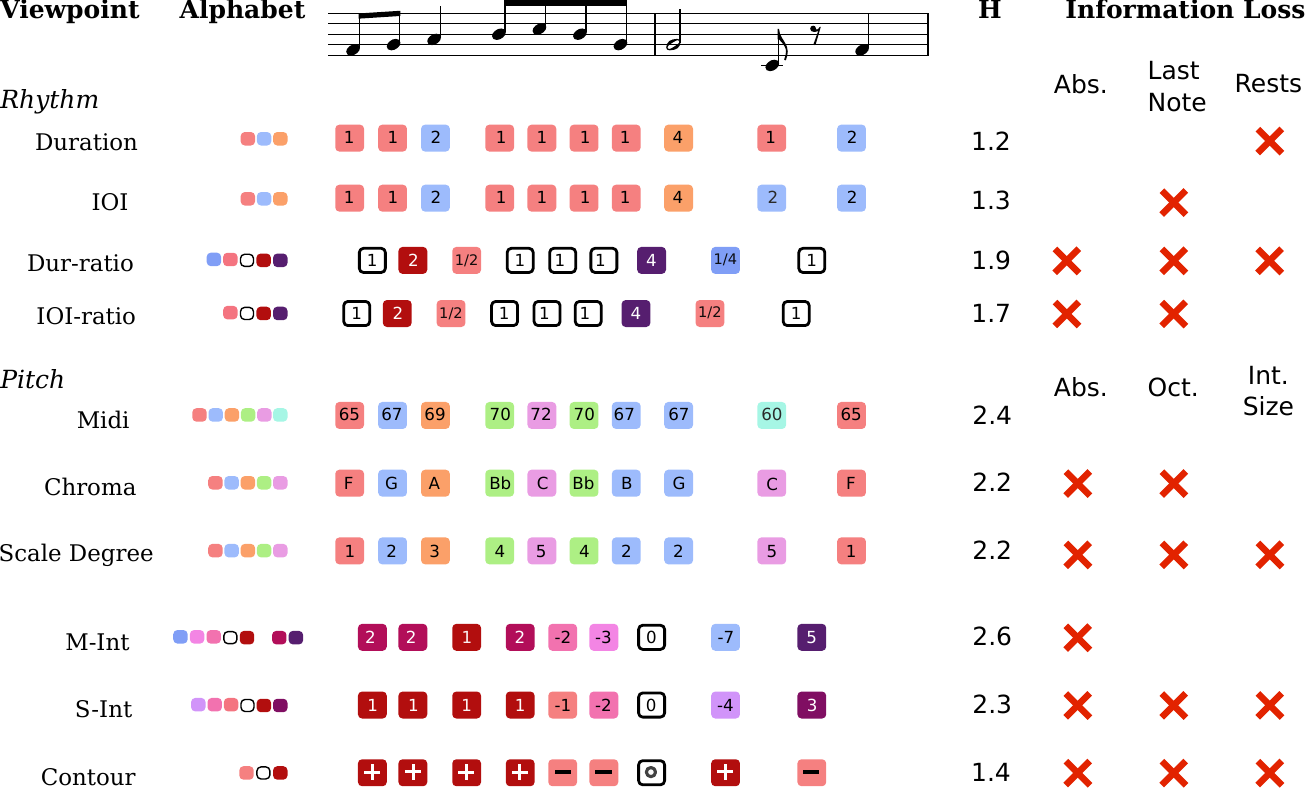}
  \caption{\label{fig:si1a}
  Illustrative example of a melody, the different viewpoints, and some of their information
  properties. Alphabet size is the number of unique letters / values; entropy, $\HH$,
  is a measure of information, or complexity; different viewpoints differ in what information
  they contain, which is represented via columns on Information Loss.
  Duration describes the length of time a note is held; it ignores the value of rests.
  IOI describes the length of time in between note onsets; it is not defined for the last note.
  IOI-ratio (and $\rDr$) is the ratio of consecutive IOI ($\rD$) values; this is a time-invariant
  representation.
  Pitch describes the absolute pitch in Midi units. Chroma is $\pM \mod 12$; it is restricted to a single
  octave range so absolute pitch is lost, but it still retains some information about pitch
  position. Scale Degree is equivalent to Chroma, but on an ordinal scale so that the relative
  size between notes is lost.
  Melodic Interval ($\pMi$) is the difference between successive $\pM$ notes.
  Scale Degree Interval ($\pSi$) is the difference between successive $\pS$ notes.
  Contour describes whether the consecutive pitch is higher, lower or the same as the preceeding pitch.
  }
  \end{figure*}

\subsection{Overview}
  Viewpoints are representations of melodies,
  and fall into three categries: pitch sequences, rhythm sequences, and 
  joint sequences encoding both pitch and rhythm. We can further subdivide
  viewpoints into either 1st order, or 2nd order (time or log-frequency
  invariant) representations. We here define the different viewpoints,
  and compare them in terms of information efficiency and information loss.

  For Rhythm, first order viewpoints describe either the duration ($\rD$) that a note is held for,
  or the time between successive note onsets $i$ and $i+1$,
  \begin{equation}\label{eq:ioi}
  \rI_{i} = t_{i+1} - t_i~,
  \end{equation}
  known as the inter-onset-interval ($\rI$).
  $\rI$ is more commonly used in cognitive experiments, while $\rD$ is used in music notation.
  The main difference between these two is that $\rI$ accounts for the periods of silence in between
  notes, while $\rD$ sequences ignores the duration value of rests. Use of $\rI$ is problematic in some corpora when there are
  multiple parts, which can entail long periods of silence due to, e.g. call and response singing.
  Second order viewpoints describe the characteristic timescale of a note in relation to its successive
  note, and are therefore invariant to the absolute time of notes (tempo):
  $\rI$-ratio ($\rIr$) is commonly used to describe the $\rI$ of a pair of notes $i$ and $i+1$,
  in relation to the subsequent (overlapping) pair $i+1$ and $i+2$,
  \begin{equation}\label{eq:ioir}
  \rIr_{i} = \frac{\rI_{i+1}}{\rI_{i}} = \frac{t_{i+2} - t_{i+1}}{t_{i+1} - t_{i}}~;
  \end{equation}
  $\rIr$ is typically normalized so that it exists between 0 and 1 by dividing by the sum,
  $\rI_{i} + \rI_{i+1}$, although this does not affect the information properties so we do not
  follow this step. Similarly, one can define a duration ratio ($\rDr$).

  For the example melody shown in \fref{fig:si1a}, $\rI$ has a smaller alphabet size, $\AL$, than $\rIr$,
  and has lower entropy, $\HH$. In this case, converting $\rI$ to $\rIr$ results in a loss of information,
  as we need to know the durational value of one of the notes to reconstruct $\rI$ from $\rIr$.
  So, despite $\rIr$ having greater $\HH$, it has less unique information, and is thus
  less efficient. It is not as clear whether duration or $\rI$ is more efficient, since they
  have similar $\HH$, and they both contain unique information that the other does not
  (duration ignores rests, while $\rI$ ignores the duration of the final note).
  
  For pitch, first order viewpoints describe the position of pitch within
  some frame of reference:
  Western staff notation describes music on a 12-note chromatic scale, and this is captured
  in the commonly used computer \textit{midi} respresentation. This describes the 
  absolute frequency of a note in Hz, such that a midi value of 69 is concert pitch, 440 Hz.
  The Chroma representation is a transformation of midi that collapses pitch
  to any single octave range, $pC \equiv \pM \pmod {12}$; in \fref{fig:si1a} it is
  shown using the \textit{solf\`{e}ge} notation. 
  Scale Degree is the collapse of $\pC$ onto an ordinal scale.
  Second order viewpoints describe the changes in pitch (intervals) between
  successive notes, and are therefore invariant to pitch position (key):
  Melodic Interval ($\pMi$) is the difference between succesive $\pM$ values;
  Scale Degree Interval ($\pSi$) is the difference between successive Scale Degrees;
  Contour is the simplest description, whereby only the direction of the change
  in pitch is recorded (up, down, or same).

  $\pM$ is the most informative representation, and as one goes from $\pM$ to other representations
  there is typically some loss of information. $\pC$ loses the octave; $\pS$ loses
  the octave, and the intervallic relation between scale degrees, although this can
  be recovered if one knows the scale.
  $\pMi$ loses the relative position, and the absolute position, although this can be
  recovered by knowing the absolute pitch of a single note.
  $\pSi$ additionally loses the size of intervals, although this can again be recovered
  by knowing the scale. Contour is by far the most compressed representation,
  but unlike the other representations, most of this information is irretrievably lost,
  as up / down could have many possible meanings.
  Of the other representations, only $\pMi$ is clearly less efficient, as it has higher
  $\HH$ than $\pM$, yet it has less information.
  It is difficult to determine the relative efficiency of the remaining representations,
  as one has to consider the cost of additional information (scale, starting position)
  needed to convert to $\pM$.

  Joint viewpoints can be any combination of pitch and rhythm viewpoints.
  We report exclusively on ``$\pC$:$\rI$'' in the main text, but alternative
  viewpoints give similar results.

  Beyond the simple example in \fref{fig:si1a}, it is not clear which pitch representations
  are most informatic / efficient in real melodies, and how this varies within and between
  cultures. For a better understanding of this we next compare $\HH$ of different
  representations for real melodies.

  \begin{figure*}  \centering
  \includegraphics[width=0.90\linewidth]{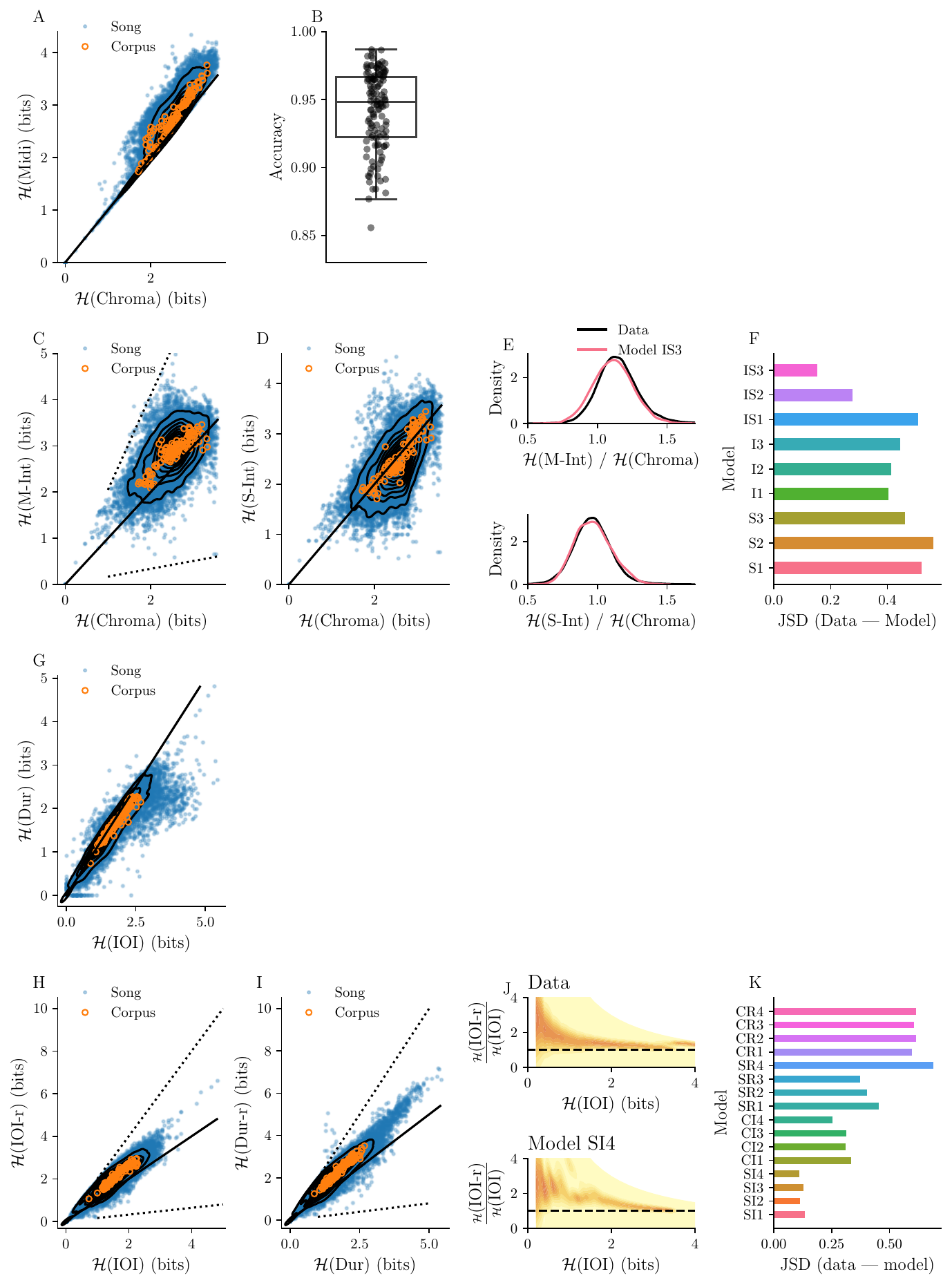}
  \caption{\label{fig:si1b}
  \textbf{Comparing information content of different viewpoints for empirical melodies.}
  A: Empirical scaling of entropy between melodic viewpoints: $\pC$ vs. $\pM$.
  Contour lines show kernel-density estimates;
  blue circles are shown for individual melodies;
  orange circles are corpus averages.
  B: Accuracy of predicting changes in octave using the $\pC$ viewpoint,
  by assuming scalar motion. 
  C-D: Empirical scaling of entropy between melodic viewpoints: 
  $\pC$ vs. $\pMi$ (C), $\pC$ vs. $\pSi$ (D).
  Approximate theoretical upper and lower bounds are indicated in C.
  E: Empirical and model distributions of ratios $\HH(\pMi) / \HH(\pC)$
  and $\HH(\pSi) / \HH(\pC)$ for the best-fitting model ``IS3''.
  F: Jensen-Shannon divergence (JSD) between empirical and model distributions
  for all pitch sequence models.
  G: Empirical scaling of entropy between melodic viewpoints: $\rD$ vs. $\rI$.
  H-I: Empirical scaling of entropy between melodic viewpoints: 
  $\rI$ vs. $\rIr$ (H), $\rD$ vs. $\rDr$ (I);
  Approximate theoretical upper and lower bounds are indicated.
  J: Empirical and model conditional probability distribution of
  the entropy ratio $\HH(\rIr) / \HH(\rI)$ given $\HH(\rI)$,
  $\PP(\HH(\rIr) / \HH(\rI) | \HH(\rI))$.
  K: JSD between empirical and model distributions
  for all rhythm sequence models.
  }
  \end{figure*}

\subsection{Pitch: Pitch vs Chroma}
\begin{figure}  \centering
  \includegraphics[width=0.95\linewidth]{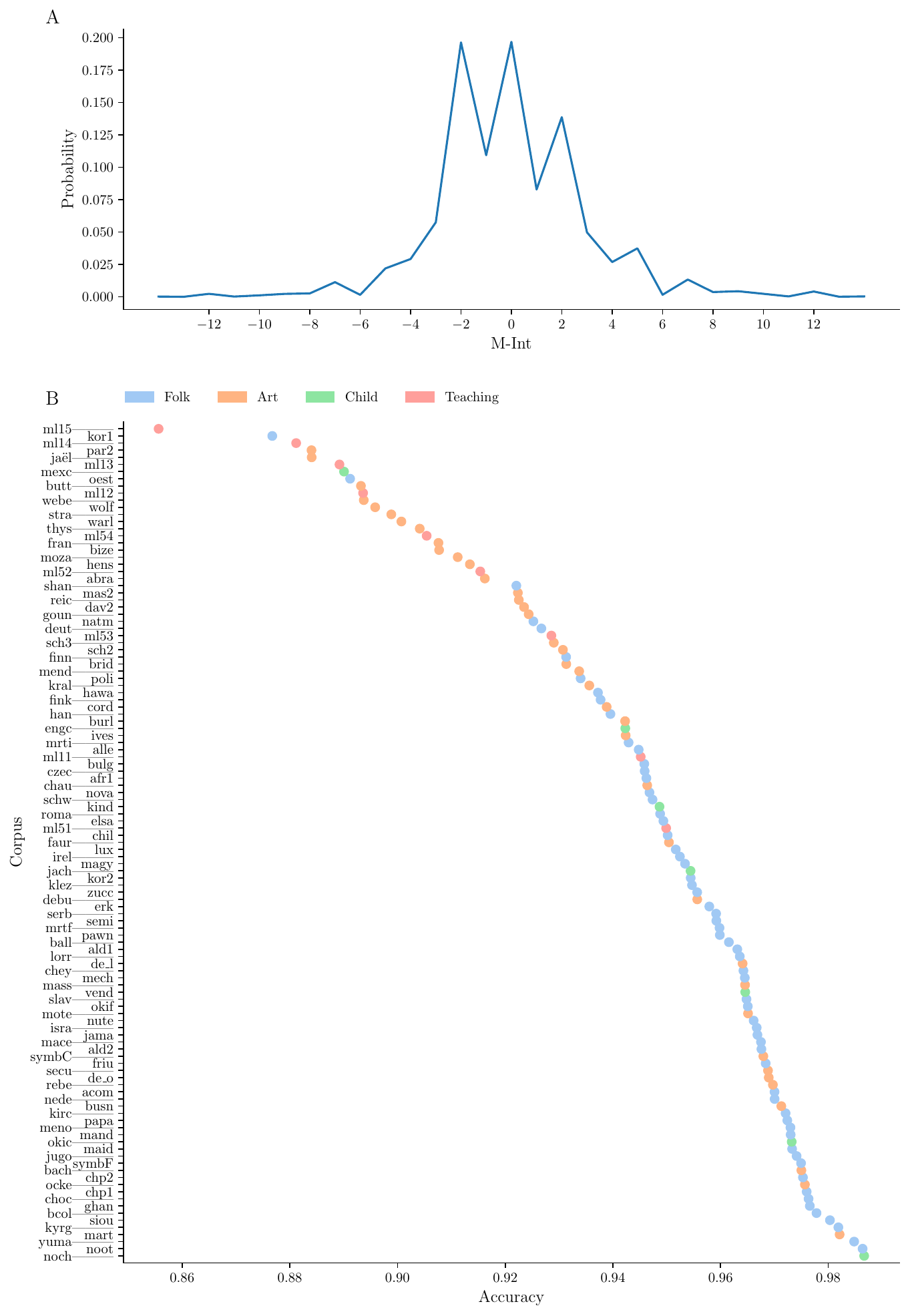}
  \caption{\label{fig:si1}
  A: Melodic-Interval ($\pMi$) probability distribution.
  B: Accuracy of an algorithm that predicts changes in octave in $\pC$ sequences
  by always assuming that the smallest interval is correct, for each
  corpus. Corpora are coloured according to type. \\\\
  The corpora with the lowest accuracy ("ml14" and "ml15") are the Teaching corpora
  that correspond to the highest levels of singing difficulty, which corresponds
  to a higher proportion of large melodic intervals.
  }
  \end{figure}

  $\pM$ is the most informative, but is it efficient? We first compare $\pM$ to $\pC$.
  By definition, $\HH(\pC) < \HH(\pM)$, with the mean difference being about 0.2 bits (\fref{fig:si1b}A).
  However, how much information is irretrievably lost?
  We reason that if melodic pitch progresses predominantly by small changes (scalar motion \fref{fig:si1}A),
  then the lost information can be easily recovered by assuming that out of two possibilities
  (upward or downward motion), the one with the smaller interval is most likely.
  For example, given a change from $\pC=2$ to $\pC=10$, one may predict that
  the interval is $\pMi=-4$ rather than $\pMi=8$.
  This prediction is correct on average \SI{95}{\%} of the time when considering
  all songs from all corpora (\fref{fig:si1b}B); accuracy for individual corpora ranges from
  \SIrange{85}{99}{\%}~(\fref{fig:si1}B).
  Thus, the $\pM$ representation is less efficient than $\pC$.

\subsection{Pitch: first vs second order representations}
  At the level of a single melody, $\AL$ and $\HH$ are the same for $\pC$
  and $\pS$, so we will henceforth disregard $\pS$.
  This leaves us with three pitch viewpoints: $\pC$, $\pMi$, $\pSi$.
  Looking at all melodies from all corpora, there are clear linear relations between
  $\pC$, $\pMi$ and $\pSi$. We find approximate relations of $\HH(\pMi) \sim 1.1 \HH(\pC)$,
  and $\HH(\pSi) \sim 0.9 \HH(\pC)$ (\fref{fig:si1b}C-D).
  While corpora differ in their mean values, they follow these trends (\fref{fig:si8}).

  These correlations between entropy of 1st and 2nd order pitch representations
  is not necessarily expected. For example, we can plot approximate limits on what possible
  ratios exist for $\HH(\pMi) / \HH(\pC)$ (\fref{fig:si1b}C).
  For example, melodies that only consist of up/down semitone steps can have low $\HH(\pM)$ and
  high $\HH(\pC)$ (Section \textit{Approximate bounds on entropy ratios}, \fref{fig:si2}). 
  Thus, this consistency across different societies and musical traditions (\fref{fig:si1b}C-D)
  suggests that there is some underlying mechanism that constrains melodies.

\begin{figure*}  \centering
  \includegraphics[width=0.70\linewidth]{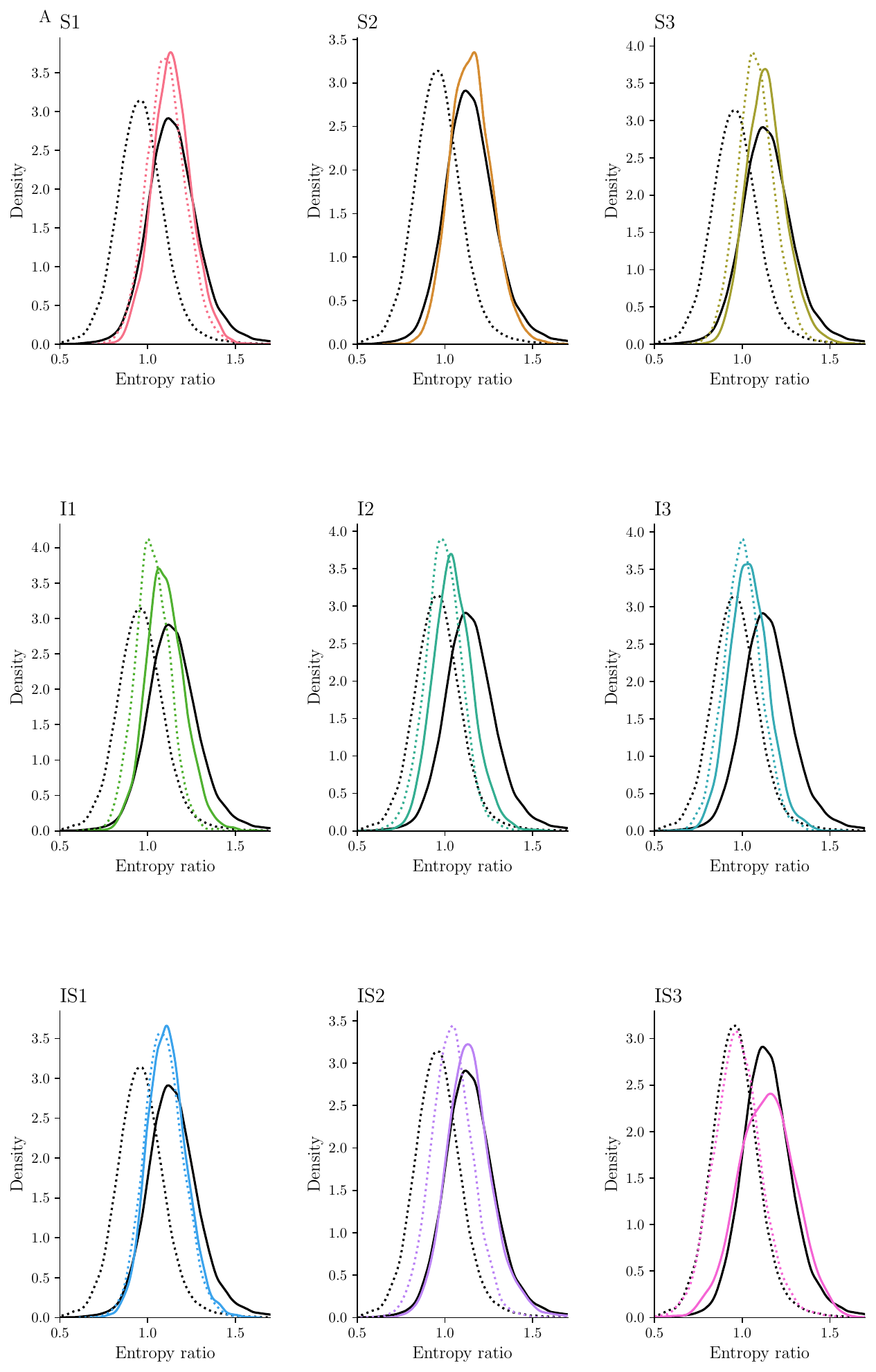}
  \caption{\label{fig:si3}
  Distributions of $\HH(\pMi) / \HH(\pC)$ (solid lines) and $\HH(\pSi) / \HH(\pC)$
  (dotted lines) for empirical melodies (black) and model-generated pitch sequences (colour).
  }
  \end{figure*}

  To understand the regularities in $\HH$ across melodic representations, we
  study nine stochastic models of melody generation (SI Section \textit{Generative model
  of pitch sequences}). The models are described by
  a prefix (`S', `I', or `IS') according to what pitches are able to be drawn,
  and a suffix (`1', `2' or `3') according to how those pitches are randomly selected.
  The models generate either: melodic intervals (`I'), scale degrees (`S'),
  or melodic intervals with an additional constraint that they correspond to a pre-determined
  random scale (`IS'). The pitches are drawn from either: a uniform distribution (`1'),
  a power-law distribution, with probabilities randomly assigned to letters (`2'),
  or a power-law distribution with probabilities assigned according to proximity
  to the median value (`3'; for intervals, this corresponds to scalar motion).
  The models are evaluated by how well they match the empirical distributions of
  $\HH(\pMi) / \HH(\pC)$ and $\HH(\pSi) / \HH(\pC)$ (\fref{fig:si1b}E), \fref{fig:si3}).
  This is captured by the Jensen-Shannon divergence between the empirical and model
  distributions (\fref{fig:si1b}F). The model that best describes the empirical
  correlations only assumes that melodies are composed using scalar motion,
  and that they use a reduced set of pitches (scales).

\subsection{Rhythm: Duration vs IOI}
\begin{figure*}  \centering
  \includegraphics[width=0.70\linewidth]{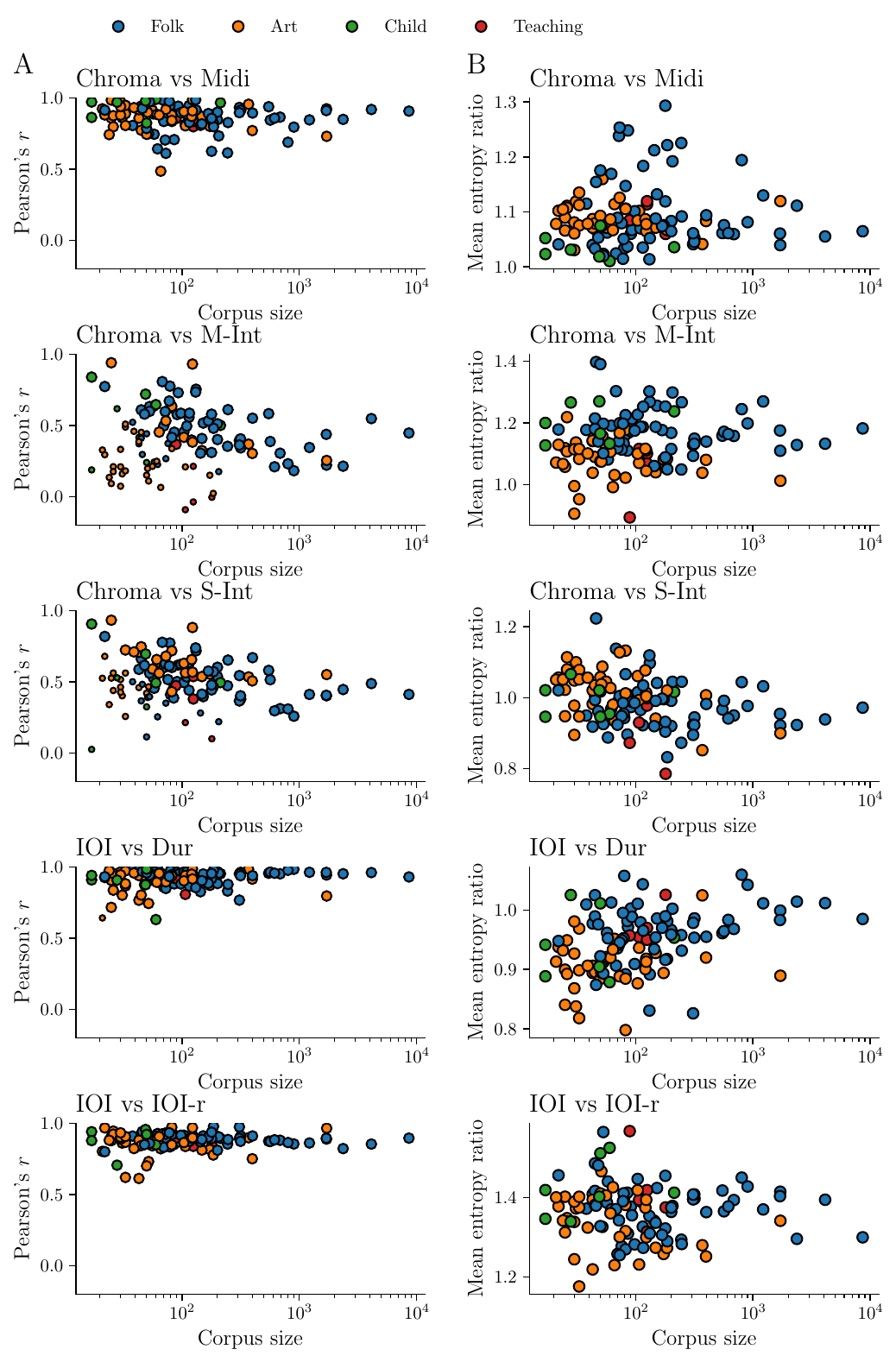}
  \caption{\label{fig:si8}
  A: Pearson's correlation coefficient between entropy of viewpoints within
  a corpus against number of melodies in a corpus.
  Large circles indicate that $p<0.05/N_{corpora}$, where we have applied
  a Bonferroni correction to account for multiple comparisons.
  B: Mean entropy ratio between viewpoints within
  a corpus against number of melodies in a corpus.
  }
  \end{figure*}

  We typically find that $\HH(\rD) < \HH(\rI)$ (\fref{fig:si1b}G, \fref{fig:si8}). This is due to the loss of information
  in ignoring rests. However, we consider that for the purposes of studying melodies, it
  may be appropriate to ignore rests. In many corpora, durations of rests are not necessarily the
  real sources of information, as there are multiple parts (solo vs chorus,
  solo vs instrumental), and the information about when to start can come from these
  other parts or from metrical structure, rather than counting rest durations. For example, in our corpora of melodies from Mozart operas, the singer will not necessarily be expected to count long rests in between singing sections, when they can instead follow cues from other instruments.
  In the corpora that are purely monophonic, rests
  are uncommon so little information is lost.

\subsection{Rhythm: first vs second order representations}
  The relation between $\rI$ ($\rD$) and the time-invariant $\rIr$ ($\rDr$) tends to follow a consistent
  trend within and between cultures (\fref{fig:si1b}H-I, \fref{fig:si8}). Following the same arguments outlined for 
  pitch, we show that many different values of $\HH(\rIr)/\HH(\rI)$
  are possible (\fref{fig:si1b}H-I; SI Section \textit{Approximate bounds on entropy ratios}).
  Thus, there may be some underlying mechanistic reason for this consistency.

\begin{figure*}  \centering
  \includegraphics[width=0.70\linewidth]{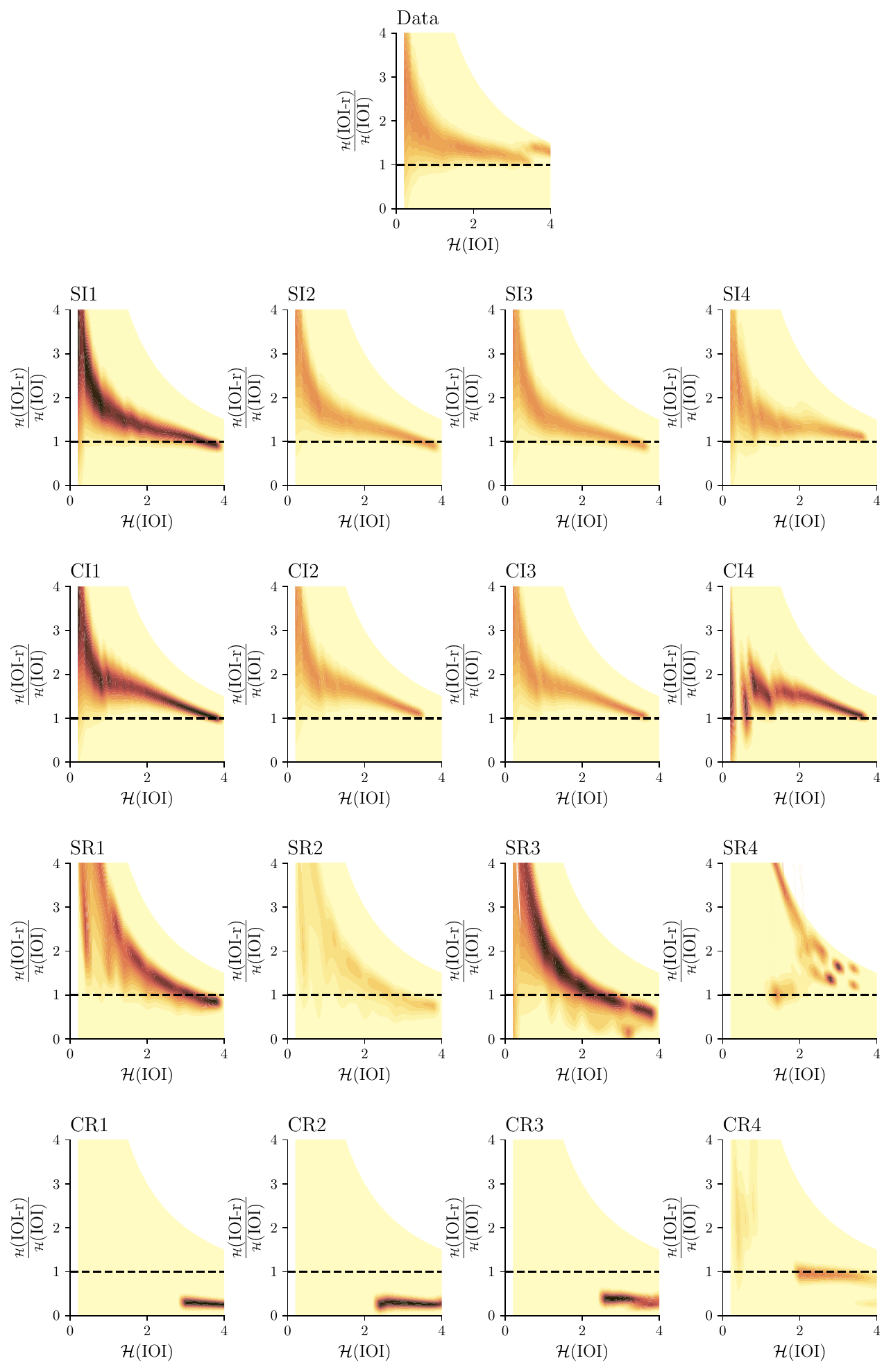}
  \caption{\label{fig:si4}
  Conditional probability distribution of $\HH(\rIr) / \HH(\rI)$ given
  $\HH(\rI)$ for empirical melodies (Data, top) and model-generated rhythm sequences.
  }
  \end{figure*}

  To understand the regularities in $\HH$ across rhythm representations, we study
  16 stochastic models of rhythm generation (SI Section
  \textit{Generative model of rhythm sequences}, \fref{fig:si4}). We find that the informational
  properties can be mostly replicated by simply generating $\rI$ values
  from a set that are related by simple (e.g. 1:2) ratios (Model names `S**'), as opposed to
  using prime number ratios (Model names `C**'). Ratios of primes are unique, which results in $\AL(\rIr) \gg \AL(\rI)$,
  while using only factors of 2 will result in many combinations of $\rI$ values
  having the same $\rIr$ value. Generating $\rIr$ values (Model names `*R*')
  performs significantly worse than generating $\rI$ values (Model names `*I*').
  We modelled four methods of choosing the pitches in a set:
  from a uniform distribution (`1'); from a power-law distribution with probabilities
  assigned randomly to letters (`2'); a power-law distribution with probabilities
  assigned to median values (`3'); or else values are chosen according
  to how well they fit into a metrical hierarchy (`4').
  The best results are obtained using a model that generates simple
  $\rI$ values that fit into a metrical hierarchy (`SI4', \fref{fig:si1b}J-K, \fref{fig:si4}).

  \begin{figure*}  \centering
  \includegraphics[width=0.70\linewidth]{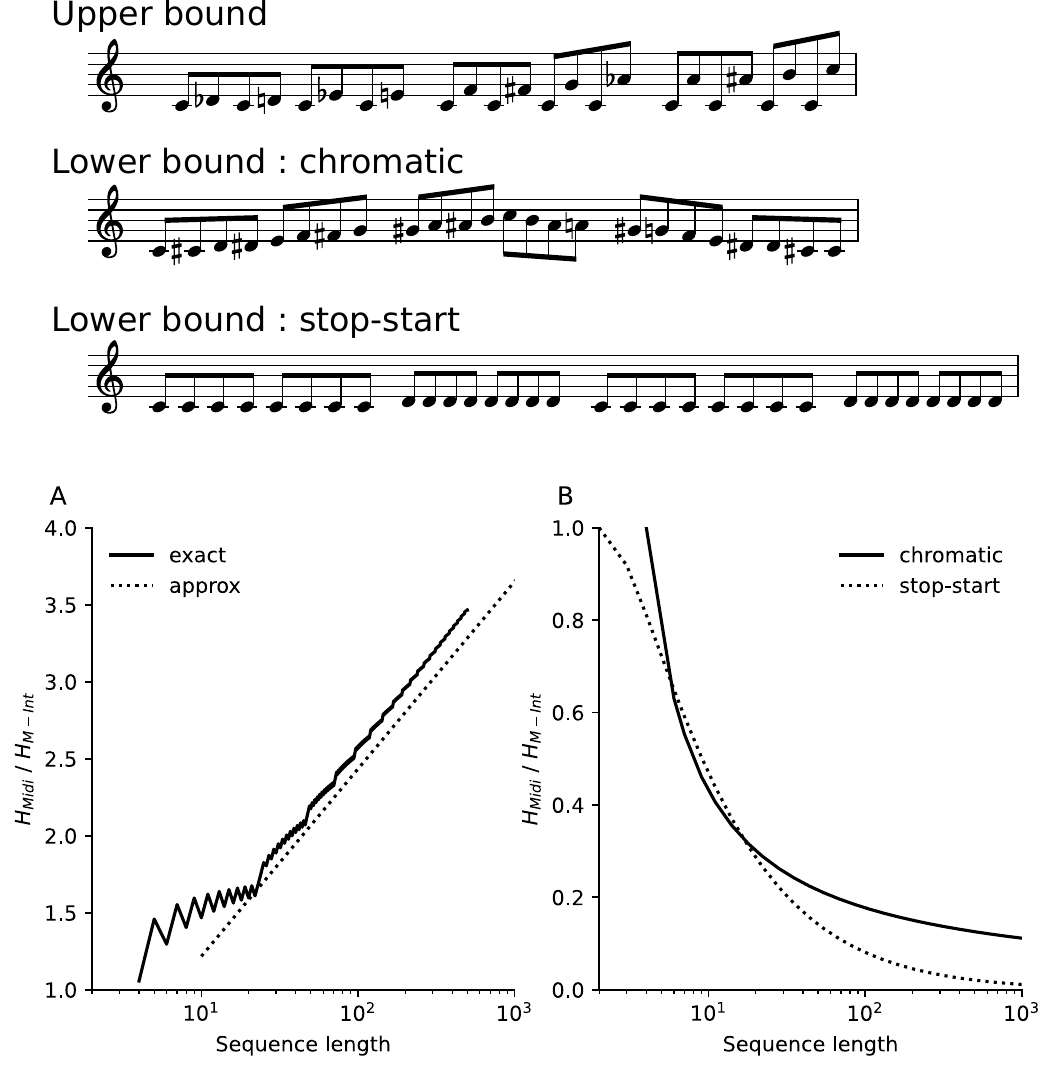}
  \caption{\label{fig:si2}
  Top: Examples of melodies that approach upper and lower bounds on the entropy ratio between first and second degree pitch viewpoints
  ratio $\HH(\pM) / \HH(\pMi)$.
  A: Scaling of the upper bound on $\HH(\pM) / \HH(\pMi)$ with sequence length;
  solid line is an exact solution; dotted line is an analytical approximation.
  B: Scaling of lower bounds on $\HH(\pM) / \HH(\pMi)$ with sequence length;
  solid line is for the `chromatic' example; dotted line is for the `stop-start' example.
  }
  \end{figure*}

\section{Approximate bounds on entropy ratios}
  For pitch, 1st order viewpoints are transformed into 2nd order viewpoints by
  the linear difference between sequential notes, while for rhythm, it is the logarithmic
  difference. At first glance, there appear to be no hard limits to the entropy ratios of
  1st and 2nd order viewpoints, apart from the fact that they have to be positive,
  and these limits are equivalent for both pitch and rhythm viewpoints.
  One can achieve $\HH(\pC) / \HH(\pMi) = \infty$ if one steadily climbs in pitch
  in a fixed interval size (\ie, $\HH(\pC)>0$ and $\HH(\pMi)=0$;
  one can achieve $\HH(\rI) / \HH(\rIr) = \infty$ if
  one keeps doubling the duration of notes in a sequence (\fref{fig:si2}).
  Less obvious is the fact that is possible that $\HH(\pC) / H(\pMi) \to 0$
  as $\LL \to \infty$. This is true in the case of 
  a melody that looks like a wave with an amplitude that grows with time;
  the melody alternates between one regular pitch, and other pitches that are only
  ever heard once, such that $\HH(\pMi) = \log \LL$ and $\HH(\pC)\to \log{2}$ as $L\to\infty$.
  A similar case can be made for rhythmic viewpoints.

  In practice, the absolute pitch range of melodies is often limited
  to that of typical vocal range (within two octaves), $\rI$ values within
  a song do not differ by more than a maximum ratio of $\sim8$ (between shortest
  and longest notes), and melodies have fixed length.
  In the main text we report approximate bounds that correspond to melody lengths
  of $\LL=100$ (\fref{fig:si2}.

\section{Generative models}

\subsection{Generative model of pitch sequences.~}
  The models generate pitch sequences by randomly drawing $L$ letters from an alphabet
  composed of either: $\pM$ values limited to a predetermined set of pitches (scale), $\pMi$,
  or $\pMi$ that are limited by a predetermined scale. 
  Correspondingly, the model names are prefixed by `S', `I', and `IS'.
  For each of these three approaches, we either draw letters 
  with either: a uniform probability, a power-law distribution with probabilities randomly
  assigned to letters, or a power-law distribution with probabilities assigned to be
  lowest at the extremes of the pitch / interval range, and highest in the middle;
  this last case is akin to biasing towards scalar motion, such that small intervals
  (both ascending and descending) are picked with higher probability.
  The model names are suffixed according to probability distribution by either
  `1', `2', or `3'.
  The combination of three types of alphabet and three types of probability distributions
  results in nine models: `S1', `S2', `S3', `I1', `I2', `I3', `IS1', `IS2', `IS3'.

  For `S' models, a scale is first randomly fixed as a subset of 
  $\AL$ scale degrees drawn from a set of pitches on an equidistantly-spaced
  set of intervals from 1 to 12. We assume octave equivalence, such that pitches $P$ outside
  of this set are equivalent to $P \mod 12$; this corresponds
  to choosing a scale out of 12 possible notes per octave $\OC$. We allow
  the total pitch range to vary in our model, in the region of $1\leq \OC \leq3$.
  After drawing a sequence of $\pM$, we convert this to $\pS$, $\pMi$ and $\pSi$.

  For `I' models, we allow all $2\AL + 1$ intervals from
  $-\AL \leq I \leq \AL$. We keep track of $\pM$ (starting from zero),
  and only allow intervals that result in the pitch being $-\OC/6 \leq P \leq \OC/6$;
  in this way $\OC$ fixes the pitch range. After drawing a sequence of $\pMi$.
  Although we do not constrain the pitches to a scale in this case,
  we do assign a scale to the sequence by collapsing the pitches in the sequence
  to a single octave (\ie, $P \mod 12$) and assigning a scale degree to each
  unique pitch. This allows us to convert this to sequences of $\pS$ and $\pSi$.

  For `IS' models, we first choose $\AL$ scale degrees, and only allow intervals that
  lead to a pitch that is in the predetermined scale,
  and falls within the pitch range $-\OC/6 \leq P \leq \OC/6$. Melodies start at zero,
  which allows us to convert $\pMi$ to $\pS$ and $\pSi$.

  For each model we vary the alphabet size $\AL$, sequence length $L$,
  pitch range $\OC$, and the power law exponent $n$.
  We generate 100 sets of sequences ($\pC$, $\pMi$ and S-int) and calculate
  the ratios of $\HH(\pMi) / \HH(\pS)$, and $\HH(\pSi) / \HH(\pS)$.
  Note that we report ratios for $\HH(\pC)$, which is exactly equivalent to $\HH(\pS)$.
  For each model we find the optimal parameters ($3\leq \AL \leq 12$, $15 \leq L \leq 50$,
  and $1 \leq \OC \leq 3$) by minimizing the sum of 
  the Jensen-Shannon divergence (JSD) between the empirical and model distributions
  of, respecively, $\HH(\pMi) / \HH(\pC)$, and $\HH(\pSi) / \HH(\pC)$.
%  The best-fitting parameters and model results are presented in Table XXX.

\subsection{Generative model of rhythm sequences.~}
\begin{figure*}  \centering
  \includegraphics[width=0.70\linewidth]{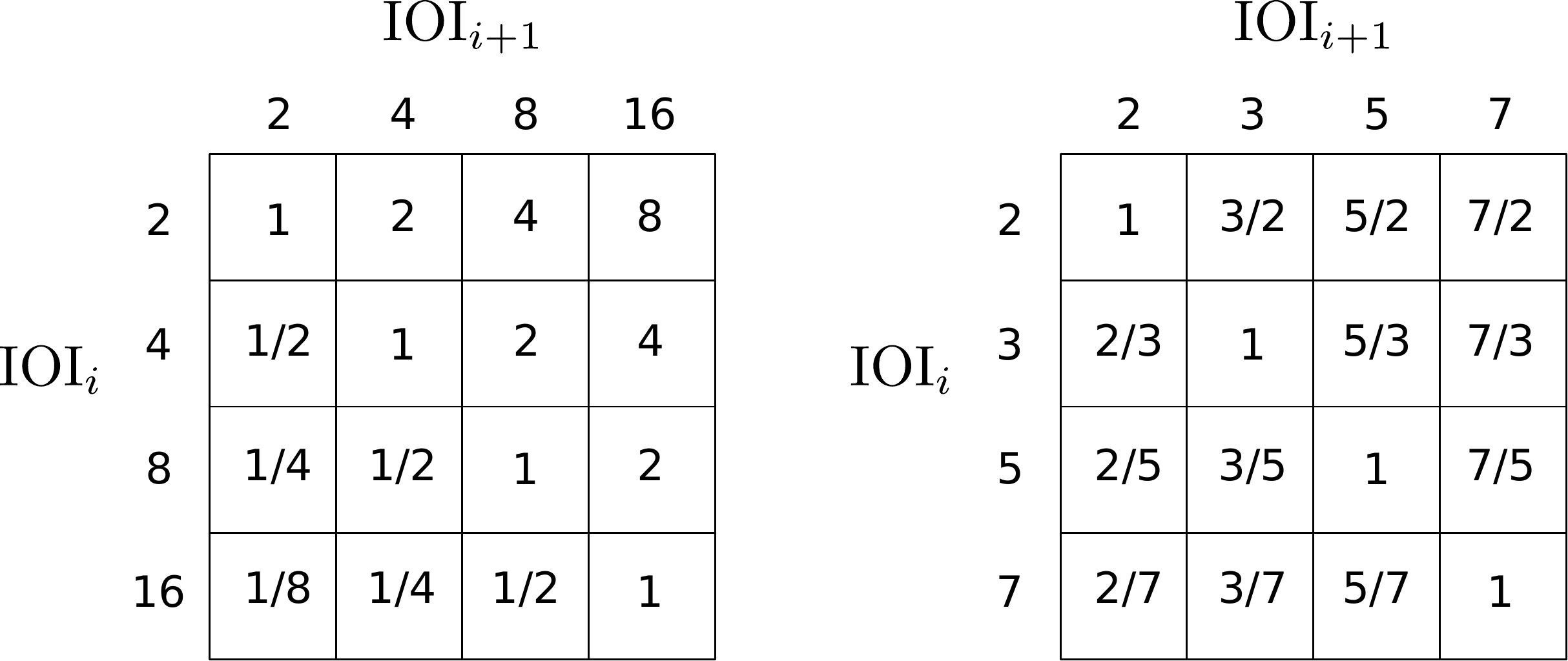}
  \caption{\label{fig:si15}
  Illustration of how simple $\rI$ values (left) produce fewer unique $\rIr$
  values (inside the square), compared to complex $\rI$ values (right).
  }
   \end{figure*}

  The models generate rhythm sequences by randomly drawing $L$ letters from an alphabet
  composed of either: simple $\rI$ (`SI') or complex $\rI$ (`CI');
  simple $\rIr$ (`SR') or complex $\rIr$ (`CR') (\fref{fig:si15}).
  For complex $\rI$ values, each combination of values leads to a unique $\rIr$;
  using primes and reciprocals of primes is one way of achieving this outcome.
  For simple $\rI$ values, different combinations of values can lead to the same
  $\rIr$; using a series of $\rI \in {x^{i-k},...,x^{n-k}}$ is a limiting case
  of simplicity, as it leads to the smallest possible $\AL(\rIr)$ for a given
  $\AL(\rI)$. We choose $x=2$ (although the results are independent of this choice),
  and choose $k$ such that the middle size is $x^0=1$.

  These values are drawn either: from a uniform distribution (`1'),
  a power-law distribution (exponent $n$) with probabilities randomly assigned
  to letters (`2'), a power-law distribution with intermediate
  values being assigned highest probability (`3') (similar to how scalar motion is modelled in the
  pitch generation models), or else subsequent letters are selected according to 
  how well they fit into a metrical hierarchy (`4'). To bias letter selection according
  metrical hierarchy, we choose a 4/4 meter, and assign probabilities to values that
  are proportional to $b^n$,
  \begin{equation}
    P(\rI) = \frac{b(\rI)^n}{\sum\limits_{i}^{\AL} b(\rI_i)^n}~;
  \end{equation}
  $b=4$ if a $\rI$ value leads to a downbeat (if the onset of the
  next note is in the series $4i + 1$), $b=3$ if the onset is in the series $4i+3$,
  $b=2$ if the onset is in the series $4i + 2$ or $4i + 4$, and $b=1$ otherwise.
  We run all 16 possible models by combinatorially choosing all of the above rules.

  For each model, we generate 100 sequences and produce $\rI$ and $\rIr$ sequences.
  To convert from $\rI$ to $\rIr$ we use Eq.~\ref{eq:ioir}.
  To convert from $\rIr$ to $\rI$, we take the starting $\rI$ value to be 1.
  We calculate $\HH(\rI)$ and $\HH(\rIr)$ for each sequence.
  Since the ratio of $\HH(\rI)$ to $\HH(\rIr)$ is not constant, we find the optimal free parameters
  (power-law or metrical hierarchy bias exponent $n$, melody length $L$)
  by minimizing the expected value of the JSD between the conditional probability
  of $\rIr$ given $\rI$, given the empirical $\rI$ probability.
  To achieve a wide range of $\HH(\rI)$, we group the melodies generated
  using different alphabet sizes, $2 \leq \AL \leq 20$.

\section{Pitch-Rhythm Covariance}

\subsection{Null model for pitch-rhythm covariance}
  The entropy of the joint representation of both pitch ($\Pi$)
  and rhythm ($\Rh$), $\HH(\Pi, \Rh)$, is bounded by
  incontrovertible information-theoretic constraints:
  $\min\{\HH(\Pi), \HH(\Rh)\} \leq \HH(\Pi, \Rh) \leq \HH(\Pi) + \HH(\Rh)$.
  The lower bound can only be achieved if there is an direct mapping between
  rhythmic values and pitch values (\eg, crotchets
  are always on C, quavers on D, etc.), and is never achieved in real melodies.
  The difference between the upper bound and the true joint entropy is equivalent
  to the mutual information between rhythm and pitch,
  $\MI(\Pi, \Rh) = \HH(\Pi, \Rh) - (\HH(\Pi) + \HH(\Rh))$ (\ie, how much do you know about
  the pitch, if you know the rhythm).
  Even if the underlying processes generating pitch and rhythm are independent ($\MI=0$), by measuring
  $\HH$ using finite sequences we will find that $\MI>0$. Furthermore, the
  degree to which this happens depends on the entropy of the sequence,
  such that higher entropy signals require increasingly long sequences in order
  to reliably measure the mutual information. To control for this, 
  we measure instead $\MI^* = \MI(\Pi, \Rh) - \MI_{\textrm{ran}}(\Pi, \Rh)$,
  where $\MI_{\textrm{ran}}(X, Y)$ is the mutual information of a pair of sequences $X$
  and $Y$, with one of the sequences randomly shuffled.
  In this way we can measure the mutual information in short sequences, while
  accounting for difference in length and unigram distributions. In practice we
  do this 10 times for each set of sequences and use the average $\MI_{\textrm{ran}}$.

\subsection{Musical interpretation of covariance}
\begin{figure*}  \centering
  \includegraphics[width=0.90\linewidth]{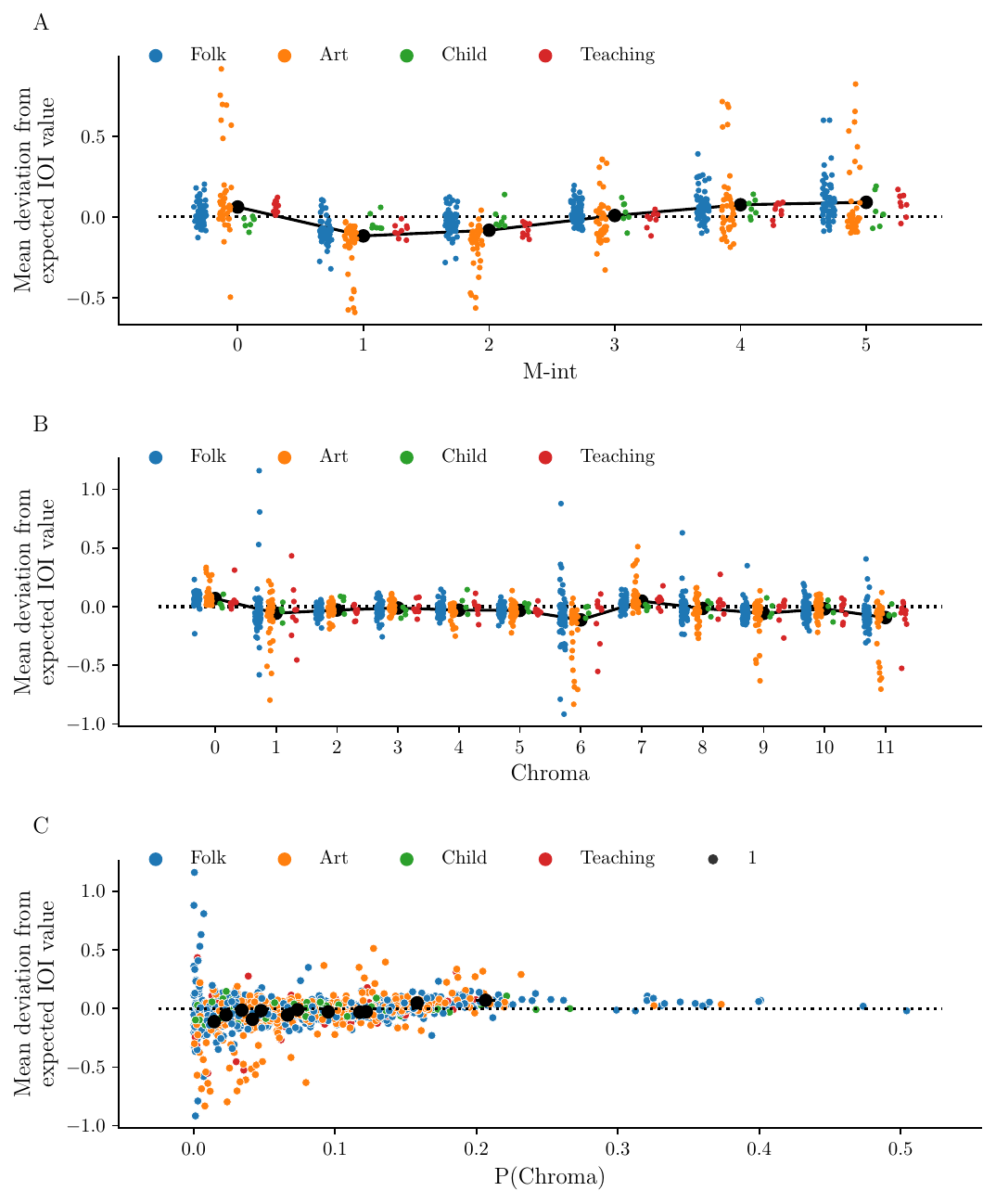}
  \caption{\label{fig:si11}
  A-C: Mean deviation from expected IOI value per corpus as a function of
  Melodic interval (A: $\pMi$), Chroma (B: $\pC$), and probability of Chroma
  (C: $P(\pC)$). Black circles show the average across all corpora;
  coloured circles show values for each corpus, coloured according to type.
  }
  \end{figure*}

  To understand how pitch and rhythm covary, we look at how much the mean
  $\rI$ value depends on pitch ($\pC$ and $\pMi$). We measure the mean
  $\rI$ value for a corpus, and then measure the mean $\rI$ value that co-occurs
  with each value of $\pC$ and $\pMi$, and plot the difference \fref{fig:si11}.
  For $\pC$, it is more meaningful to first transpose every melody in a corpus
  to the same key. This is easy when the corpus includes key annotations,
  but most of them do not. Therefore we employ a simple algorithm to identify
  the tonic, which allows us to transpose the melodies. We tried several
  approaches to estimate the tonic: modal $\pC$, first $\pC$, final $\pC$.
  We evaluated each algorithm by comparing with the melodies which have
  key annotations, finding that the final note is most indicative of
  the tonic (\SI{62}{\%} accuracy), followed by the first note (\SI{30}{\%} accuracy)
  and the modal note (\SI{22}{\%} accuracy). Despite the higher accuracy of
  the algorithm using the tonic, using the other algorithms leads to similar
  conclusions about how pitch and rhythm covary.

  \begin{algorithm*}

  \KwData{$A, \LL_{min}$} 
  \KwResult{$matchListDict$ (Dictionary: key = matched sequence; value = list of (list indices, sequence indices and match length))}

  $N \gets length(A)$ ;

  \Begin{
    \For{$i_1=1,...,N$}{
      $L_i \gets length (A[i_1])$ ;

      \For{$j_1=1,...,N$}{
        $L_j \gets length (A[j_1])$ ;

        \For{$i_2=1,...,L_i - \LL_{min}$}{
          \eIf{$i_i = j_1$}{
            $j_{min} = \LL_{min}$ ;
          }{
            $j_{min} = 0$ ;
          }

          \For{$j_2=j_{min},...,L_j - \LL_{min}$}{
            \If{$A[i_1][i_2] = A[j_1][j_2]$}{
              $match \gets A[i_1][i_2]$ ;
              $matchLength \gets 1 $ ;

              \eIf{$i_1 = i_2$}{
                $maxWidth = min(j_2 - i_2, L_i - j_2)$ ;
              }{
                $maxWidth = min(L_i - i_2, L_i - j_2)$ ;
              }

              \For {$k=1,...,maxWidth$}{
                \If{$A[i_1][i_2+k] = A[j_1][j_2+k]$}{
                  $match \gets match + A[i_1][i_2+k]$ ;
                  $matchLength \gets m_{len} + 1$ ;

                  \eIf{$m_{len} \geq \LL_{min}$}{
                    $matchListDict[match] = matchListDict[match] + (i_1,~i_2,~j_1,~j_2,~matchLength)$ ;
                  }{
                    $break$ ;
                  }

                }
              }
            }
          }
        }
      }
    }
  }
  \caption{Find all repeated substrings, $match$, longer than $\LL_{min} - 1$ in a list of strings $A$}
  \label{alg:alg1}
  \end{algorithm*}

  \begin{algorithm*}
  \label{alg:alg2}
  \KwData{$matchListDict$}
  \KwResult{$count$}

  \Begin{
    \For{$(key, value) \gets matchListDict$}{
      $alreadyCounted \gets empty set$;

      \For{$(i_1,~i_2,~j_1,~j_2,~matchLength) \gets value$}{
        \If{$(i_1, i_2) \not \subset alreadyCounted$}{
          $alreadyCounted \gets alreadyCounted + (i_1, i_2)$;
          $count[key] \gets matchLength $;
        }

        \If{$(j_1, j_2) \not \subset alreadyCounted$}{
          $alreadyCounted \gets alreadyCounted + (j_1, j_2)$;
          $count[key] \gets matchLength $;
        }
      }
    }
  }
  \caption{Remove overlapping substrings}
  \label{alg:alg2}
  \end{algorithm*}

  \begin{algorithm*}

  \KwData{$A$}
  \KwResult{$nonrepString$}

  \Begin{
    $matchListDict \gets alg1(A, \LL_{min})$ ;
    \If {$length(matchListDict) = 0$}{
      $return A$ ;
    }

    $count \gets alg2(A, \LL_{min})$ ;
    $match \gets getKeyOfHighestValue(count)$ ;
    $A \gets divideStringsByMatch(A, match) + match$ ;
    return $alg3(A, \LL_{min})$ ;

  }
  \caption{Recursively remove repeated substrings with length greater than $\LL_{min}-1$}
  \label{alg:alg3}
  \end{algorithm*}

  \begin{figure*}  \centering
  \includegraphics[width=0.90\linewidth]{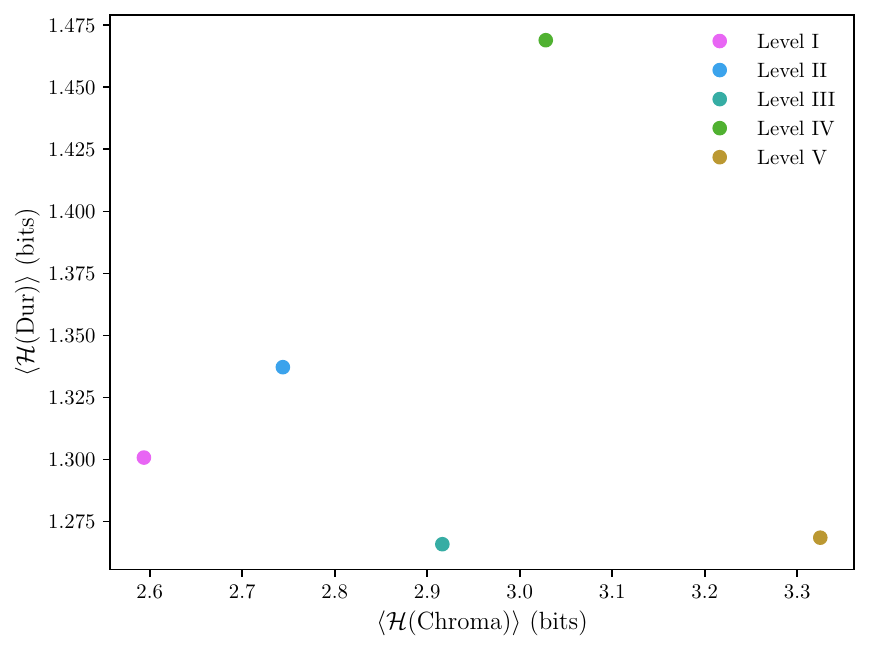}
  \caption{\label{fig:si_teaching}
  \textbf{Entropy across singing instruction books of different levels.} Mean entropy per corpus is plotted for Duration vs Chroma.
  }
  \end{figure*}

  \begin{figure*}  \centering
  \includegraphics[width=0.70\linewidth]{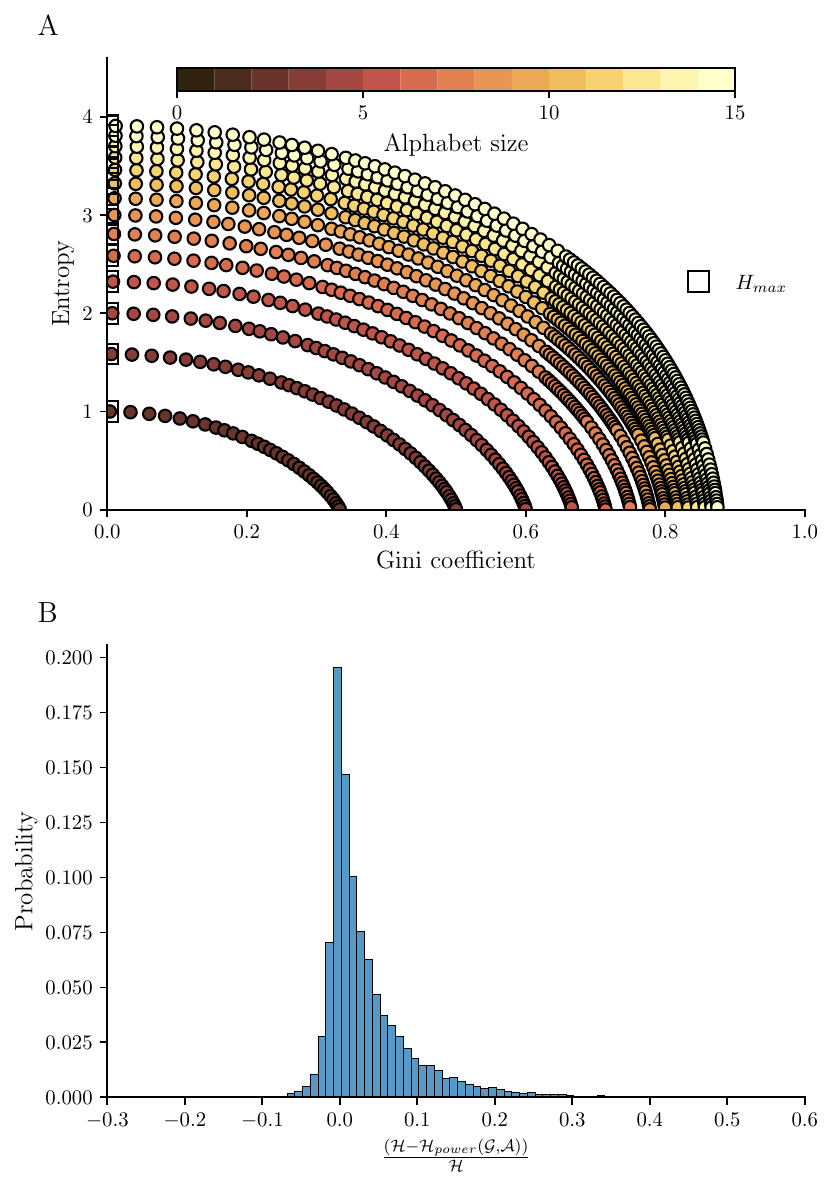}
  \caption{\label{fig:si5}
  A: Entropy against Gini coefficient for a power-law distribution as a function
  of alphabet size (color) and power-law exponent (different circles). For any known
  distribution (such a power-law), $\HH$ and $\GI$ can be calculated analytically.
  B: To see whether $\AL$ and $\GI$ are sufficient to specify the entropy
  of a sequence, we calculate $\HH(\pC)$ ($\HH$ in the axis label for brevity) for
  all sequences and calculate the entropy of a power-distribution $\HH_{power}$
  that has the same value of $\GI$ and $\AL$. The relative difference
  between $\HH(\pC)$ and $\HH_{power}$ is plotted as a histogram with bin size 0.01.
  Although there are some outliers where values of $\GI$ and $\AL$ can lead to
  very different values of $\HH$, the interquartile range is \SIrange{-0.001}{0.051}.
  }
  \end{figure*}

% \begin{figure*}  \centering
% \includegraphics[width=0.70\linewidth]{si6.pdf}
% \caption{\label{fig:si6}
% Correlation between $\AL$ and $\GI$ as a function of corpus size
% for each corpus, for $\pC$ and $\rI$ viewpoints. Colours indicate corpus type.
% Large circles indicate that $p<0.05/N_{corpora}$, where we have applied
% a Bonferroni correction to account for multiple comparisons.
% }
% \end{figure*}

  \begin{figure*}  \centering
  \includegraphics[width=0.70\linewidth]{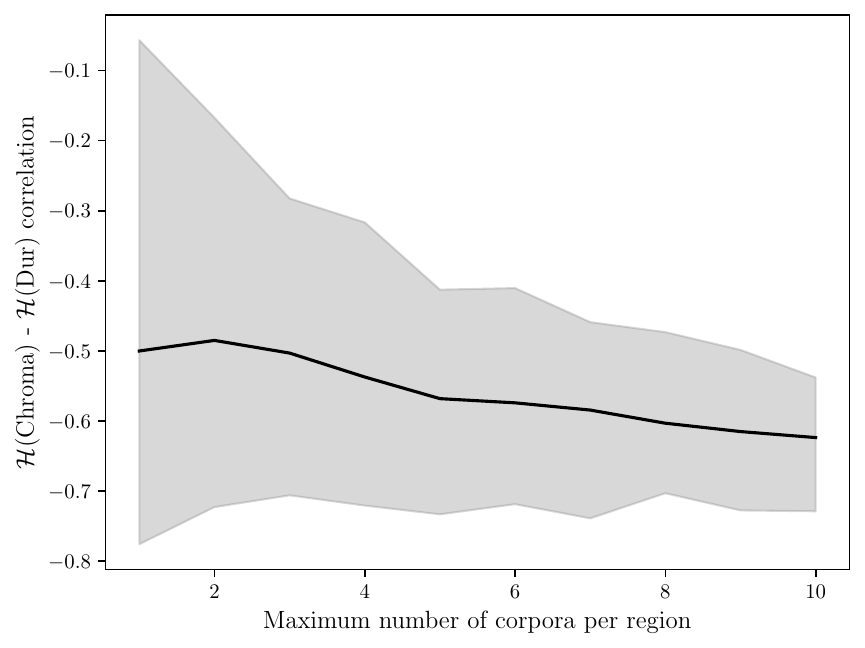}
  \caption{\label{fig:si5b}
  Pearson's correlation between mean $\pC$ and $\rD$ entropy per corpus,
  as a function of the maximum number of corpora per region (Europe, North
  America, East Asia, Africa, Middle East, Central Asia, Central America, Pacific Ocean).
  Shaded area indicates \SI{95}{\%} CI from sampling.
  }
  \end{figure*}

  \begin{figure*}  \centering
  \includegraphics[width=0.70\linewidth]{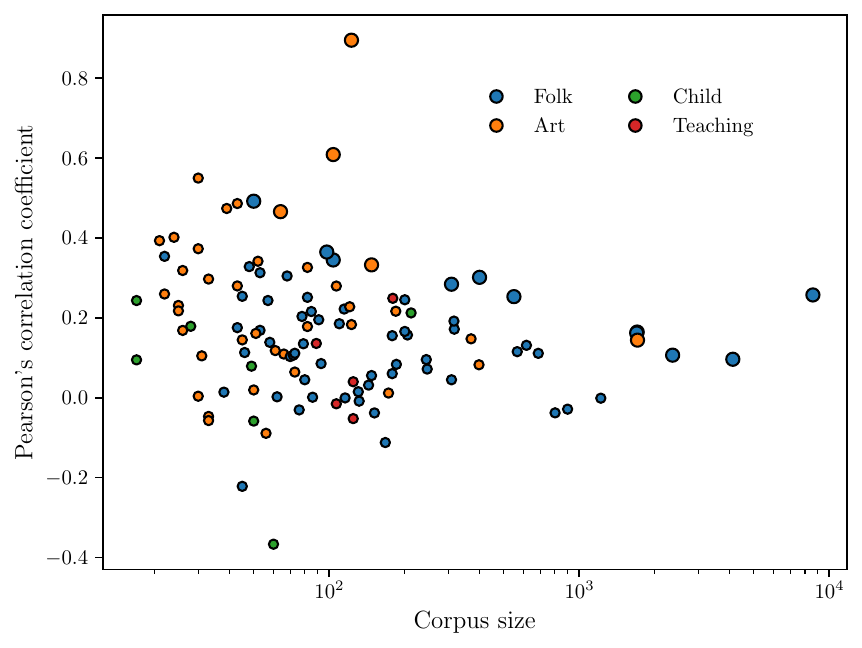}
  \caption{\label{fig:si10}
  Correlation between $\HH(\pC)$ and $\HH(\rI)$ as a function of corpus size
  for each corpus. Colours indicate corpus type.
  Large circles indicate that $p<0.05$, after using the Benajamini-Hochberg
  procedure to account for multiple comparisons. \\
  Pitch and rhythm entropy tend to be weakly, positively correlated within
  corpora, indicating that songs within societies tend to differ in terms of overall complexity
  more than along a pitch-rhythm trade-off.
  }
  \end{figure*}

  \begin{figure*}  \centering
  \includegraphics[width=0.90\linewidth]{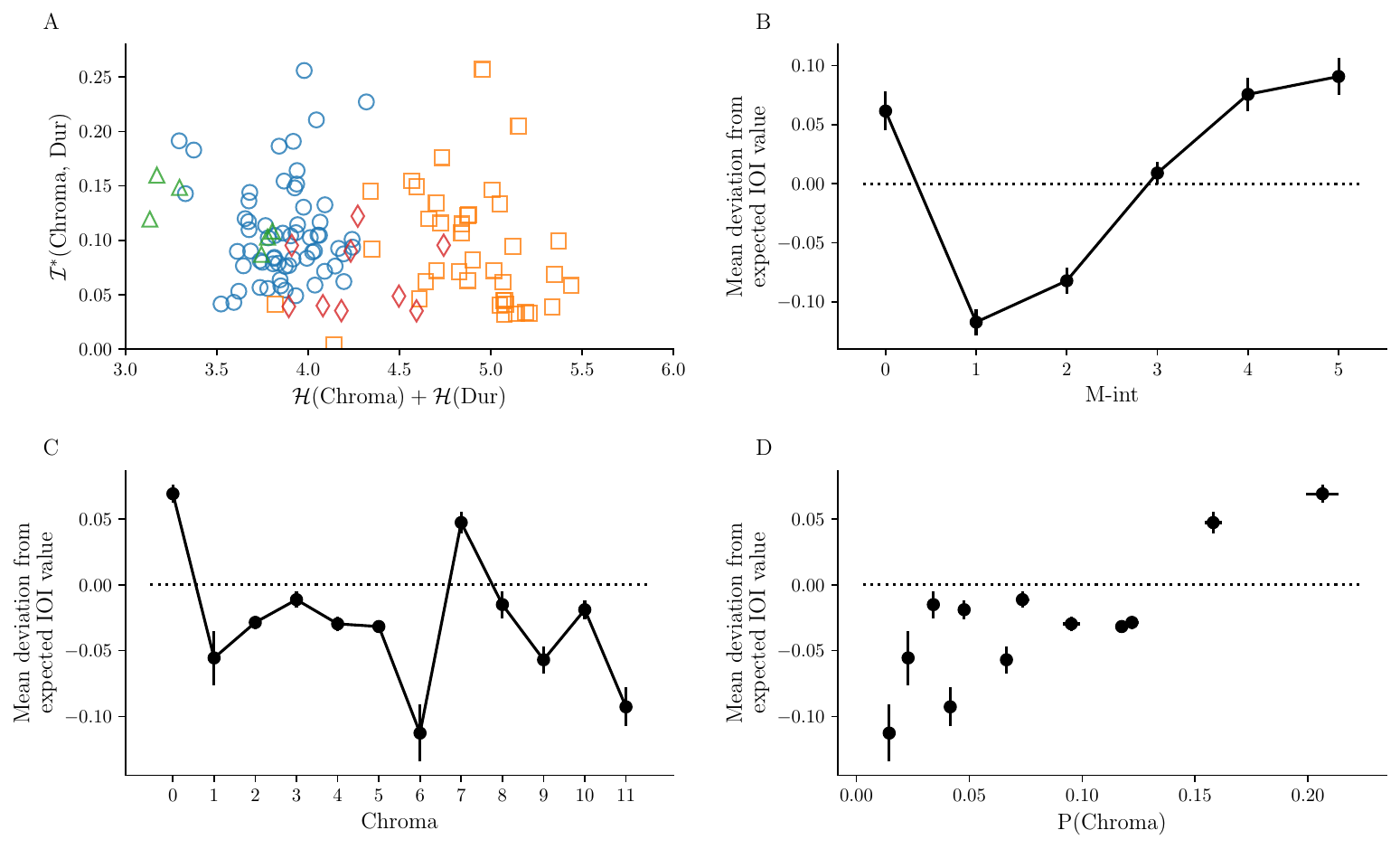}
  \caption{\label{fig:si10b}
  A: Effective mutual information $\MI^*(\pC, \rD)$ as a function of
  $\HH(\pC) + \HH(\rD)$.
  B-D: Mean deviation from the expected IOI value compared to:
  (C) the average IOI associated with each value of absolute interval size, $|\pMi|$.
  (D) the average IOI associated with each value of $\pC$.
  (E) the average IOI associated with each value of $\pC$, plotted 
  against the empirical probability of $\pC$. Error bars show
  standard error (B-D).
  }
  \end{figure*}

  \begin{figure*}  \centering
  \includegraphics[width=0.90\linewidth]{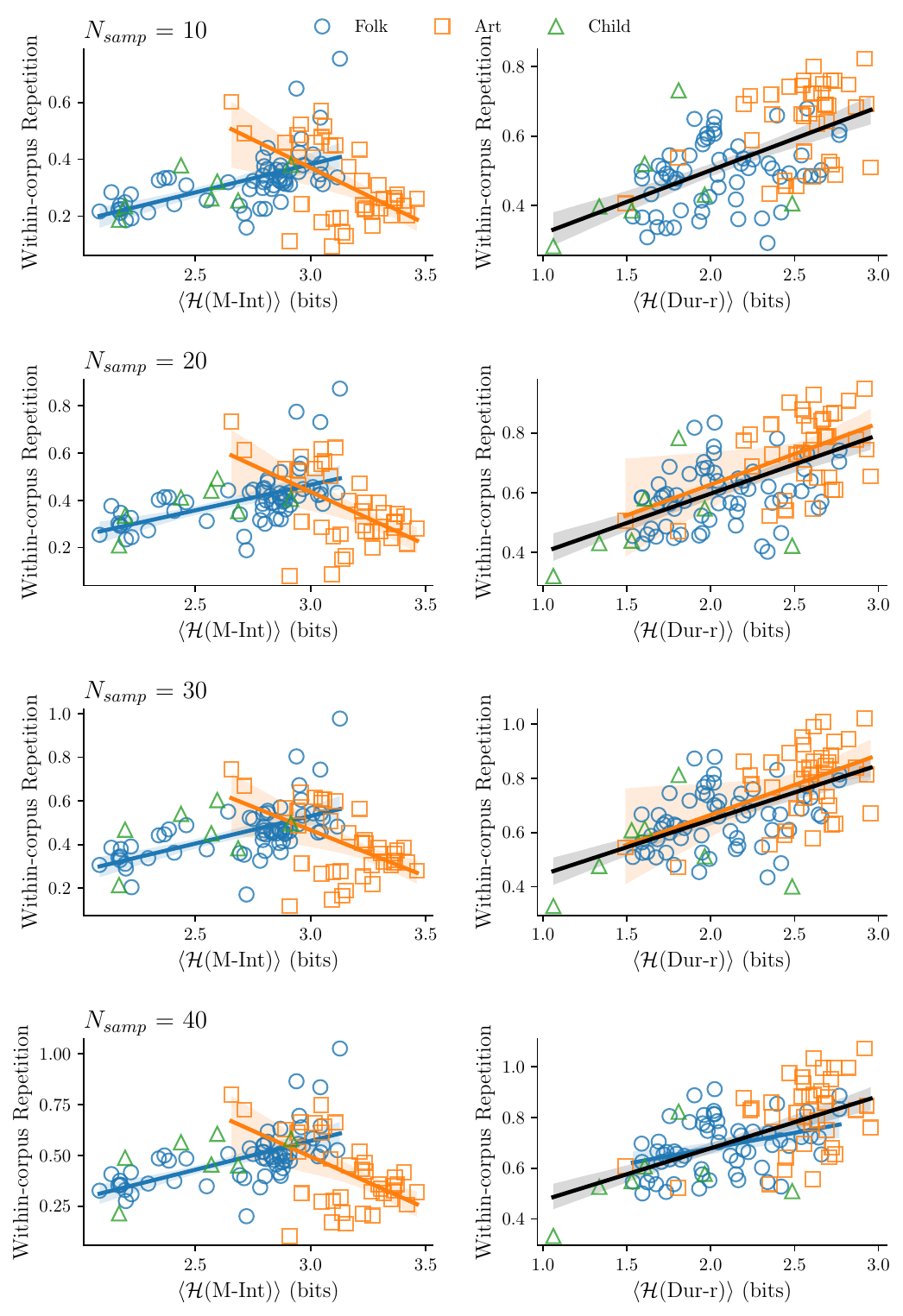}
  \caption{\label{fig:si_ltm_nsamp}
  \textbf{Effect of training set size on within-corpus repetition.} Within-corpus repetition vs mean entropy per corpus for M-Int (left) and Duration-ratio (right), for different training set sizes, $N_{samp}$. Melodies are truncated at 50 notes.
  }
  \end{figure*}

  \begin{figure*}  \centering
  \includegraphics[width=0.90\linewidth]{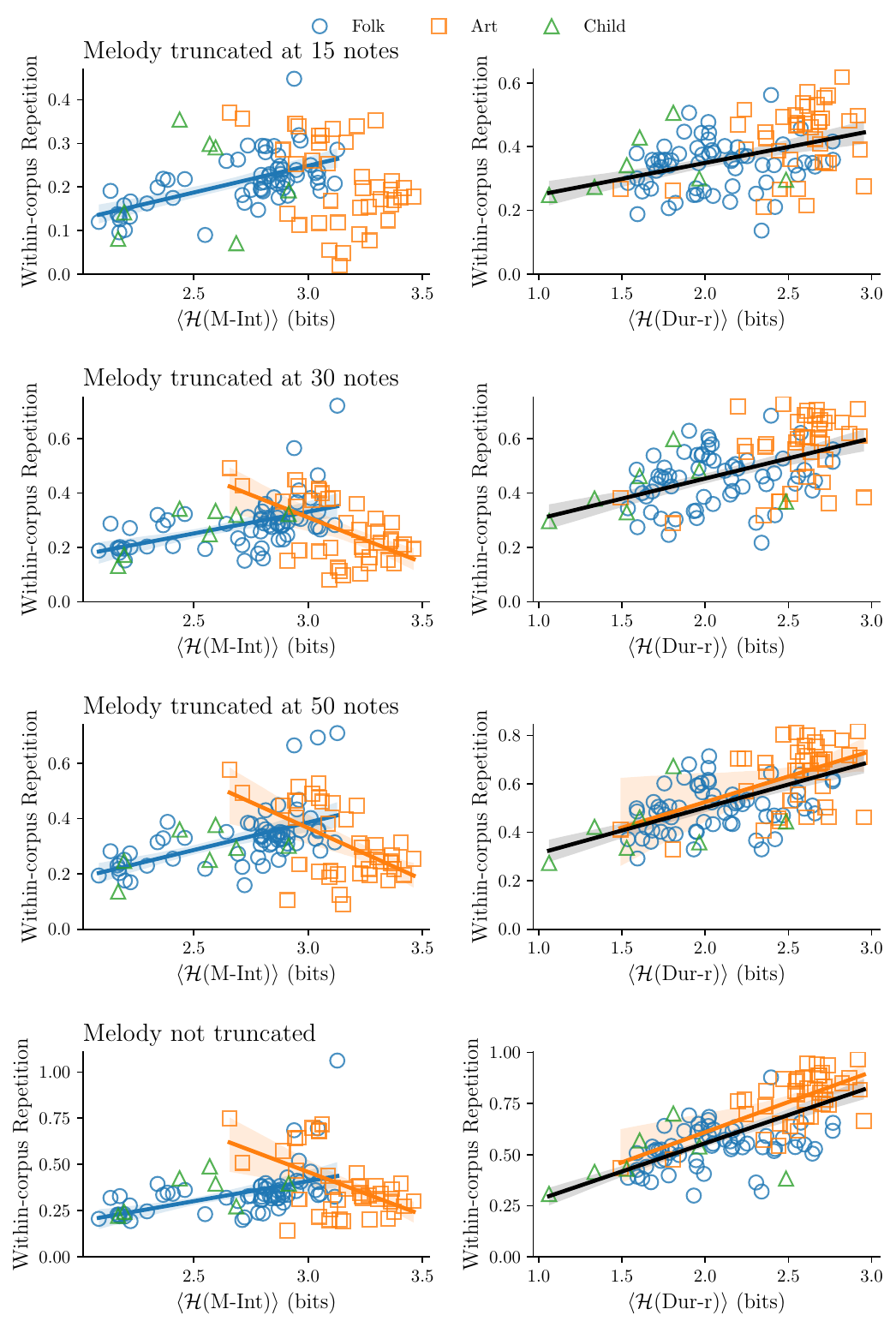}
  \caption{\label{fig:si_ltm_mellen}
  \textbf{Effect of melody truncation length on within-corpus repetition.} Within-corpus repetition vs mean entropy per corpus for M-Int (left) and Duration-ratio (right), for melodies that are truncated at 15, 30 or 50 notes, and for melodies without truncation. Number of samples, $N_{samp} = 10$.
  }
  \end{figure*}

  \begin{figure*}  \centering
  \includegraphics[width=0.90\linewidth]{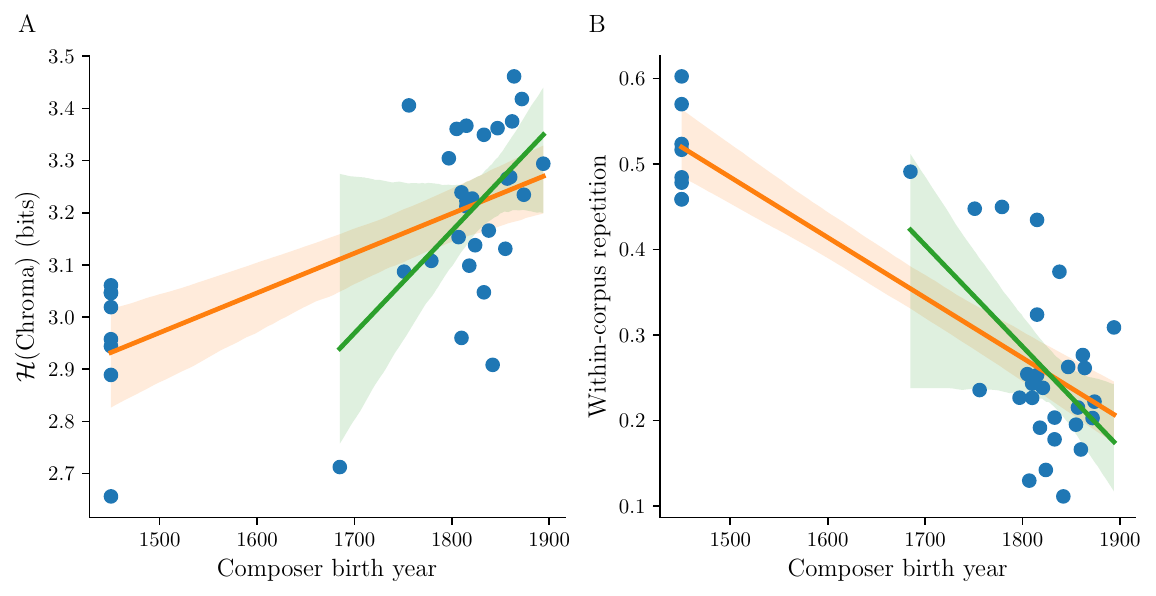}
  \caption{\label{fig:si11b}
  Pitch entropy ($\HH(\pC)$) and within-corpus repetition for Art corpora
  as a function of composers' birth years. Correlations are shown for the
  full set (A, $r = 0.64$, $p < 10^{-4}$; B, $r = -0.82$, $p < 10^{-8}$)
  and using only the composers from the common practice period
  (A, $r = 0.50$, $p = 0.006$; B, $r = -0.52$, $p = 0.005$).
  }
  \end{figure*}

  \begin{figure*}  \centering
  \includegraphics[width=0.70\linewidth]{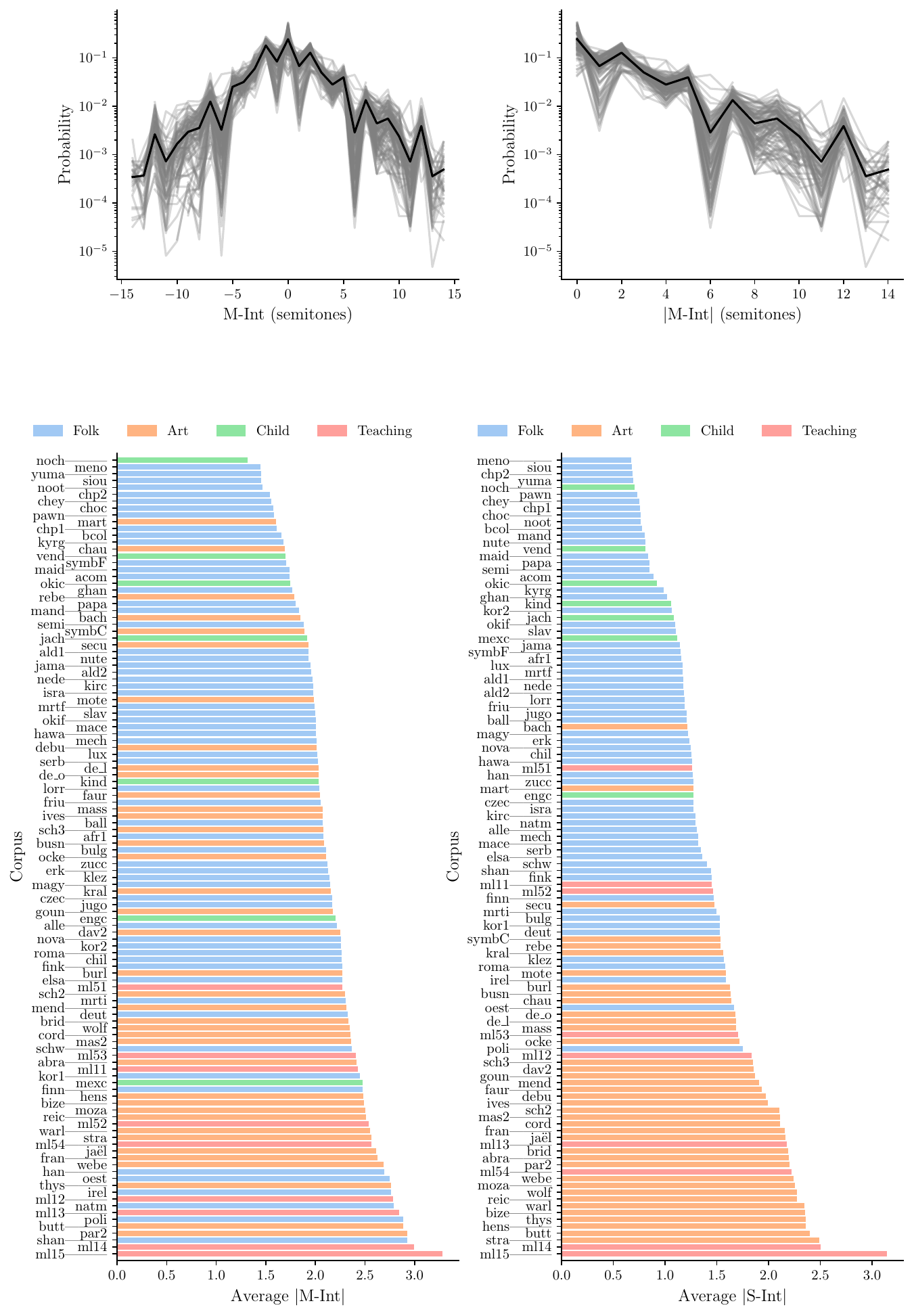}
  \caption{\label{fig:si12}
  A: Probability distribution of $\pMi$ across all corpora (black line)
  and for individual corpora (grey lines).
  B: Probability distribution of $|\pMi|$ across all corpora (black line)
  and for individual corpora (grey lines).
  C-D: Average $|\pMi|$ (C) and average $|\pSi|$ (D) for individual corpora, coloured by
  corpus type.\\
  The inclusion of $|\pSi|$ shows a much clearer difference between Folk and Art corpora,
  since it controls for the fact that some Folk corpora predominantly use pentatonic
  scales.
  }
  \end{figure*}

  \begin{figure*}  \centering
  \includegraphics[width=0.70\linewidth]{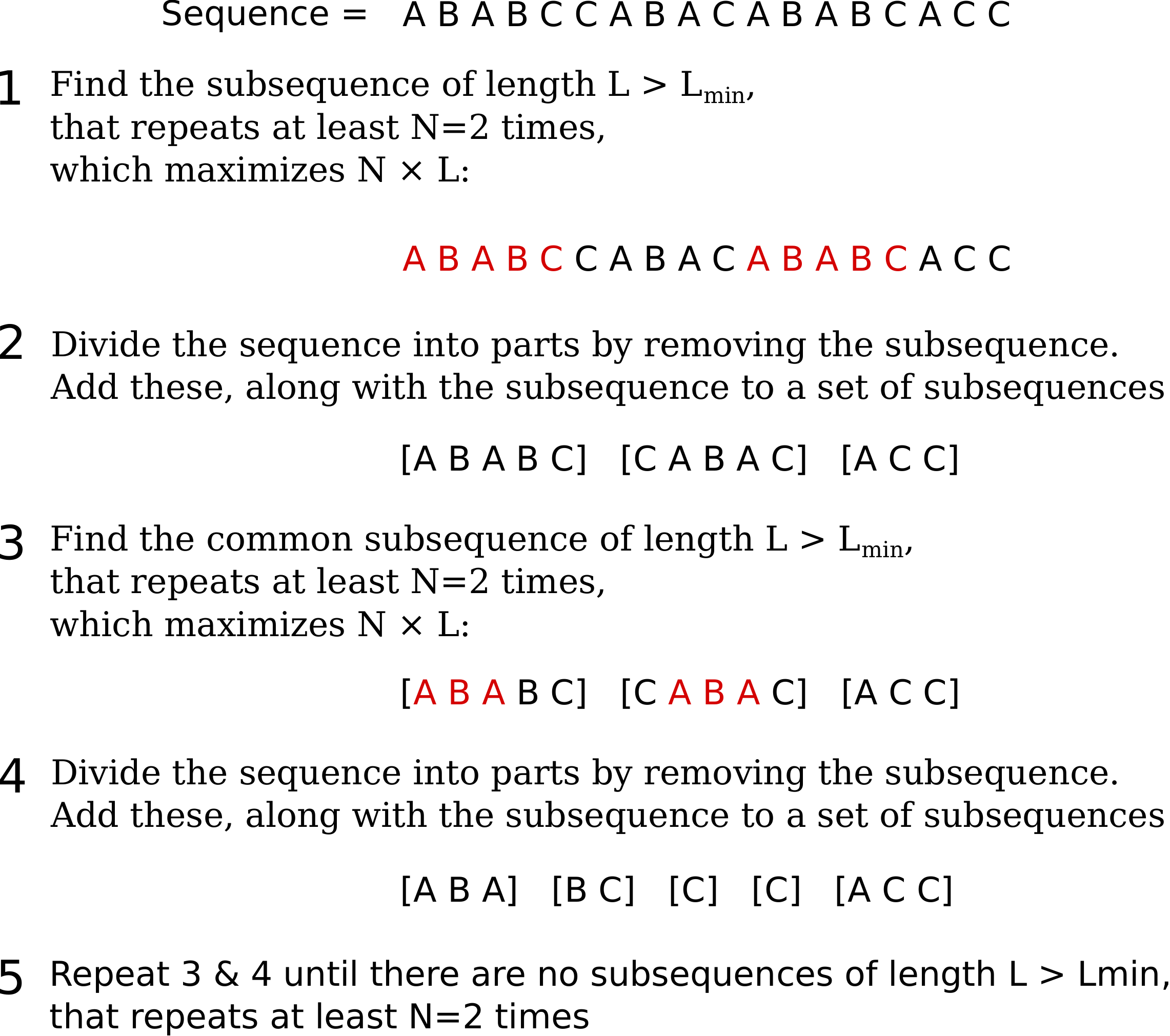}
  \caption{\label{fig:si16}
  Visualization of the steps in Alg. \ref{alg:alg3}.
  Starting from an input sequence, the algorithm recursively identifies repeated
  subsequences, divides the sequence(s) and repeats until no repeated subsequence
  has length $\LL_{min}$.
  }
  \end{figure*}

  \begin{figure*}  \centering
  \includegraphics[width=0.70\linewidth]{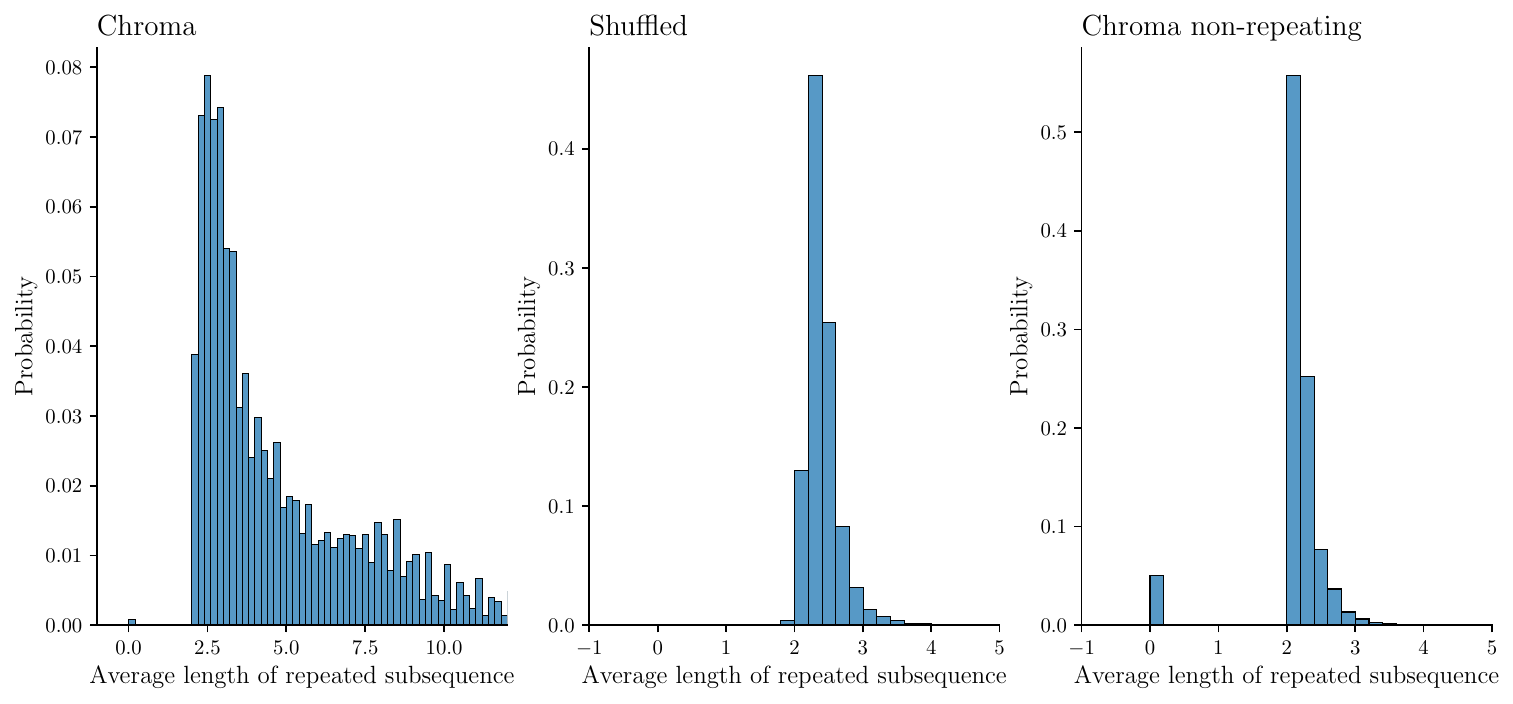}
  \caption{\label{fig:si14}
  Distribution of average length of all repeated subsequences (output of Alg. \ref{alg:alg1}) for
  all $\pC$ sequences, shuffled $\pC$ sequences, and $\pC$ sequences after removing
  repetition with $\LL_{min}=2$ (output of Alg. \ref{alg:alg3}).\\
  This shows that the algorithm to remove repeated substrings produces an output
  with similar levels of repetition as random sequences.
  }
  \end{figure*}

  \begin{figure*}  \centering
  \includegraphics[width=0.70\linewidth]{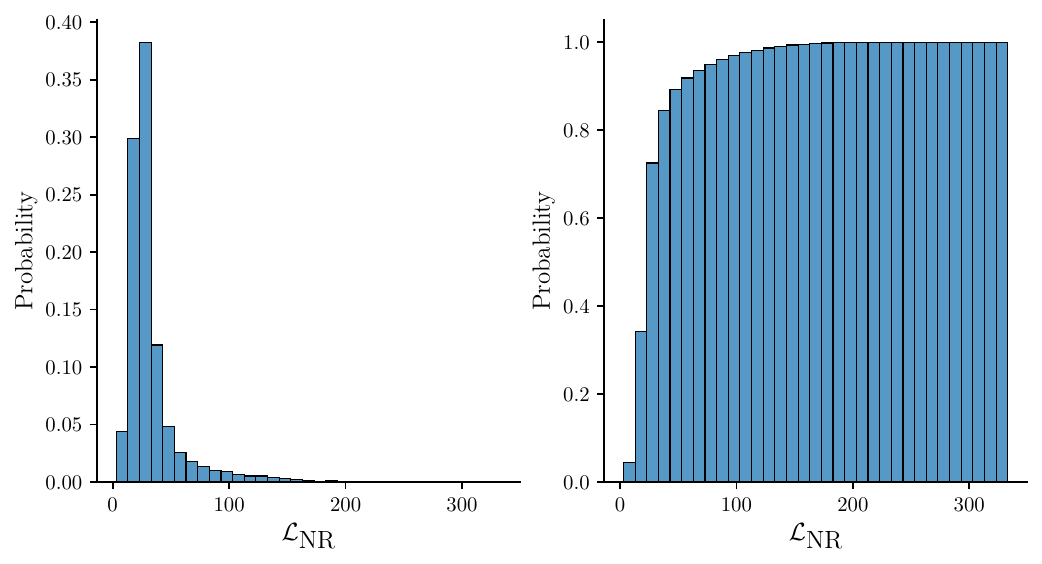}
  \caption{\label{fig:si13}
  Probability distribution (A) and cumulative distribution (B)
  of the length of melodies in Folk corpora after removing repeating
  substrings, $\LN$.
  }
  \end{figure*}

  \clearpage

  \begin{longtable*}{lllcc}
  \caption{Description of melodic corpora.}\label{tab:tab1} \\
  Name & Source & Region / Society & Type & \# Songs \\
  \toprule
  bulg & ABC \cite{abc22} & Bulgaria & Folk & 79 \\
  isra & ABC \cite{abc22} & Israel & Folk & 201 \\
  klez & ABC \cite{abc22} & Klezmer & Folk & 400 \\
  mace & ABC \cite{abc22} & Macedonia & Folk & 58 \\
  roma & ABC \cite{abc22} & Romania & Folk & 76 \\
  serb & ABC \cite{abc22} & Serbia & Folk & 70 \\
  mech & BethNotes \cite{Www2023} & Mexico & Folk & 56 \\
  hawa & BethNotes \cite{Www2023} & Hawai'i & Folk & 22 \\
  noch & Densmore \cite{shapi14} & Nootka, Quileute & Child & 17 \\
  siou & Densmore \cite{shapi14} & Sioux & Folk & 245 \\
  pawn & Densmore \cite{shapi14} & Pawnee & Folk & 86 \\
  acom & Densmore \cite{shapi14} & Acomi, Isleta, Cochiti, Zuni & Folk & 82 \\
  bcol & Densmore \cite{shapi14} & British Colombia area & Folk & 98 \\
  chey & Densmore \cite{shapi14} & Cheyenne, Arapaho & Folk & 72 \\
  chp1 & Densmore \cite{shapi14} & Chippewa & Folk & 186 \\
  chp2 & Densmore \cite{shapi14} & Chippewa & Folk & 179 \\
  choc & Densmore \cite{shapi14} & Choctaw & Folk & 68 \\
  maid & Densmore \cite{shapi14} & Maidu & Folk & 53 \\
  mand & Densmore \cite{shapi14} & Mandan & Folk & 73 \\
  meno & Densmore \cite{shapi14} & Menominee & Folk & 144 \\
  noot & Densmore \cite{shapi14} & Nootka, Quileute & Folk & 116 \\
  nute & Densmore \cite{shapi14} & Northern Ute & Folk & 116 \\
  papa & Densmore \cite{shapi14} & Tohono O'odham & Folk & 168 \\
  semi & Densmore \cite{shapi14} & Seminole & Folk & 247 \\
  yuma & Densmore \cite{shapi14} & Yuman & Folk & 134 \\
  kind & Essen \cite{sch95} & German & Child & 213 \\
  czec & Essen \cite{sch95} & Czechia & Folk & 43 \\
  magy & Essen \cite{sch95} & Hungary & Folk & 45 \\
  nede & Essen \cite{sch95} & Netherlands & Folk & 85 \\
  elsa & Essen \cite{sch95} & Alsace & Folk & 91 \\
  jugo & Essen \cite{sch95} & Yugoslavia & Folk & 115 \\
  schw & Essen \cite{sch95} & Switzerland & Folk & 93 \\
  oest & Essen \cite{sch95} & Austria & Folk & 104 \\
  fink & Essen \cite{sch95} & German & Folk & 566 \\
  erk & Essen \cite{sch95} & German & Folk & 1700 \\
  ald1 & Essen \cite{sch95} & German & Folk & 309 \\
  ald2 & Essen \cite{sch95} & German & Folk & 316 \\
  ball & Essen \cite{sch95} & German & Folk & 687 \\
  alle & Essen \cite{sch95} & German & Folk & 110 \\
  zucc & Essen \cite{sch95} & German & Folk & 616 \\
  han & Essen \cite{sch95} & Han & Folk & 1222 \\
  natm & Essen \cite{sch95} & Natmin & Folk & 206 \\
  shan & Essen \cite{sch95} & Shanxi & Folk & 802 \\
  finn & Finnish \cite{eerolaSuomen2004} & Finland & Folk & 8613 \\
  mass & Josquin \cite{rod22} & Josquin masses & Art & 398 \\
  mote & Josquin \cite{rod22} & Josquin motets & Art & 173 \\
  secu & Josquin \cite{rod22} & Josquin secular & Art & 148 \\
  de\_l & Josquin \cite{rod22} & de la Rue & Art & 185 \\
  mart & Josquin \cite{rod22} & Martini & Art & 123 \\
  ocke & Josquin \cite{rod22} & Ockeghem & Art & 107 \\
  busn & Josquin \cite{rod22} & Busnoys & Art & 66 \\
  de\_o & Josquin \cite{rod22} & de Orto & Art & 43 \\
  bach & KernScores \cite{sapis05} & Bach & Art & 370 \\
  nova & KernScores \cite{sapis05} & Nova Scotia & Folk & 152 \\
  poli & KernScores \cite{sapis05} & Poland & Folk & 900 \\
  lux & KernScores \cite{sapis05} & Luxembourg & Folk & 549 \\
  lorr & KernScores \cite{sapis05} & Lorraine & Folk & 317 \\
  friu & KernScores \cite{sapis05} & Friuli & Folk & 80 \\
  irel & KernScores \cite{sapis05} & Ireland & Folk & 62 \\
  chil & KernScores \cite{sapis05} & England & Folk & 38 \\
  deut & KernScores \cite{sapis05} & Germany & Folk & 201 \\
  kirc & KernScores \cite{sapis05} & Germany & Folk & 1708 \\
  abra & Lieder \cite{gothamScores2018} & Cornelius, Peter & Art & 90 \\
burl & Lieder \cite{gothamScores2018} & Schr\"{o}te, Corona & Art & 25 \\
  butt & Lieder \cite{gothamScores2018} & Warlock, Peter & Art & 22 \\
  warl & Lieder \cite{gothamScores2018} & Faisst, Clara Mathilda & Art & 26 \\
  kral & Lieder \cite{gothamScores2018} & Reichardt, Louise & Art & 43 \\
  reic & Lieder \cite{gothamScores2018} & Brahms, Johannes & Art & 104 \\
  brid & Lieder \cite{gothamScores2018} & Viardot, Pauline & Art & 21 \\
  thys & Lieder \cite{gothamScores2018} & Holm\`{e}s, Augusta Mary Anne & Art & 73 \\
  jaël & Lieder \cite{gothamScores2018} & Chaminade, C\'{e}cile & Art & 31 \\
  webe & Lieder \cite{gothamScores2018} & Lang, Josephine & Art & 52 \\
  par2 & Lieder \cite{gothamScores2018} & Lehmann, Liza & Art & 26 \\
  cord & Lieder \cite{gothamScores2018} & Kinkel, Johanna & Art & 45 \\
  mrtf & Meertens \cite{van19} & Netherlands & Folk & 4120 \\
  mrti & Meertens \cite{van19} & Netherlands & Folk & 2367 \\
  ml11 & MeloSol \cite{bakerMeloSol2021} & Singing & Teaching & 180 \\
  ml12 & MeloSol \cite{bakerMeloSol2021} & Singing & Teaching & 107 \\
  ml13 & MeloSol \cite{bakerMeloSol2021} & Singing & Teaching & 125 \\
  ml14 & MeloSol \cite{bakerMeloSol2021} & Singing & Teaching & 125 \\
  ml15 & MeloSol \cite{bakerMeloSol2021} & Singing & Teaching & 89 \\
  moza & Mozart Opera \cite{kunst} & Mozart & Art & 82 \\
  afr1 & South Africa \cite{eerolaPerceived2006} & South Africa & Folk & 90 \\
  symbC & SymbTr \cite{karaosmanogluTurkish2012} & Turkey & Art & 1713 \\
  symbF & SymbTr \cite{karaosmanogluTurkish2012} & Turkey & Folk & 309 \\
  slav & \cite{allenSlave1867} & African-American & Folk & 135 \\
  vend & \cite{blackingVenda1967} & Venda & Child & 60 \\
  kor1 & \cite{chunhyangka77} & Korea & Folk & 46 \\
  kor2 & \cite{folksongkorea69} & Korea & Folk & 50 \\
  ives & \cite{ives1141922} & Ives, Charles & Art & 31 \\
  jach & \cite{lewinRock2000} & Jamaica & Child & 17 \\
  jama & \cite{lewinRock2000} & Jamaica & Folk & 57 \\
  engc & \cite{mof04} & England & Child & 50 \\
  mexc & \cite{mon07} & Mexico & Child & 28 \\
  okic & \cite{nishikawaCultural2022} & Okinawa & Child & 49 \\
  okif & \cite{nishikawaCultural2022} & Okinawa & Folk & 179 \\
  ghan & \cite{nke63} & Ghana & Folk & 58 \\
  kyrg & \cite{sipos14} & Kyrgyzstan & Folk & 85 \\
  hens & van Handel \cite{vanjn10} & Hensel, Fanny (Mendelssohn) & Art & 50 \\
  wolf & van Handel \cite{vanjn10} & Wolf, Hugo & Art & 82 \\
  stra & van Handel \cite{vanjn10} & Strauss, Robert & Art & 33 \\
  sch2 & van Handel \cite{vanjn10} & Schubert, Franz & Art & 121 \\
  mend & van Handel \cite{vanjn10} & Mendelssohn, Felix & Art & 56 \\
  sch3 & van Handel \cite{vanjn10} & Schumann, Robert & Art & 123 \\
  fran & van Handel \cite{vanjn10} & Franz, Robert & Art & 61 \\
  faur & van Handel \cite{vanjn10} & Faur\'{e}, Gabriel & Art & 64 \\
  debu & van Handel \cite{vanjn10} & Debussy, Claude & Art & 33 \\
  goun & van Handel \cite{vanjn10} & Gounod, Charles & Art & 51 \\
  rebe & van Handel \cite{vanjn10} & Reber, Napol\'{e}on Henri & Art & 30 \\
  chau & van Handel \cite{vanjn10} & Chausson, Ernest & Art & 30 \\
  dav2 & van Handel \cite{vanjn10} & David, F\'{e}licien & Art & 33 \\
  mas2 & van Handel \cite{vanjn10} & Massenet, Jules & Art & 39 \\
  bize & van Handel \cite{vanjn10} & Bizet, Georges & Art & 24 \\
  \bottomrule
  \end{longtable*}
  \clearpage

  \begin{longtable*}{lcccccc}
  \caption{Average information properties of melodic corpora.}\label{tab:tab2} \\
  Name & H\_chroma & H\_dur & H\_chroma\_dur & Length & Length (no repeat) & Total Info \\
  \toprule
  jama & 2.47 & 1.73 & 3.74 & 54 & 34 & 130.5 \\
  siou & 2.08 & 2.16 & 3.83 & 65 & 41 & 161.0 \\
  pawn & 1.89 & 1.77 & 3.29 & 56 & 29 & 102.4 \\
  bach & 2.62 & 1.20 & 3.51 & 49 & 34 & 121.7 \\
  nova & 2.37 & 1.44 & 3.49 & 56 & 35 & 127.6 \\
  mrtf & 2.49 & 1.51 & 3.60 & 52 & 34 & 125.7 \\
  mrti & 2.65 & 1.55 & 3.88 & 75 & 46 & 183.8 \\
  poli & 2.61 & 1.16 & 3.38 & 39 & 25 &  87.7 \\
  lux & 2.42 & 1.27 & 3.37 & 52 & 31 & 110.9 \\
  lorr & 2.47 & 1.37 & 3.50 & 47 & 32 & 116.0 \\
  friu & 2.38 & 1.56 & 3.43 & 43 & 24 &  84.3 \\
  irel & 2.53 & 1.52 & 3.74 & 78 & 42 & 162.5 \\
  chil & 2.43 & 1.51 & 3.51 & 48 & 30 & 110.5 \\
  deut & 2.62 & 1.53 & 3.79 & 63 & 41 & 160.5 \\
  kirc & 2.60 & 1.28 & 3.53 & 45 & 31 & 111.3 \\
  czec & 2.46 & 1.19 & 3.27 & 31 & 22 &  76.7 \\
  magy & 2.39 & 1.21 & 3.24 & 35 & 25 &  84.8 \\
  nede & 2.48 & 1.45 & 3.57 & 44 & 30 & 111.9 \\
  elsa & 2.47 & 1.43 & 3.51 & 49 & 29 & 107.4 \\
  jugo & 2.24 & 1.29 & 3.10 & 22 & 17 &  55.3 \\
  schw & 2.46 & 1.38 & 3.48 & 49 & 30 & 109.4 \\
  oest & 2.50 & 1.28 & 3.43 & 51 & 30 & 108.6 \\
  fink & 2.60 & 1.49 & 3.73 & 58 & 37 & 141.6 \\
  erk & 2.45 & 1.29 & 3.40 & 46 & 28 &  99.1 \\
  ald1 & 2.49 & 1.36 & 3.52 & 48 & 32 & 116.2 \\
  ald2 & 2.49 & 1.24 & 3.43 & 48 & 31 & 110.2 \\
  ball & 2.46 & 1.32 & 3.43 & 40 & 28 &  99.5 \\
  alle & 2.49 & 1.39 & 3.53 & 53 & 32 & 119.4 \\
  zucc & 2.49 & 1.32 & 3.48 & 51 & 31 & 113.0 \\
  kind & 2.13 & 1.00 & 2.82 & 39 & 21 &  61.5 \\
  han & 2.24 & 1.62 & 3.52 & 73 & 42 & 151.8 \\
  natm & 2.26 & 1.75 & 3.62 & 68 & 39 & 147.8 \\
  shan & 2.33 & 1.42 & 3.38 & 47 & 30 & 105.3 \\
  ml11 & 2.59 & 1.30 & 3.49 & 34 & 25 &  90.3 \\
  ml12 & 2.74 & 1.34 & 3.68 & 44 & 32 & 120.8 \\
  ml13 & 2.92 & 1.27 & 3.82 & 54 & 38 & 150.3 \\
  ml14 & 3.03 & 1.47 & 4.03 & 53 & 39 & 161.9 \\
  ml15 & 3.32 & 1.27 & 4.07 & 42 & 36 & 149.8 \\
  ml51 & 2.49 & 1.42 & 3.51 & 46 & 29 & 104.1 \\
  ml52 & 2.54 & 1.70 & 3.78 & 48 & 31 & 121.4 \\
  ml53 & 2.72 & 1.55 & 3.75 & 52 & 31 & 122.6 \\
  ml54 & 3.04 & 1.70 & 4.13 & 53 & 38 & 159.7 \\
  mass & 2.82 & 2.25 & 4.87 & 312 & 194 & 955.3 \\
  mote & 2.79 & 2.27 & 4.86 & 340 & 201 & 990.8 \\
  secu & 2.78 & 2.12 & 4.57 & 127 & 84 & 391.0 \\
  de\_l & 2.83 & 2.22 & 4.86 & 310 & 191 & 940.8 \\
  mart & 2.46 & 1.89 & 4.12 & 269 & 130 & 584.3 \\
  ocke & 2.85 & 2.22 & 4.84 & 288 & 173 & 853.9 \\
  busn & 2.87 & 2.27 & 4.85 & 154 & 113 & 557.4 \\
  de\_o & 2.84 & 2.24 & 4.87 & 287 & 179 & 885.1 \\
  bulg & 2.55 & 1.32 & 3.52 & 110 & 46 & 170.7 \\
  isra & 2.59 & 1.47 & 3.71 & 79 & 44 & 170.7 \\
  klez & 2.68 & 1.56 & 3.89 & 101 & 56 & 231.2 \\
  mace & 2.62 & 1.32 & 3.58 & 86 & 41 & 155.5 \\
  roma & 2.63 & 1.42 & 3.62 & 93 & 40 & 150.6 \\
  serb & 2.53 & 1.31 & 3.48 & 86 & 40 & 150.3 \\
  moza & 2.90 & 2.15 & 4.68 & 274 & 155 & 781.7 \\
  symbF & 2.64 & 1.68 & 4.00 & 394 & 98 & 416.9 \\
  symbC & 2.92 & 1.81 & 4.45 & 347 & 128 & 594.7 \\
  acom & 1.98 & 1.69 & 3.41 & 129 & 61 & 219.9 \\
  bcol & 1.99 & 2.04 & 3.61 & 56 & 35 & 129.3 \\
  chey & 1.94 & 1.74 & 3.36 & 69 & 36 & 126.1 \\
  chp1 & 2.06 & 1.56 & 3.30 & 47 & 28 & 103.8 \\
  chp2 & 2.00 & 2.16 & 3.72 & 52 & 33 & 126.7 \\
  choc & 1.73 & 1.56 & 2.97 & 98 & 34 & 105.9 \\
  maid & 1.87 & 1.50 & 3.03 & 69 & 28 &  90.3 \\
  mand & 1.99 & 2.07 & 3.61 & 53 & 33 & 124.1 \\
  meno & 2.00 & 1.82 & 3.45 & 52 & 30 & 110.2 \\
  noot & 1.78 & 2.04 & 3.42 & 57 & 35 & 126.8 \\
  nute & 1.89 & 2.08 & 3.53 & 53 & 32 & 118.2 \\
  papa & 2.04 & 1.99 & 3.62 & 54 & 35 & 132.0 \\
  semi & 1.95 & 1.97 & 3.48 & 56 & 32 & 117.3 \\
  yuma & 1.71 & 1.62 & 3.05 & 98 & 39 & 123.6 \\
  afr1 & 2.31 & 1.67 & 3.40 & 50 & 25 &  88.7 \\
  abra & 2.97 & 1.91 & 4.37 & 112 & 71 & 328.1 \\
  burl & 2.77 & 1.37 & 3.83 & 55 & 42 & 165.3 \\
  butt & 2.93 & 1.91 & 4.39 & 132 & 74 & 332.5 \\
  warl & 3.10 & 2.02 & 4.68 & 124 & 82 & 396.6 \\
  kral & 2.73 & 1.61 & 3.95 & 104 & 49 & 203.8 \\
  reic & 3.05 & 1.65 & 4.34 & 148 & 76 & 344.8 \\
  brid & 3.02 & 2.00 & 4.68 & 213 & 120 & 583.3 \\
  thys & 2.90 & 2.25 & 4.69 & 224 & 101 & 497.5 \\
  jaël & 2.89 & 1.70 & 4.24 & 193 & 82 & 356.6 \\
  webe & 2.92 & 2.09 & 4.57 & 175 & 79 & 371.4 \\
  par2 & 2.78 & 1.96 & 4.23 & 112 & 62 & 269.7 \\
  cord & 2.98 & 1.97 & 4.47 & 193 & 64 & 294.5 \\
  finn & 2.45 & 1.23 & 3.31 & 55 & 27 &  93.5 \\
  hens & 3.13 & 1.70 & 4.48 & 137 & 83 & 380.9 \\
  wolf & 3.20 & 1.88 & 4.70 & 143 & 100 & 486.6 \\
  stra & 3.20 & 2.14 & 4.90 & 135 & 105 & 522.0 \\
  sch2 & 2.97 & 1.90 & 4.51 & 187 & 97 & 456.7 \\
  mend & 2.88 & 1.78 & 4.27 & 125 & 71 & 311.6 \\
  sch3 & 2.93 & 1.91 & 4.44 & 138 & 79 & 361.9 \\
  fran & 2.91 & 1.65 & 4.17 & 105 & 55 & 237.6 \\
  faur & 3.13 & 2.08 & 4.80 & 126 & 93 & 460.8 \\
  debu & 3.31 & 2.03 & 4.97 & 167 & 125 & 635.9 \\
  goun & 2.93 & 1.72 & 4.34 & 154 & 88 & 396.7 \\
  rebe & 2.68 & 1.93 & 4.24 & 110 & 71 & 314.3 \\
  chau & 3.22 & 2.22 & 4.98 & 143 & 114 & 579.4 \\
  dav2 & 2.84 & 1.86 & 4.28 & 88 & 60 & 264.6 \\
  mas2 & 2.95 & 1.93 & 4.44 & 117 & 78 & 365.6 \\
  bize & 3.12 & 2.07 & 4.82 & 160 & 113 & 565.1 \\
  vend & 1.97 & 0.73 & 2.40 & 51 & 21 &  55.1 \\
  ghan & 2.45 & 1.48 & 3.56 & 70 & 43 & 161.3 \\
  slav & 2.36 & 1.54 & 3.53 & 48 & 33 & 120.5 \\
  noch & 1.86 & 1.91 & 3.39 & 53 & 32 & 113.7 \\
  jach & 2.40 & 1.41 & 3.44 & 44 & 29 & 105.2 \\
  mech & 2.47 & 1.21 & 3.35 & 64 & 32 & 114.7 \\
  hawa & 2.32 & 1.60 & 3.40 & 52 & 32 & 114.5 \\
  ives & 3.13 & 2.24 & 4.68 & 92 & 73 & 362.2 \\
  kor1 & 2.25 & 1.78 & 3.86 & 232 & 130 & 511.9 \\
  kor2 & 2.19 & 1.90 & 3.77 & 112 & 58 & 225.5 \\
  kyrg & 2.14 & 1.47 & 3.16 & 33 & 24 &  82.1 \\
  okif & 2.35 & 1.57 & 3.57 & 62 & 41 & 152.4 \\
  okic & 1.95 & 1.22 & 2.81 & 52 & 26 &  79.2 \\
  mexc & 2.10 & 1.20 & 2.89 & 38 & 19 &  58.4 \\
  engc & 2.51 & 1.24 & 3.42 & 51 & 30 & 105.8 \\
  \bottomrule
  \end{longtable*}

\bibliographystyle{unsrtnat}
\bibliography{master}